\documentclass[11pt,a4paper]{article}
\usepackage{jheppub}

\title{PDF dependence of Higgs cross sections at the Tevatron and LHC: response to recent criticism}

\author[a]{R.~S.~Thorne}
\author[b]{and G.~Watt}

\affiliation[a]{Department of Physics and Astronomy, University College London, WC1E 6BT, UK}
\affiliation[b]{Theory Group, Physics Department, CERN, CH-1211 Geneva 23, Switzerland}

\emailAdd{thorne@hep.ucl.ac.uk}
\emailAdd{Graeme.Watt@cern.ch}

\abstract{We respond to some criticism questioning the validity of the current Standard Model Higgs exclusion limits at the Tevatron, due to the significant dependence of the dominant production cross section from gluon--gluon fusion on the choice of parton distribution functions (PDFs) and the strong coupling ($\alpha_S$).  We demonstrate the ability of the Tevatron jet data to discriminate between different high-$x$ gluon distributions, performing a detailed quantitative comparison to show that fits not explicitly including these data fail to give a good description.  In this context we emphasise the importance of the consistent treatment of luminosity uncertainties.  We comment on the values of $\alpha_S$ obtained from fitting deep-inelastic scattering data, particularly the fixed-target NMC data, and we show that jet data are needed for stability.  We conclude that the Higgs cross-section uncertainties due to PDFs and $\alpha_S$ currently used by the Tevatron and LHC experiments are not significantly underestimated, contrary to some recent claims.}


\arxivnumber{1106.5789}

\begin{document}

\begin{flushright}
  CERN-PH-TH/2011-150 \\
  LCTS/2011-07 \\
  25th July 2011
\end{flushright}

\maketitle

\section{Introduction} \label{sec:introduction}

Discovery or exclusion of the Standard Model Higgs boson ($H$) at the Tevatron and Large Hadron Collider (LHC) requires precise knowledge of the theoretical cross section; see, for example, refs.~\cite{Anastasiou:2008tj,deFlorian:2009hc,LHCHiggsCrossSectionWorkingGroup:2011ti}, and references therein.  Cross-section predictions for the dominant production channel of gluon--gluon fusion ($gg\to H$)  are strongly dependent on both the gluon distribution in the proton and the strong coupling $\alpha_S$, which enters squared at leading-order (LO) with sizeable next-to-leading order (NLO) and next-to-next-to-leading order (NNLO) corrections.  In particular, the Tevatron Higgs analysis~\cite{CDF:2010ar,CDF:2011gs}, with current exclusion at 95\% confidence-level (C.L.) for a Standard Model Higgs boson mass $M_H\in[158,173]$~GeV~\cite{CDF:2011gs}, requires knowledge of the gluon distribution at relatively large momentum fractions $x\gtrsim 0.1$ where constraints from data on deep-inelastic scattering (DIS) or Drell--Yan production are fairly weak.  In this paper, which accompanies a separate paper~\cite{bench7TeV}, we respond to several (related) issues which have been raised in recent months~\cite{Baglio:2010um,Baglio:2010ae,Baglio:2011wn,Alekhin:2010dd,Alekhin:2011ey,Alekhin:2011cf}, particularly regarding the use of parton distribution functions (PDFs) determined from limited data sets in making predictions for the Tevatron (and LHC) Higgs cross sections, as alternatives to the most common choice of the MSTW 2008 PDFs~\cite{Martin:2009iq} used in the Tevatron~\cite{CDF:2010ar,CDF:2011gs} and LHC~\cite{LHCHiggsCrossSectionWorkingGroup:2011ti} Higgs analyses.

First in section~\ref{sec:dependence} we demonstrate explicitly how the $gg\to H$ cross sections depend on the Standard Model Higgs boson mass $M_H$, the gluon--gluon luminosity function and the choice of $\alpha_S(M_Z^2)$, by comparing predictions obtained using PDFs (and $\alpha_S$ values) from various different PDF fitting groups.  In section~\ref{sec:tevjets} we present a detailed quantitative comparison of the quality of the description of Tevatron jet data using different PDF sets.  The MSTW 2008 analysis~\cite{Martin:2009iq} is the only current NNLO PDF fit which includes the Tevatron jet data, providing the only direct constraint on the high-$x$ gluon distribution.  In section~\ref{sec:alphaS} we examine the different values of the strong coupling $\alpha_S$ used by the different PDF groups, particularly those values mainly extracted from DIS data, and we look at the constraints arising from different sources.  In section~\ref{sec:nmcdata} we respond to recent claims~\cite{Alekhin:2011ey} that the theoretical treatment of the longitudinal structure function $F_L$ for the NMC data~\cite{Arneodo:1996qe} can explain the bulk of the difference between predictions for Higgs cross sections calculated using either the MSTW08~\cite{Martin:2009iq} or ABKM09~\cite{Alekhin:2009ni} PDFs.  Finally we conclude in section~\ref{sec:conclusions} that MSTW08 is presently the only fully reliable PDF set for calculating Higgs cross sections at NNLO, particularly if sensitive to the high-$x$ gluon distribution, and that the recent exclusion bounds~\cite{CDF:2010ar,CDF:2011gs} obtained by the Tevatron experiments are robust based upon this choice.

\section{Dependence \texorpdfstring{of Higgs cross sections on PDFs and $\alpha_S$}{on PDFs and alphaS}} \label{sec:dependence}

\subsection{Dependence on Higgs mass}
\begin{figure}
  \centering
  \begin{minipage}{0.5\textwidth}
    (a)\\
    \includegraphics[width=\textwidth]{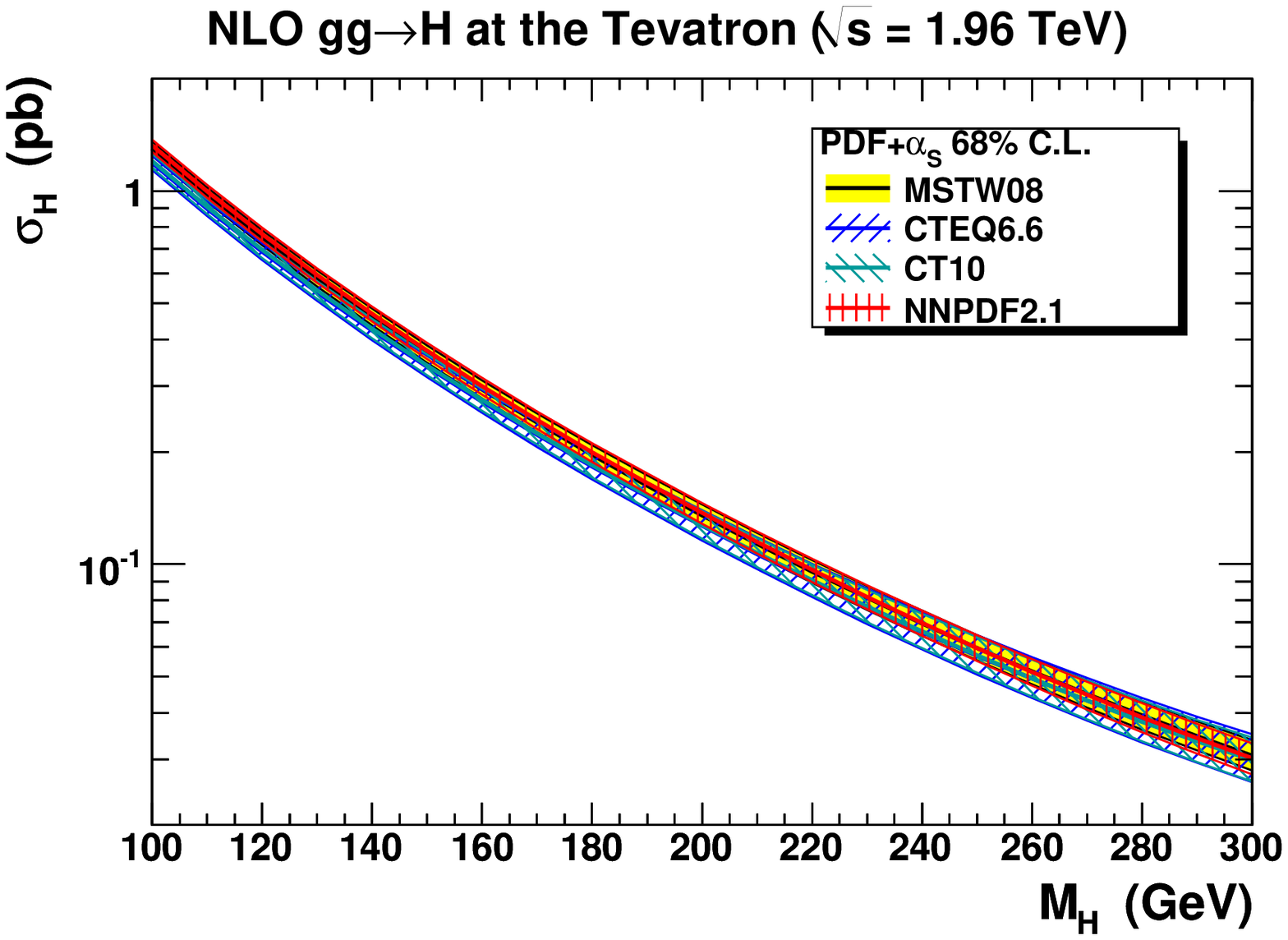}
  \end{minipage}%
  \begin{minipage}{0.5\textwidth}
    (b)\\
    \includegraphics[width=\textwidth]{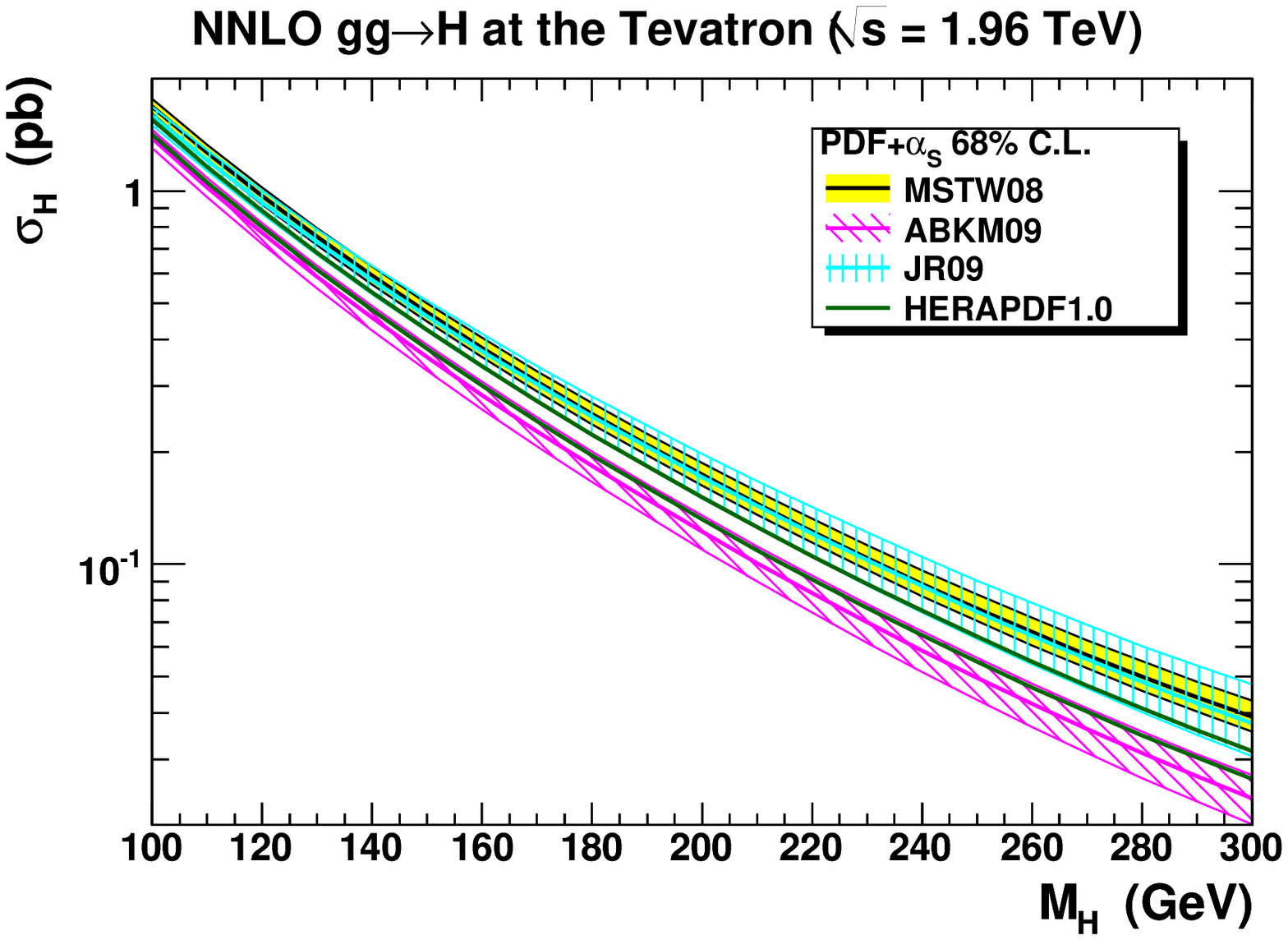}
  \end{minipage}
  \begin{minipage}{0.5\textwidth}
    (c)\\
    \includegraphics[width=\textwidth]{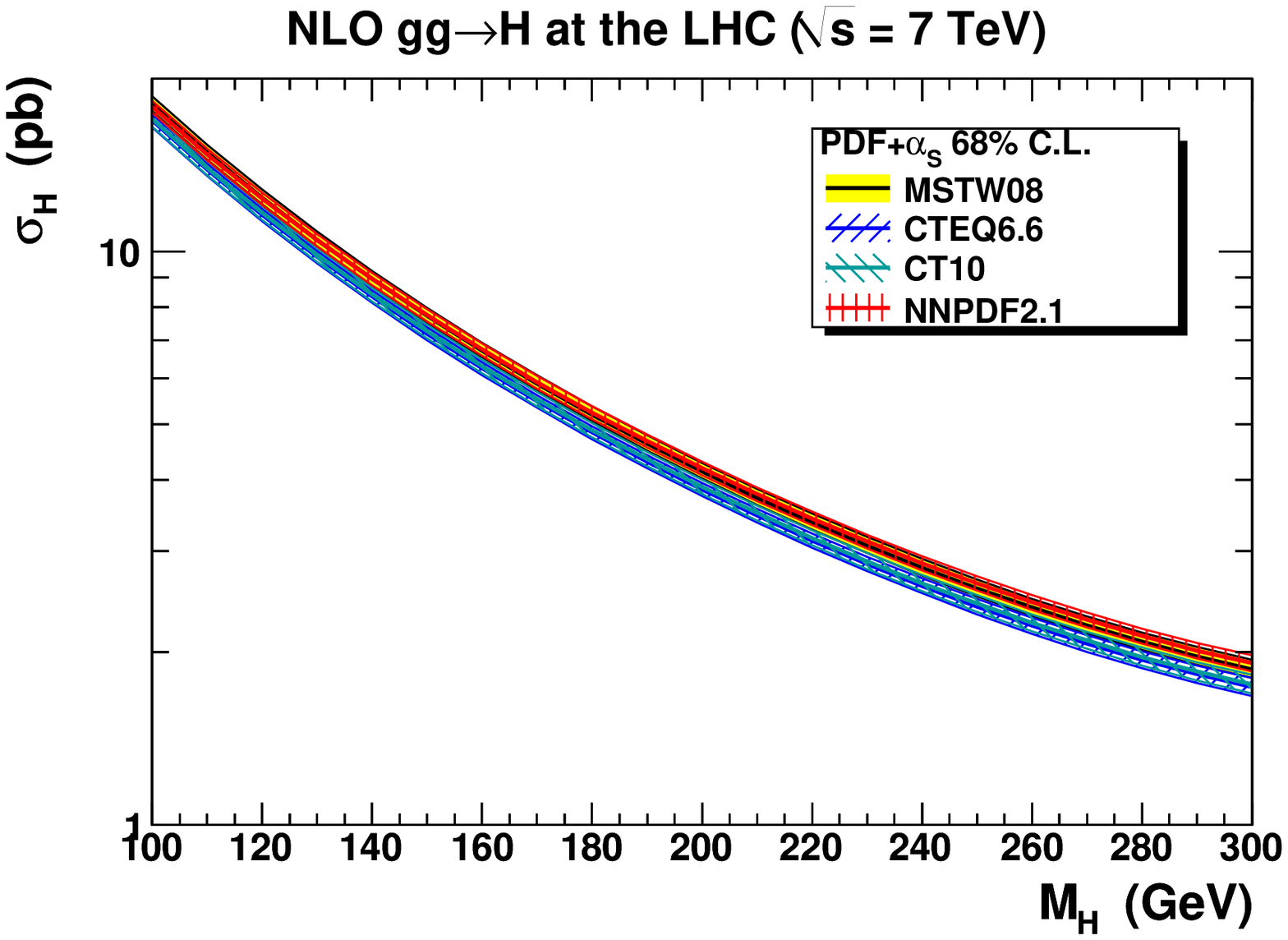}
  \end{minipage}%
  \begin{minipage}{0.5\textwidth}
    (d)\\
    \includegraphics[width=\textwidth]{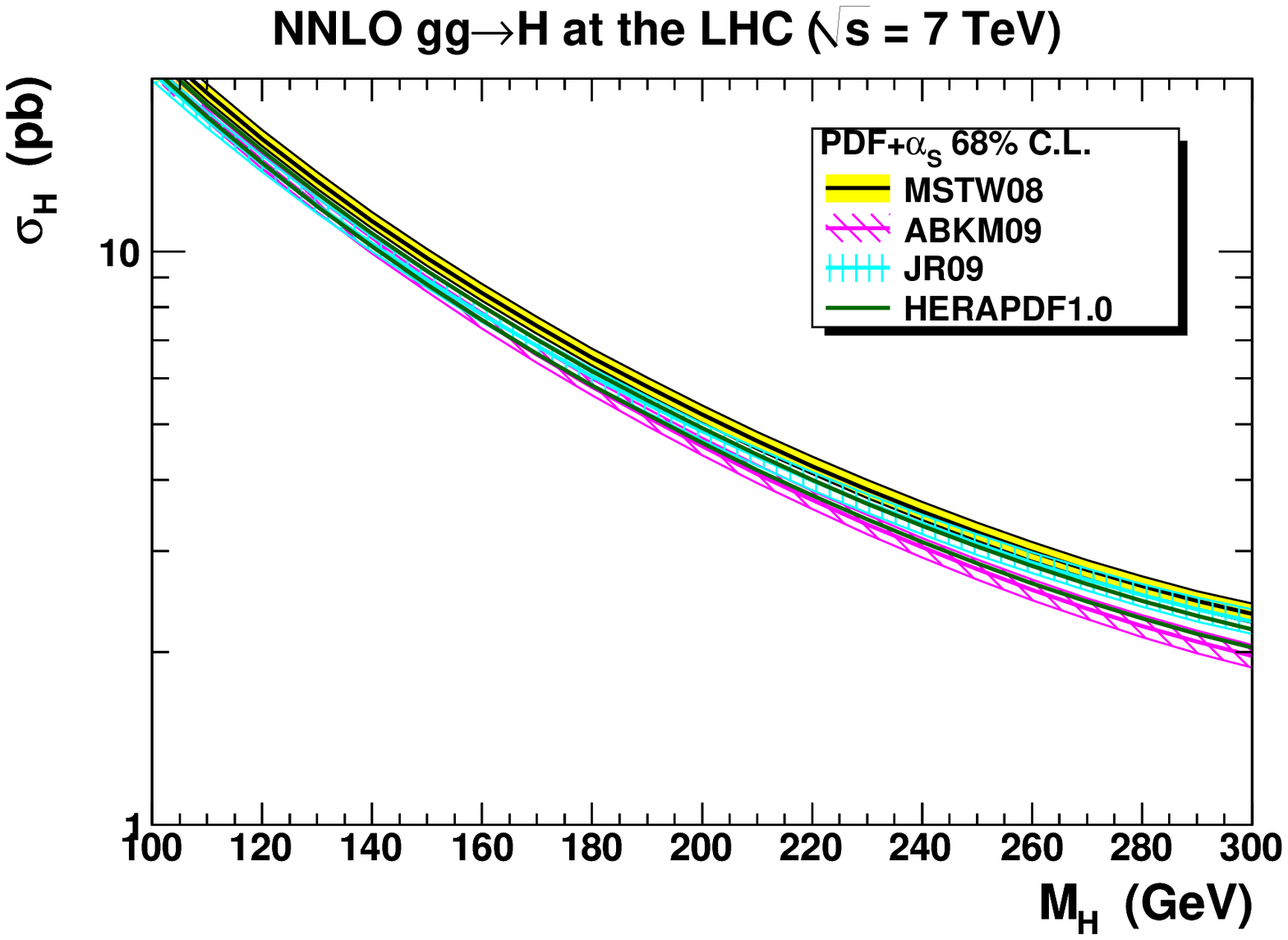}
  \end{minipage}
  \caption{$\sigma_H$ vs.~$M_H$ with PDF+$\alpha_S$ uncertainties at 68\%~C.L.~for $gg\to H$ calculated at (a)~NLO at the Tevatron, (b)~NNLO at the Tevatron, (c)~NLO at the LHC, and (d)~NNLO at the LHC.}
  \label{fig:sigmaH}
\end{figure}
We show the NLO and NNLO $gg\to H$ total cross sections ($\sigma_H$) versus the Standard Model Higgs boson mass $M_H$ in figure~\ref{fig:sigmaH} at the Tevatron (centre-of-mass energy, $\sqrt{s} = 1.96$~TeV) and the LHC ($\sqrt{s} = 7$~TeV) for different PDF sets and a fixed scale choice of $\mu_R=\mu_F=M_H$, calculated with settings given in section 4.2 of ref.~\cite{bench7TeV}.  At NLO~\cite{Djouadi:1991tka}, we use the corresponding NLO PDFs (and $\alpha_S$ values) from MSTW08~\cite{Martin:2009iq}, CTEQ6.6~\cite{Nadolsky:2008zw}, CT10~\cite{Lai:2010vv} and NNPDF2.1~\cite{Ball:2011mu}, all of which are fully \emph{global} fits to HERA and fixed-target DIS data, fixed-target Drell--Yan production, and Tevatron data on vector boson and jet production.  At NNLO~\cite{Harlander:2002wh}, we use the corresponding NNLO PDFs (and $\alpha_S$ values) from MSTW08~\cite{Martin:2009iq}, ABKM09~\cite{Alekhin:2009ni}, JR09~\cite{JimenezDelgado:2008hf,JimenezDelgado:2009tv} and HERAPDF1.0~\cite{HERA:2009wt}, where in the last case no uncertainty PDF sets are provided and the two curves correspond to $\alpha_S(M_Z^2) = 0.1145$ and $\alpha_S(M_Z^2) = 0.1176$, with the larger $\alpha_S$ value giving the larger Higgs cross section.  For the other PDF sets, we compute the ``PDF+$\alpha_S$'' uncertainty at 68\% C.L.~according to the recommended prescription of each group, summarised in ref.~\cite{bench7TeV}.  The data sets included in the MSTW08 fit at NNLO are the same as at NLO, with the omission of HERA data on jet production, while the ABKM09 and JR09 fits only include DIS and fixed-target Drell--Yan data.  The HERAPDF1.0 fit \emph{only} includes combined HERA I inclusive DIS data, while the other NNLO fits (MSTW08, ABKM09, JR09) instead include the older separate data from H1 and ZEUS.  However, including the combined HERA I data~\cite{HERA:2009wt} in a variant of the MSTW08 fit was found to have little effect on predictions for Higgs cross sections~\cite{Thorne:2010kj}.  The NNPDF fits parameterise the starting distributions at $Q_0^2=2$~GeV$^2$ as neural networks, whereas other groups all use the more traditional approach of parameterising the input PDFs as some functional form in $x$, each with a number of free parameters, which varies significantly between groups.  Contrary to the ``standard'' input parameterisation at $Q_0^2\ge1$~GeV$^2$, the JR09 set uses a ``dynamical'' parameterisation of valence-like input distributions at an optimally chosen $Q_0^2<1$~GeV$^2$, which gives a slightly worse fit quality and lower $\alpha_S$ values than the corresponding ``standard'' parameterisation, but is nevertheless favoured by the JR09 authors.  More details on differences between PDF sets are given in section 2 of ref.~\cite{bench7TeV}; see also the descriptions in refs.~\cite{Forte:2010dt,Alekhin:2011sk,DeRoeck:2011na}.

The size of the higher-order corrections to the $gg\to H$ total cross sections is substantial.  Taking the appropriate MSTW08 PDFs and $\alpha_S$ values consistently at each perturbative order for $\sigma_H$ with $M_H=160$~GeV, then the NLO/LO ratio is 2.1 (Tevatron) or 1.9 (LHC), the NNLO/LO ratio is 2.7 (Tevatron) or 2.4 (LHC), and so the NNLO/NLO ratio is 1.3 (Tevatron and LHC).  The perturbative series is therefore slowly convergent, mandating the use of (at least) NNLO calculations together with the corresponding NNLO PDFs and $\alpha_S$ values.  The convergence can be improved by using a scale choice $\mu_R=\mu_F=M_H/2$, which mimics the effect of soft-gluon resummation.  However, the goal of this paper is to study only the PDF and $\alpha_S$ dependence of the $gg\to H$ cross sections, and we do \emph{not} aim to come up with a single ``best'' prediction together with a complete evaluation of all sources of theoretical uncertainty.  We do \emph{not} consider, for example, optimal (factorisation and renormalisation) scale choices and variations, electroweak corrections, the effect of threshold resummation, $(C_A\,\pi\,\alpha_S)^n$-enhanced terms, use of a finite top-quark mass in the calculation of higher-order corrections, bottom-quark loop contributions, etc.  The PDF and $\alpha_S$ dependence roughly decouples from these other, more refined, aspects of the calculation, and therefore the findings regarding PDFs and $\alpha_S$ reported here will be relevant also for more complete calculations found, for example, in refs.~\cite{Anastasiou:2008tj,deFlorian:2009hc} or the recent \emph{Handbook of LHC Higgs Cross Sections}~\cite{LHCHiggsCrossSectionWorkingGroup:2011ti}.

\begin{figure}
  \centering
  \begin{minipage}{0.5\textwidth}
    (a)\\
    \includegraphics[width=\textwidth]{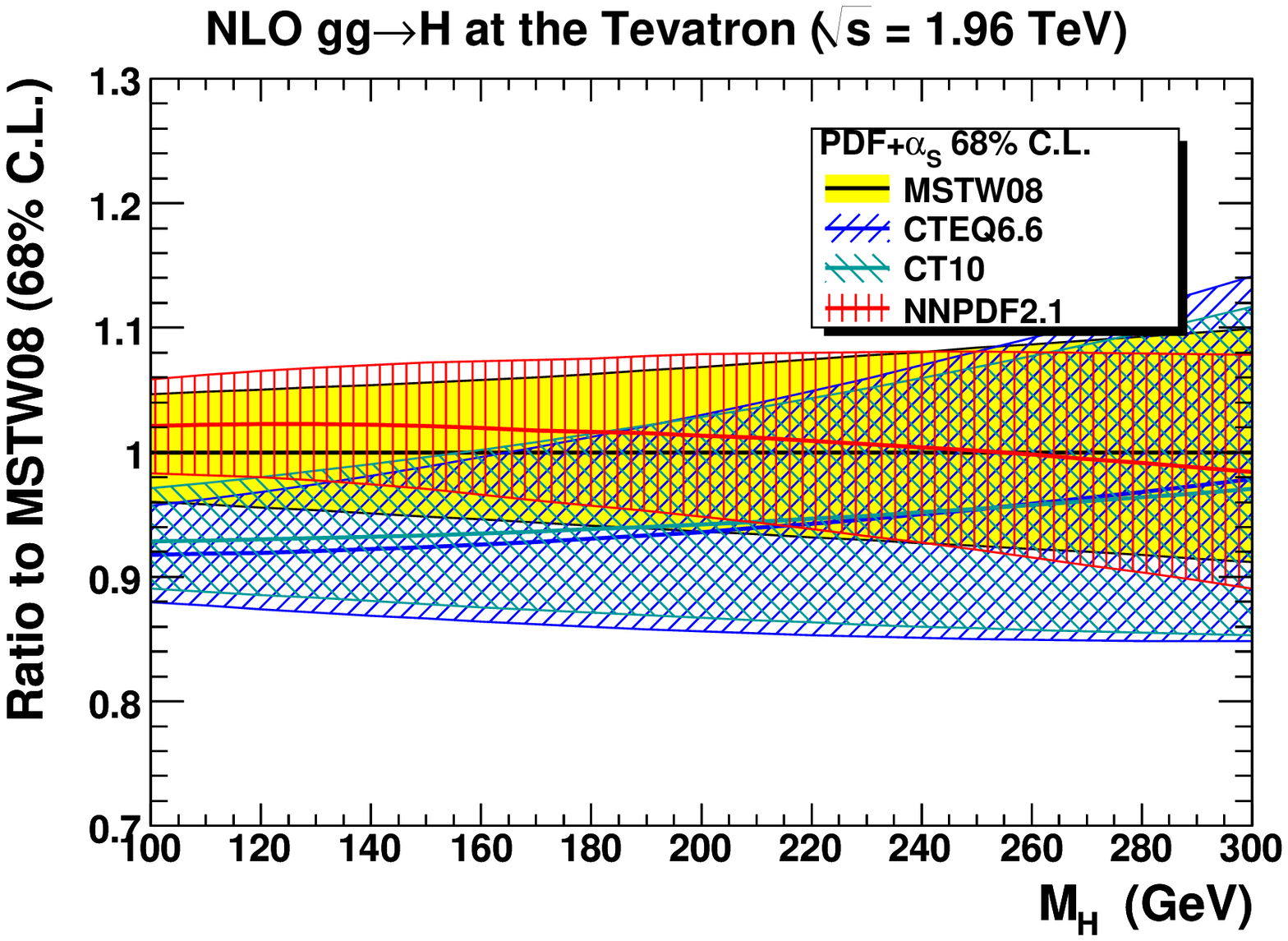}
  \end{minipage}%
  \begin{minipage}{0.5\textwidth}
    (b)\\
    \includegraphics[width=\textwidth]{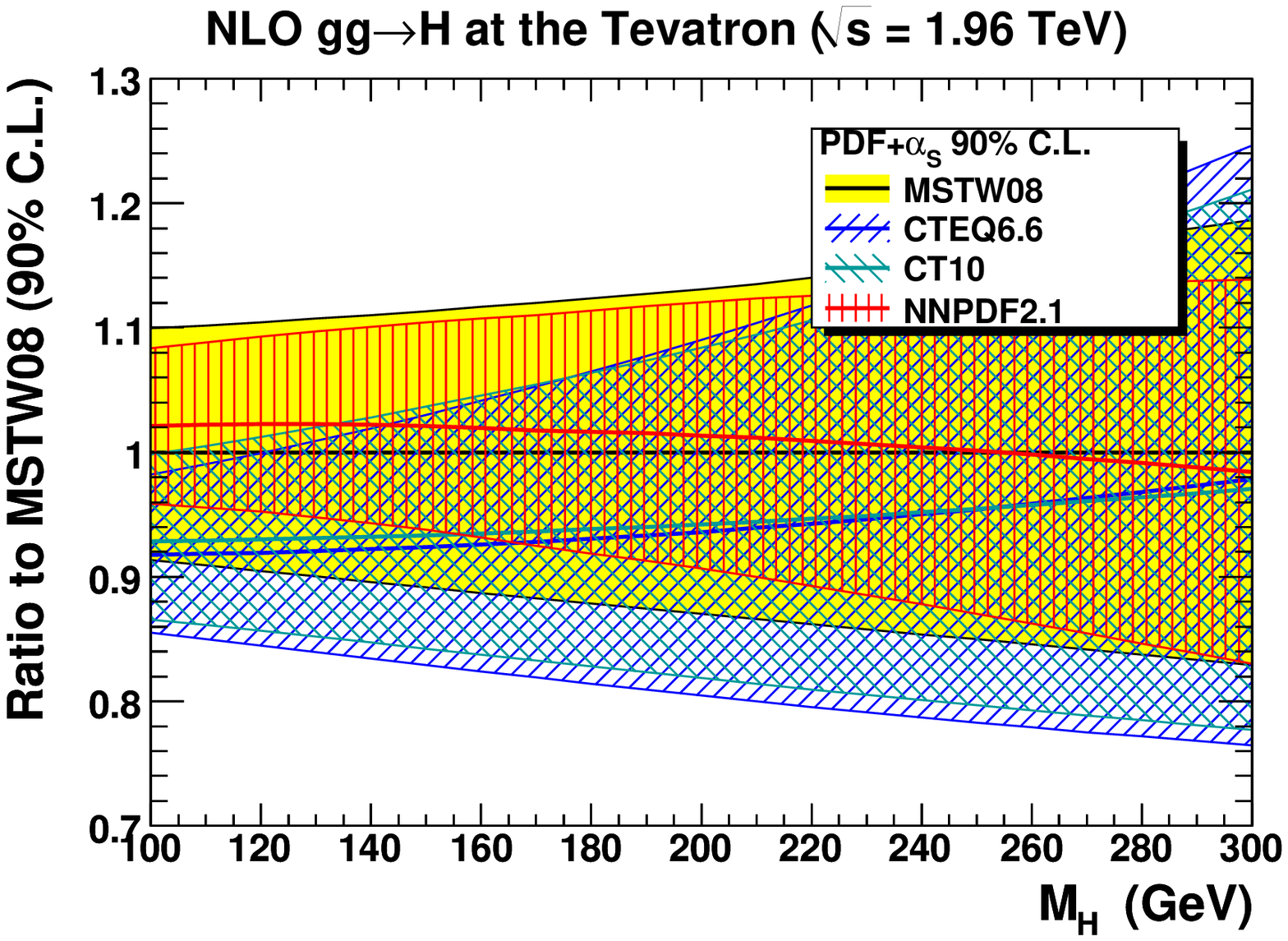}
  \end{minipage}
  \begin{minipage}{0.5\textwidth}
    (c)\\
    \includegraphics[width=\textwidth]{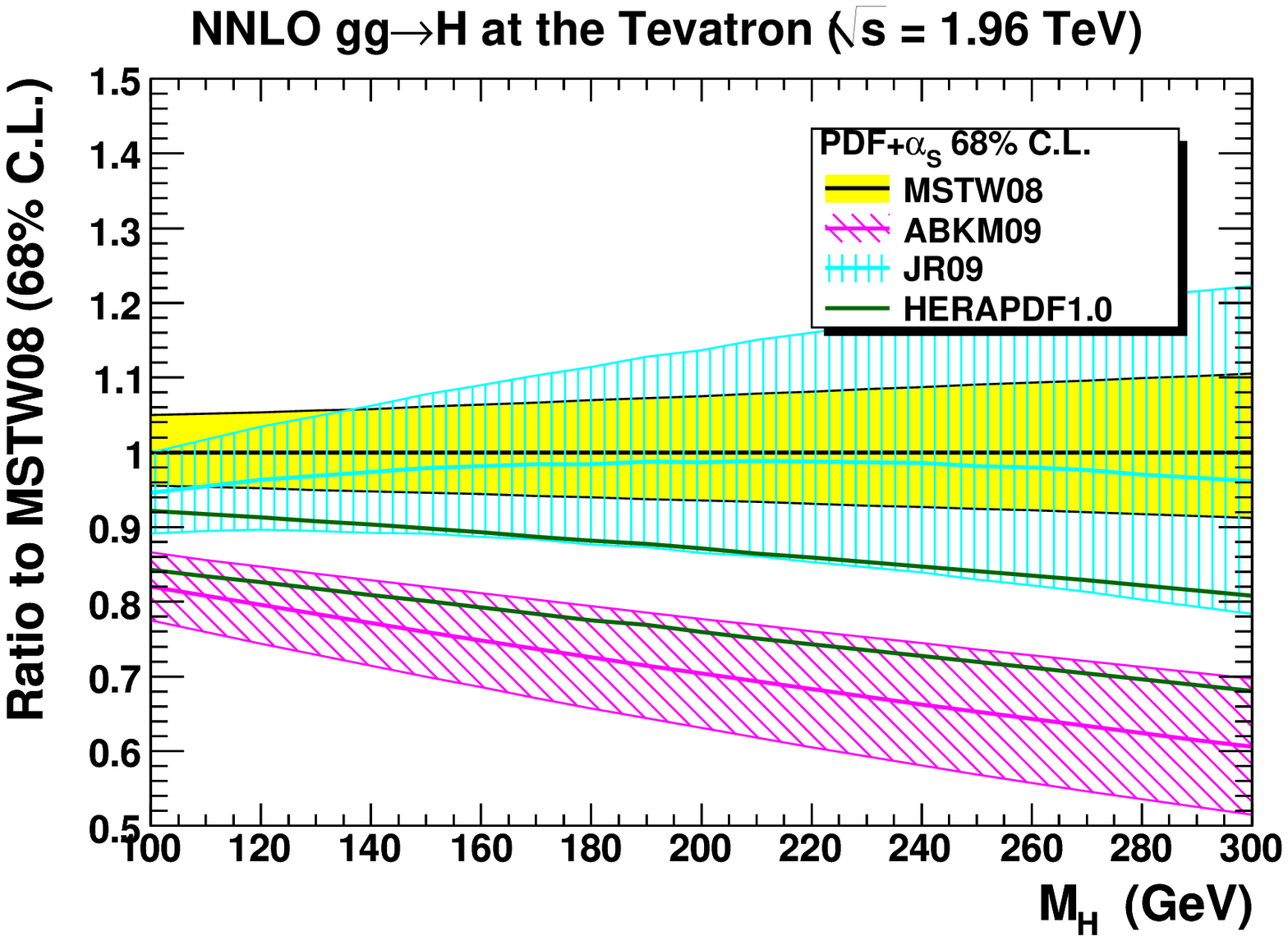}
  \end{minipage}%
  \begin{minipage}{0.5\textwidth}
    (d)\\
    \includegraphics[width=\textwidth]{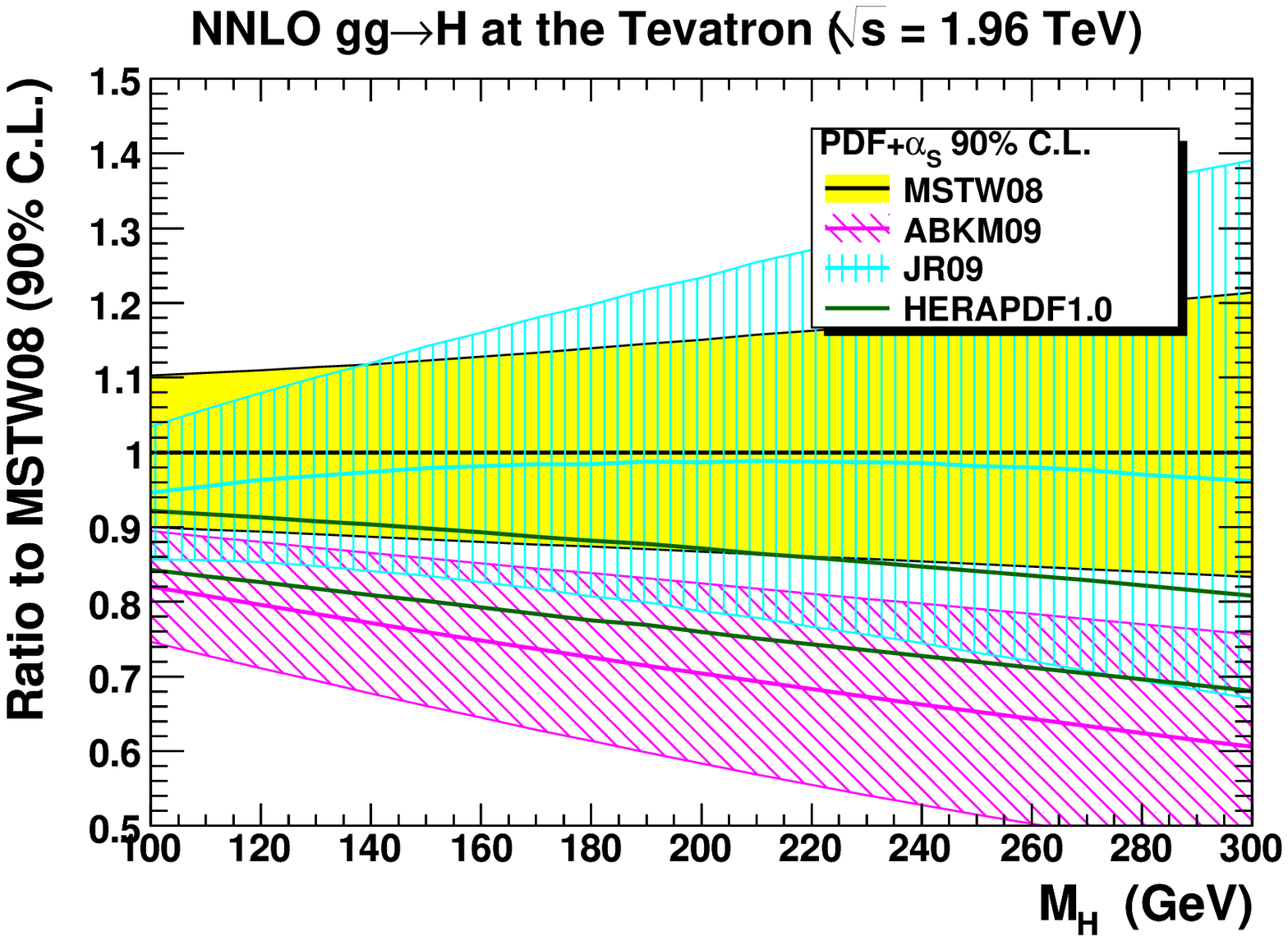}
  \end{minipage}
  \caption{Ratio to MSTW08 $gg\to H$ cross section at Tevatron with PDF+$\alpha_S$ uncertainties for (a)~NLO at 68\%~C.L., (b)~NLO at 90\%~C.L., (c)~NNLO at 68\%~C.L., (d)~NNLO at 90\%~C.L.}
  \label{fig:ratioTEV}
\end{figure}
\begin{figure}
  \centering
  \begin{minipage}{0.5\textwidth}
    (a)\\
    \includegraphics[width=\textwidth]{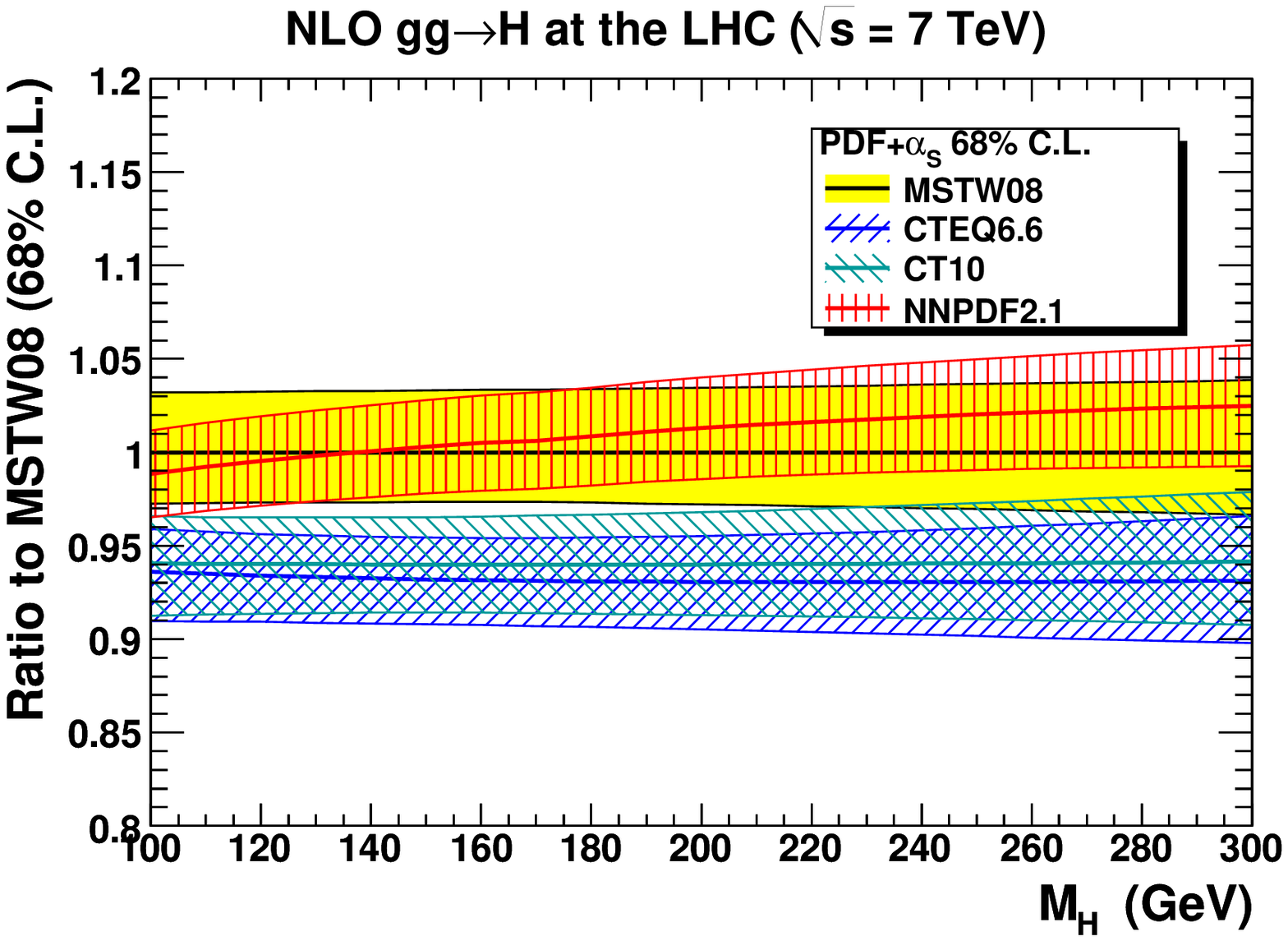}
  \end{minipage}%
  \begin{minipage}{0.5\textwidth}
    (b)\\
    \includegraphics[width=\textwidth]{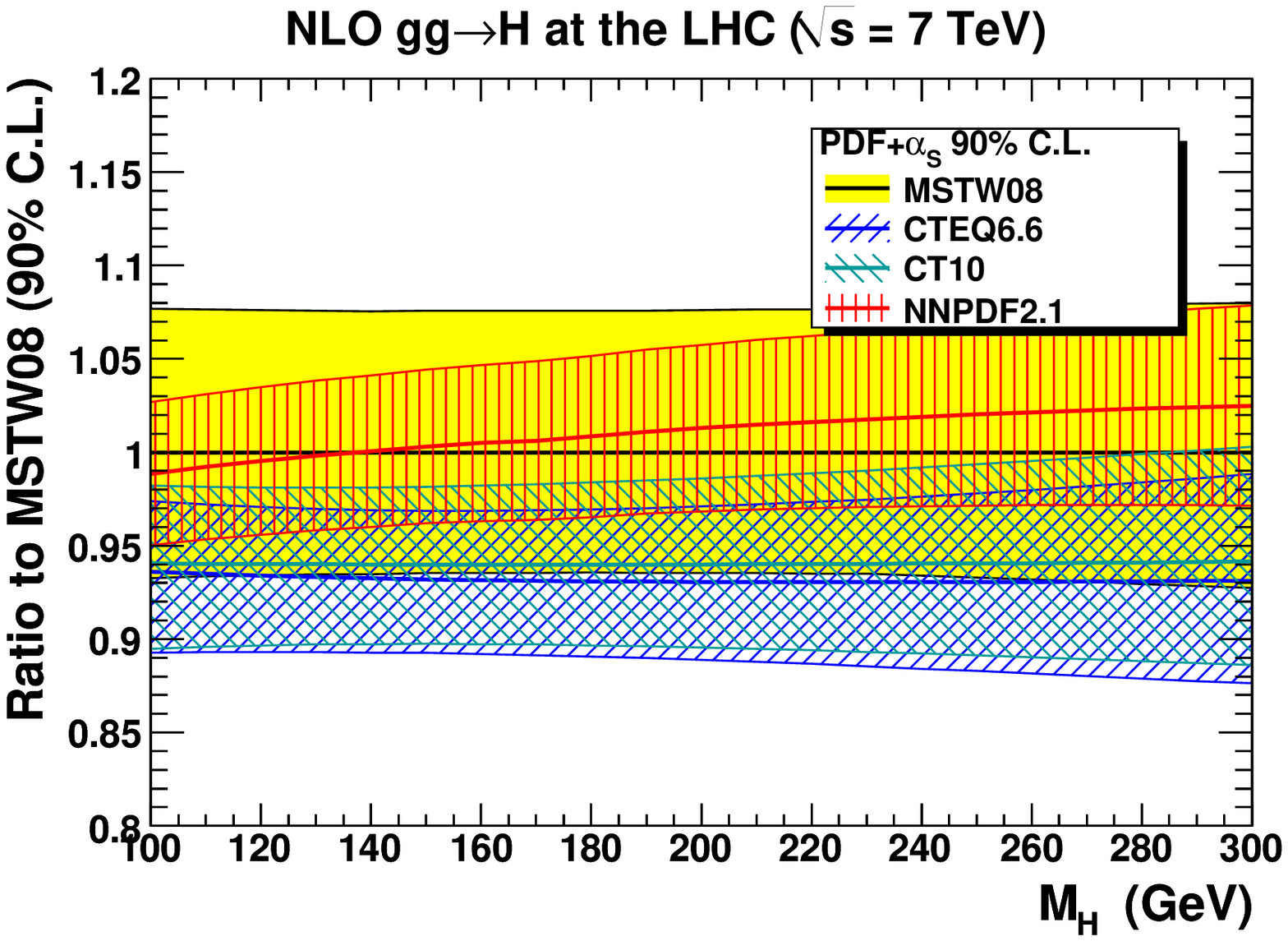}
  \end{minipage}
  \begin{minipage}{0.5\textwidth}
    (c)\\
    \includegraphics[width=\textwidth]{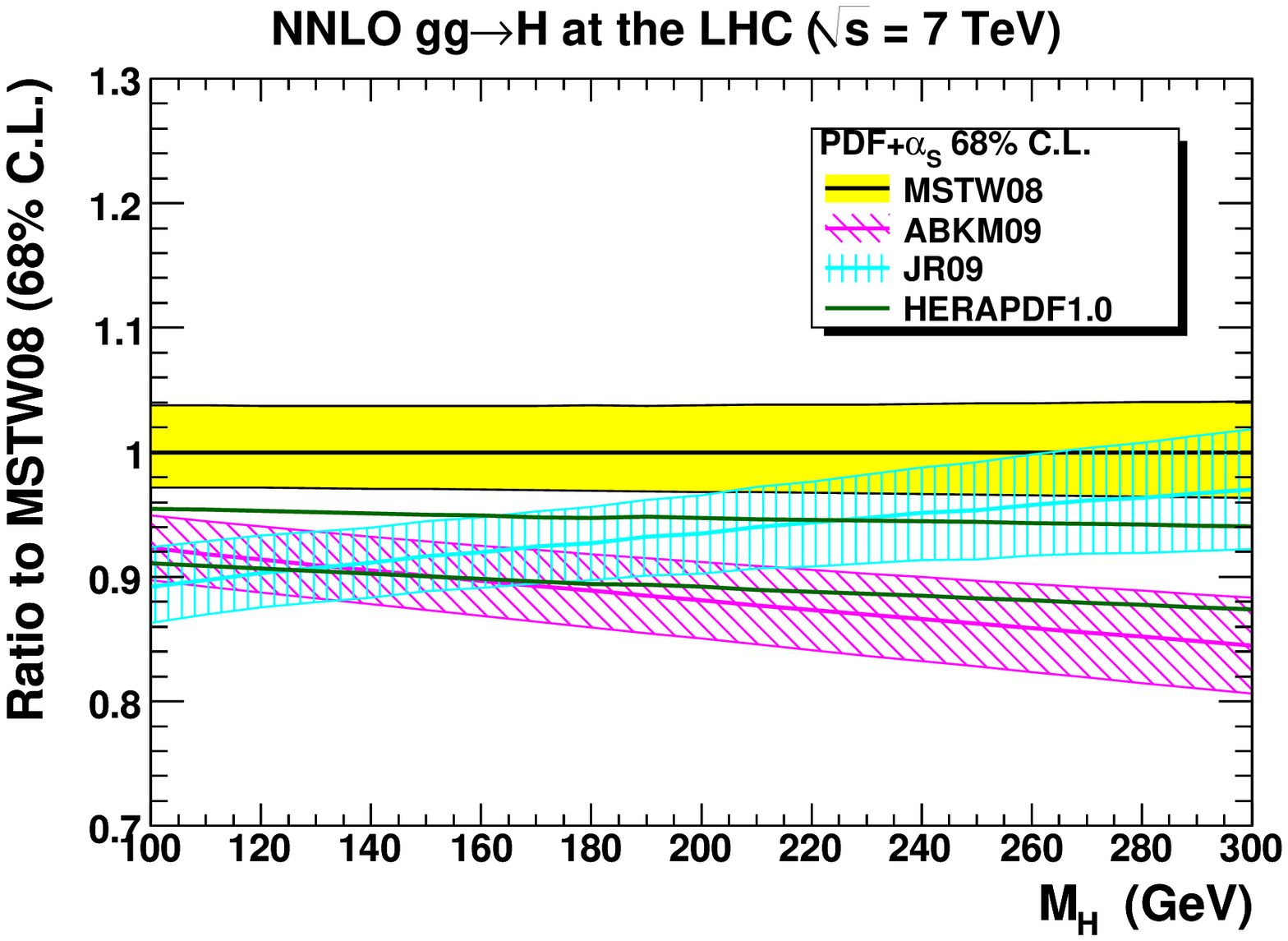}
  \end{minipage}%
  \begin{minipage}{0.5\textwidth}
    (d)\\
    \includegraphics[width=\textwidth]{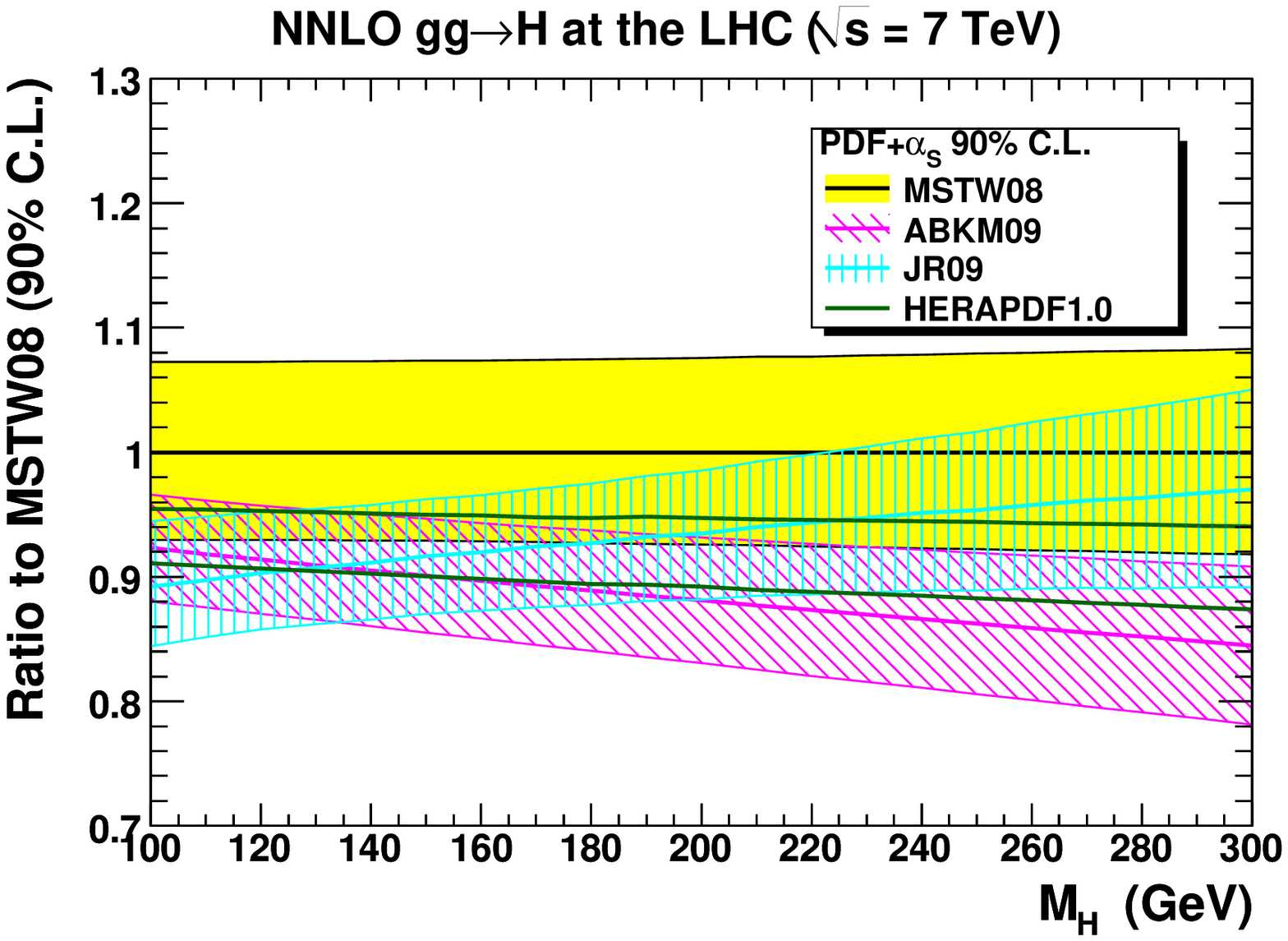}
  \end{minipage}
  \caption{Ratio to MSTW08 $gg\to H$ cross section at 7~TeV LHC with PDF+$\alpha_S$ uncertainties for (a)~NLO at 68\%~C.L., (b)~NLO at 90\%~C.L., (c)~NNLO at 68\%~C.L., (d)~NNLO at 90\%~C.L.}
  \label{fig:ratioLHC}
\end{figure}
The ratios of the cross sections with respect to the MSTW08 predictions are shown for the Tevatron in figure~\ref{fig:ratioTEV} and for the LHC in figure~\ref{fig:ratioLHC}, where PDF+$\alpha_S$ uncertainty bands at both 68\% and 90\% C.L.~are plotted.  It can be seen that there is generally good agreement between the \emph{global} fits at NLO.  However, at NNLO, the ABKM09 prediction, and the HERAPDF1.0 prediction with the lower $\alpha_S$ value, are well below MSTW08 at the Tevatron, even allowing for the 90\% C.L.~PDF+$\alpha_S$ uncertainties, with a significant discrepancy also at the LHC.

Baglio, Djouadi, Ferrag and Godbole (BDFG)~\cite{Baglio:2011wn} have claimed that some publicly available PDFs, specifically the HERAPDF1.0 NNLO set with $\alpha_S(M_Z^2)=0.1145$, can lower the Tevatron Higgs cross section by up to 40\% compared to MSTW08 for $M_H\approx 160$~GeV, requiring more than twice as much Tevatron data to recover the same sensitivity as the 2010 analysis by the Tevatron experiments~\cite{CDF:2010ar}, which used MSTW08 for the central prediction.  This is obviously potentially very worrying.  However, figure~\ref{fig:ratioTEV}(c,d) shows that the lowest cross section occurs not with either of the HERAPDF sets, but with ABKM09, where the central cross section is $\approx75\%$ that of MSTW08 at $M_H\approx 160$~GeV.  The cross-section ratios for ABKM09 and JR09 in figure~\ref{fig:ratioTEV}(d) seem close to those in the inset of figure 1 of ref.~\cite{Baglio:2011wn}, but we do not reproduce the extreme behaviour of the HERAPDF1.0 sets.  Our results are supported by those in ref.~\cite{Alekhin:2010dd} where it is also observed that ABKM09 gives lower Higgs cross sections at the Tevatron than the HERAPDF1.0 set with $\alpha_S(M_Z^2)=0.1145$.  One obvious difference is the scale choice $\mu_R=\mu_F=M_H/2$ used in ref.~\cite{Baglio:2011wn} rather than $\mu_R=\mu_F=M_H$ used here and in ref.~\cite{Alekhin:2010dd}.  However, we have checked that the ratio of cross sections with respect to MSTW08 is largely independent of the different scale choice.  The detailed arguments of ref.~\cite{Baglio:2011wn} assume the ``worst-case scenario'' of a 40\% reduction in $\sigma_H$ at $M_H\approx 160$~GeV from the central value of HERAPDF1.0 with $\alpha_S(M_Z^2)=0.1145$, and therefore the conclusions require modification if there is a mistake in their HERAPDF1.0 calculations.\footnote{We thank J.~Baglio for confirming that the HERAPDF1.0 curves in figure 1 of ref.~\cite{Baglio:2011wn} were erroneously drawn with $\mu_R=\mu_F=(3/2)M_H$, to be corrected in an erratum included in {\tt v3} of the preprint version~\cite{Baglio:2011wn}.}  Nevertheless, even the 25\% reduction in $\sigma_H$ at $M_H\approx 160$~GeV from the central value of ABKM09 is still a problem, as it lies well outside both the MSTW08 PDF+$\alpha_S$ uncertainty at 90\% C.L.~used in ref.~\cite{CDF:2010ar} and the PDF4LHC\footnote{The PDF4LHC recommendation~\cite{Botje:2011sn} is to rescale the MSTW08 NNLO PDF+$\alpha_S$ uncertainty at 68\% C.L.~by the ratio of the envelope of the MSTW08 NLO, CTEQ6.6 NLO and NNPDF2.0 NLO predictions, all including PDF+$\alpha_S$ uncertainties at 68\% C.L., to the MSTW08 NLO PDF+$\alpha_S$ 68\% C.L.~uncertainty.} uncertainty used in ref.~\cite{CDF:2011gs}.  (These two prescriptions for uncertainties give similar results, but the former is clearly much simpler; see section~5 of ref.~\cite{bench7TeV} for more discussion.)  We note that in justifying the use of the HERAPDF set, BDFG~\cite{Baglio:2011wn} make the statement: ``However, HERAPDF describes well not only the Tevatron jet data but also the $W$, $Z$ data.  Since this is a prediction beyond leading order, it has also the contributions of the gluon included.  This gives an indirect test that the gluon densities are predicted in a satisfactory way.''  This statement is very misleading: the $W$ charge asymmetry and the $Z$ rapidity distribution at the Tevatron, used as a PDF constraint, are almost insensitive to the gluon distribution, and the statement makes no reference to the quantitative comparison of PDFs to jet data.  In the rest of this paper we will present a number of arguments to show that, of all the currently available NNLO PDF sets, only MSTW08 provides a fully reliable estimate of the Higgs cross sections at the Tevatron and LHC.

\subsection{Dependence on $gg$ luminosity}

At LO, the PDF dependence of the $gg\to H$ total cross section is simply given by the gluon--gluon luminosity evaluated at a partonic centre-of-mass energy $\sqrt{\hat{s}} = M_H$,
\begin{equation}
  \frac{\partial {\cal L}_{gg}}{\partial \hat{s}} = \frac{1}{s} \int_\tau^1\frac{{\rm d}x}{x}\;f_{g}(x,\hat{s})f_{g}(\tau/x,\hat{s}), \label{eq:gglumi}
\end{equation}
where $f_g(x,\mu^2=\hat{s})$ is the gluon distribution and $\tau\equiv \hat{s}/s$.
\begin{figure}
  \centering
  \begin{minipage}{0.5\textwidth}
    (a)\\
    \includegraphics[width=\textwidth]{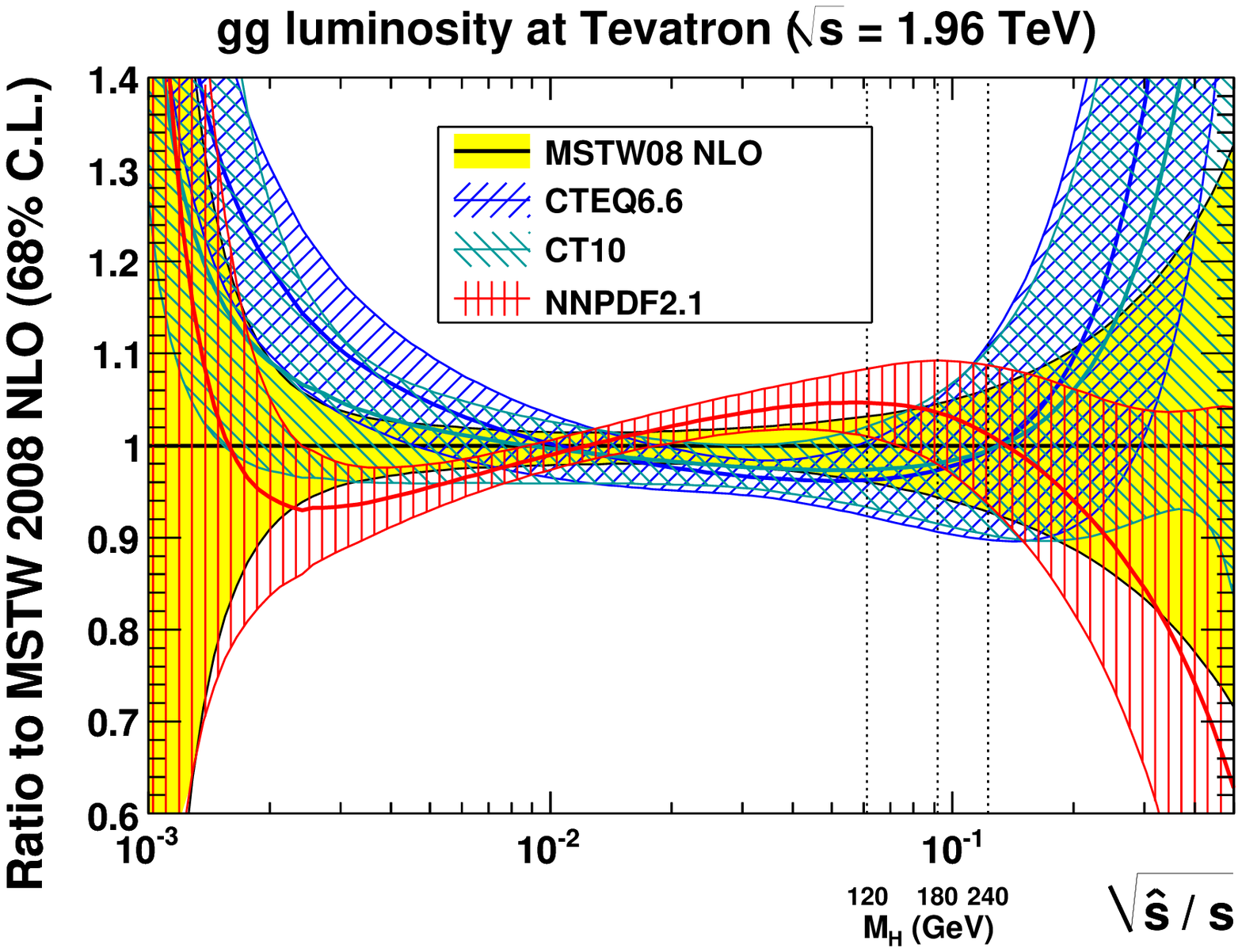}
  \end{minipage}%
  \begin{minipage}{0.5\textwidth}
    (b)\\
    \includegraphics[width=\textwidth]{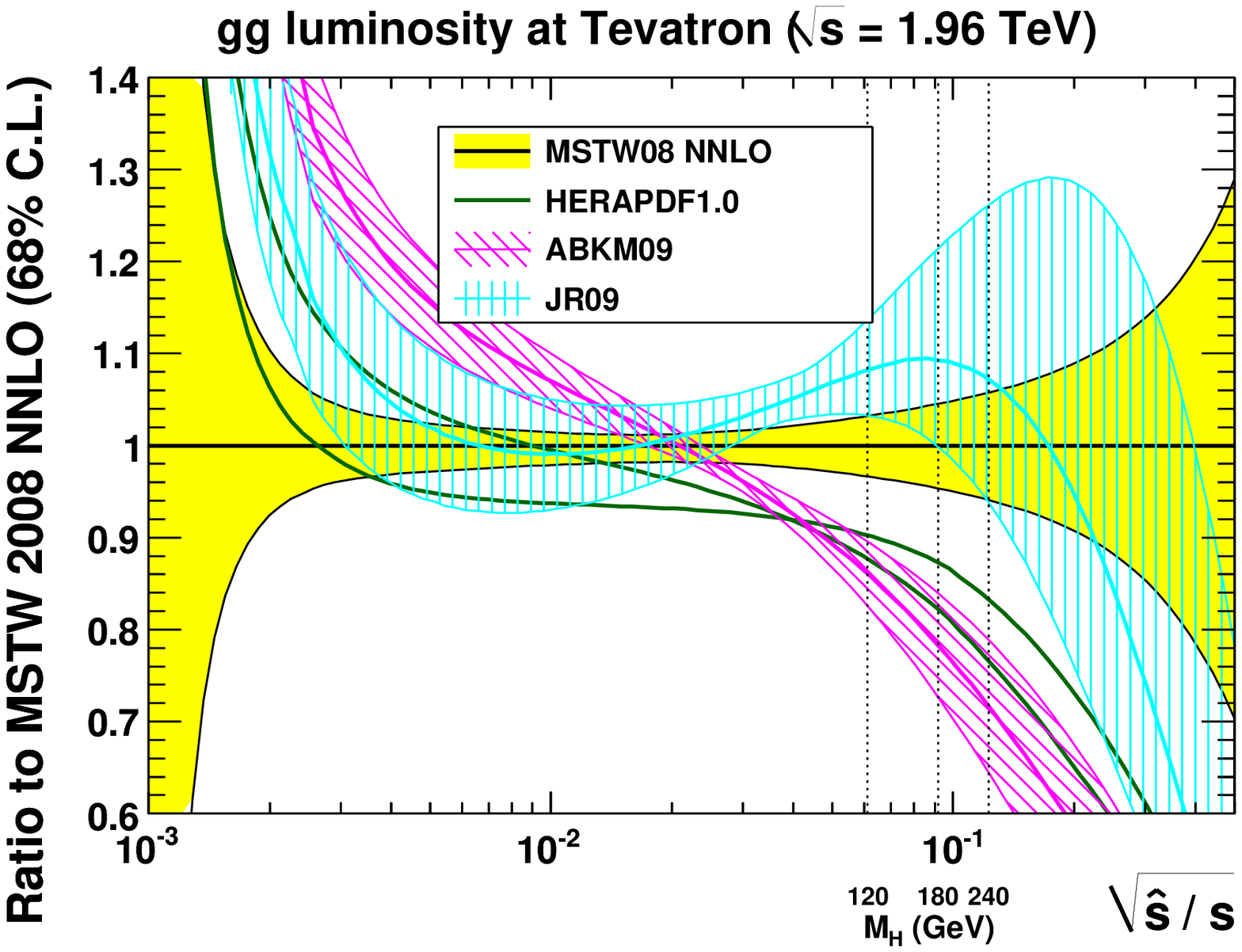}
  \end{minipage}
  \begin{minipage}{0.5\textwidth}
    (c)\\
    \includegraphics[width=\textwidth]{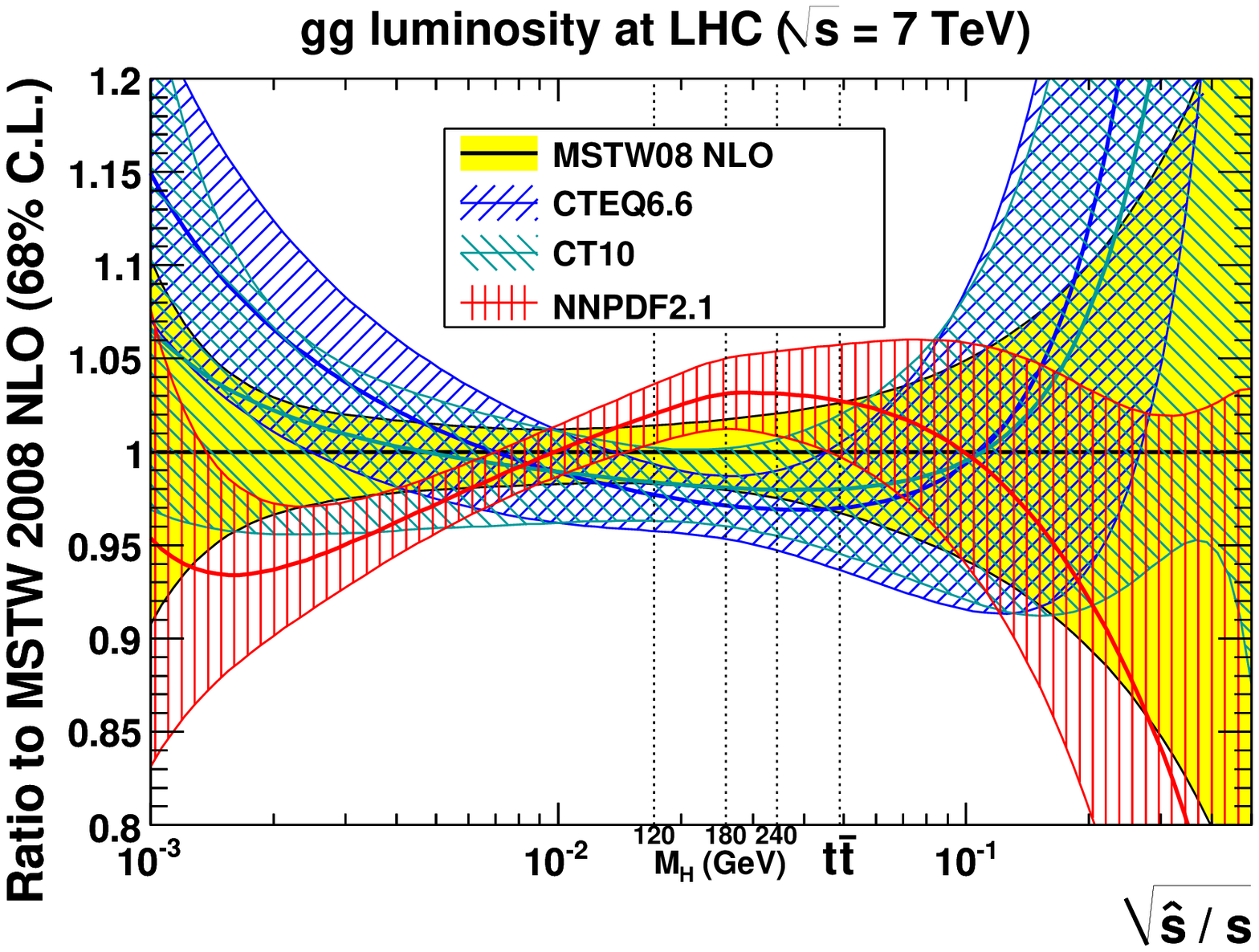}
  \end{minipage}%
  \begin{minipage}{0.5\textwidth}
    (d)\\
    \includegraphics[width=\textwidth]{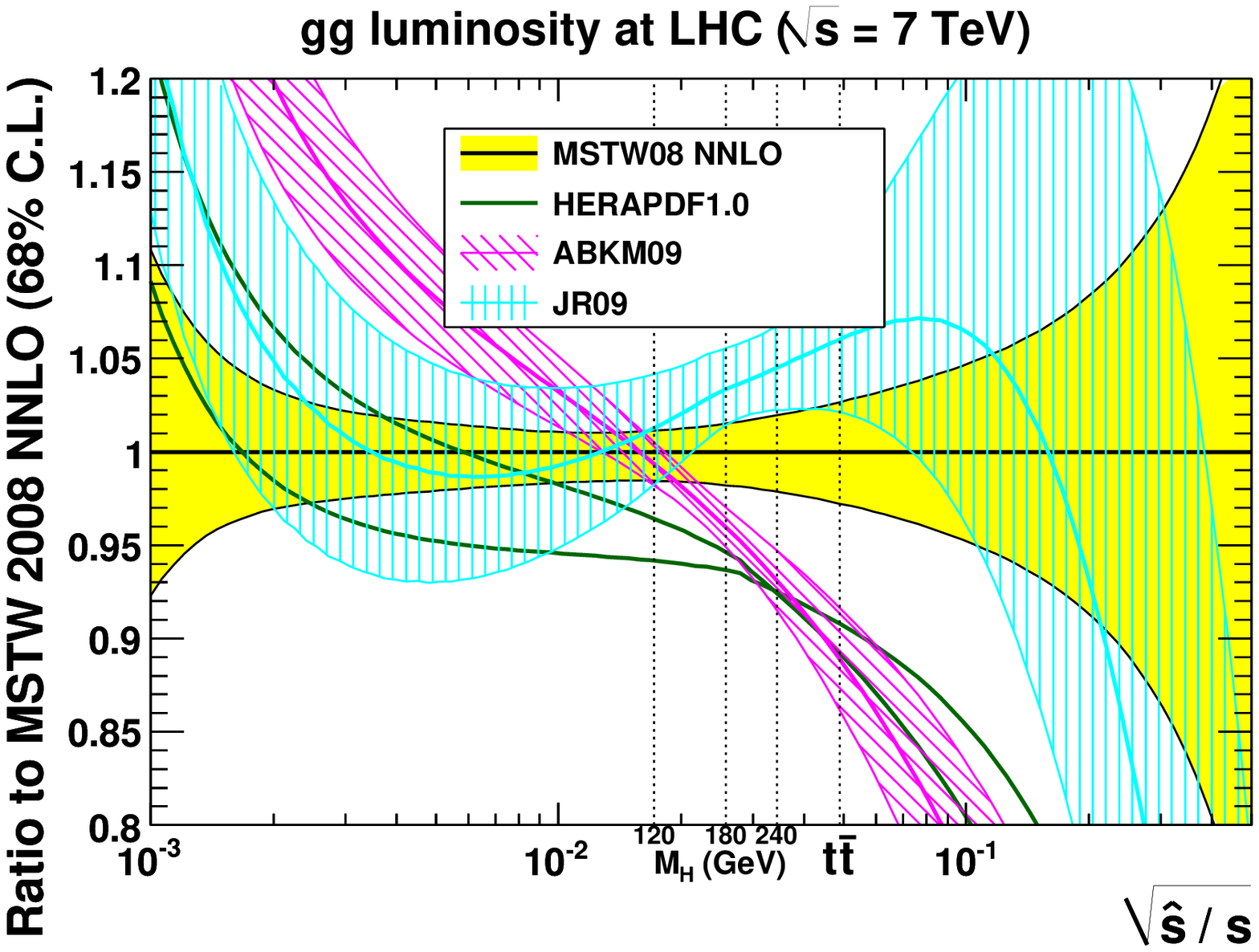}
  \end{minipage}
  \caption{Gluon--gluon luminosities as the ratio with respect to MSTW 2008 for (a)~NLO at the Tevatron, (b)~NNLO at the Tevatron, (c)~NLO at the LHC, and (d)~NNLO at the LHC.}
  \label{fig:gglumi}
\end{figure}
In figure~\ref{fig:gglumi} we show the gluon--gluon luminosities calculated using different PDF sets and taken as the ratio with respect to the MSTW 2008 value, at centre-of-mass energies corresponding to the (a,b)~Tevatron and (c,d) LHC.  The relevant values of $\sqrt{\hat{s}} = M_{H} = \{120, 180, 240\}$~GeV are indicated, along with the threshold for $t\bar{t}$ production at the LHC, $\sqrt{\hat{s}} = 2m_t$ with $m_t=171.3$~GeV, where this process is predominantly $gg$-initiated at the LHC.  Indeed, $t\bar{t}$ production at the LHC is strongly correlated with $gg\to H$ production at the Tevatron, with both processes probing the gluon distribution at similar $x$ values, as seen from figure~\ref{fig:gglumi}.  We point out in ref.~\cite{bench7TeV} that the current $t\bar{t}$ cross-section measurements at the LHC~\cite{ATLAS:ttbar,CMS:ttbar} seem to distinctly favour MSTW08 over ABKM09.

The NLO luminosities in figure~\ref{fig:gglumi}(a,c) are shown for the global fits from MSTW08~\cite{Martin:2009iq}, CTEQ6.6~\cite{Nadolsky:2008zw}, CT10~\cite{Lai:2010vv} and NNPDF2.1~\cite{Ball:2011mu}.  The NNLO luminosities in figure~\ref{fig:gglumi}(b,d) are shown for MSTW08~\cite{Martin:2009iq}, HERAPDF1.0~\cite{HERA:2009wt}, ABKM09~\cite{Alekhin:2009ni} and JR09~\cite{JimenezDelgado:2008hf,JimenezDelgado:2009tv}.  The two HERAPDF1.0 NNLO curves shown are for both $\alpha_S(M_Z^2)=0.1145$ and $0.1176$, where the latter gives the smaller $gg$ luminosity at low $\hat{s}$ values and the larger $gg$ luminosity at high $\hat{s}$ values.  The larger $\alpha_S$ value means that less gluon is required at low $x$ to fit the scaling violations of HERA data, $\partial F_2/\partial\ln(Q^2)\sim \alpha_S\,g$, therefore more gluon is required at high $x$ from the momentum sum rule.  Both these effects, larger $\alpha_S$ and more high-$x$ gluon, raise the Tevatron Higgs cross section and improve the quality of the description of Tevatron jet data, as we will see in section~\ref{sec:tevjets}.  The NNLO trend between groups is similar to at NLO~\cite{bench7TeV}.  There is reasonable agreement for the global fits, but more variation for the other sets, particularly at large $\hat{s}$, where HERAPDF1.0 and ABKM09 have much softer high-$x$ gluon distributions, and this feature has a direct impact on the $gg\to H$ cross sections, particularly at the Tevatron (see figure~\ref{fig:ratioTEV}).

\subsection{Dependence on \texorpdfstring{strong coupling $\alpha_S$}{alphaS}} \label{sec:ggHalphaS}

\begin{figure}
  \centering
  \begin{minipage}{0.5\textwidth}
    (a)\\
    \includegraphics[width=\textwidth]{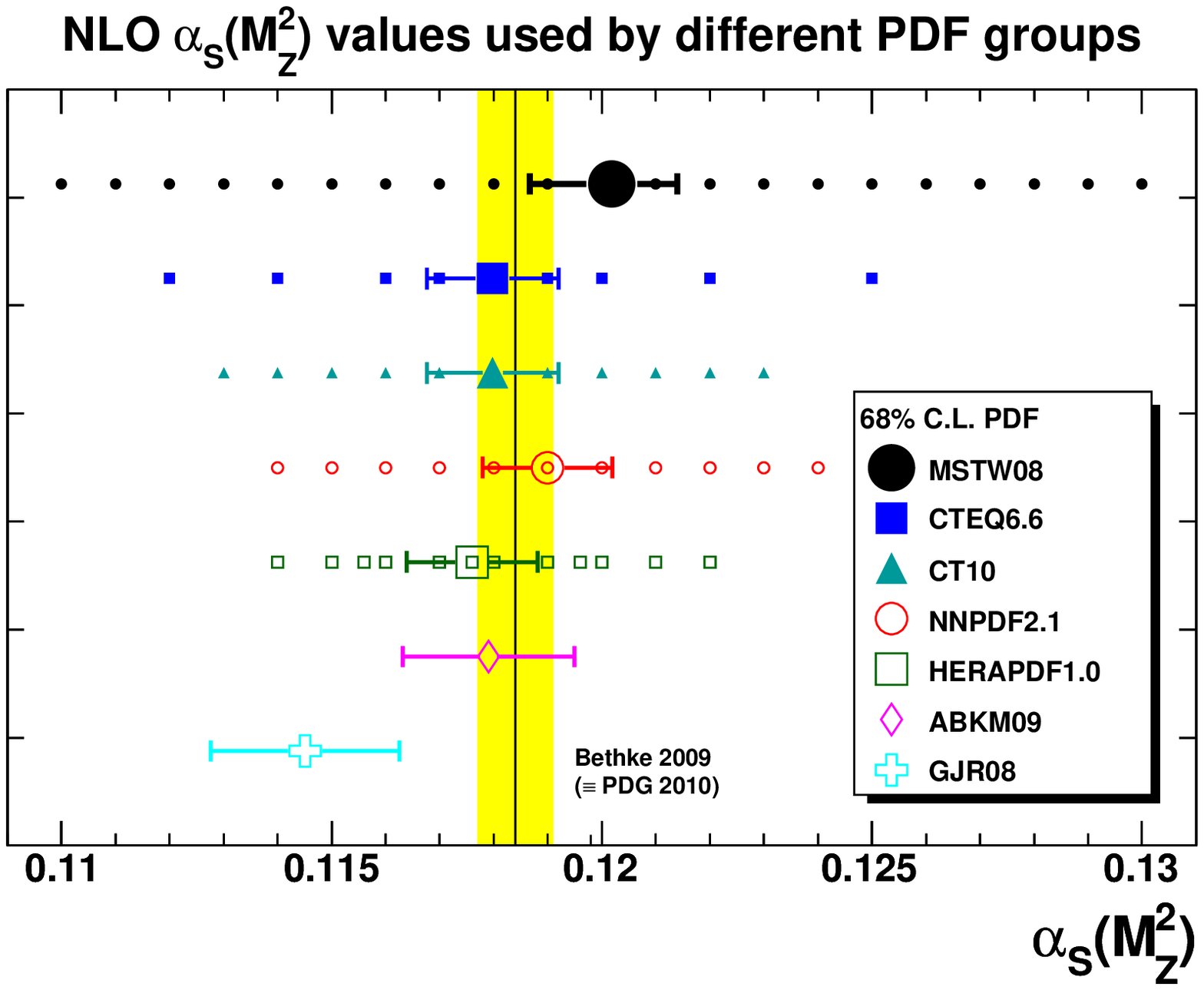}
  \end{minipage}%
  \begin{minipage}{0.5\textwidth}
    (b)\\
    \includegraphics[width=\textwidth]{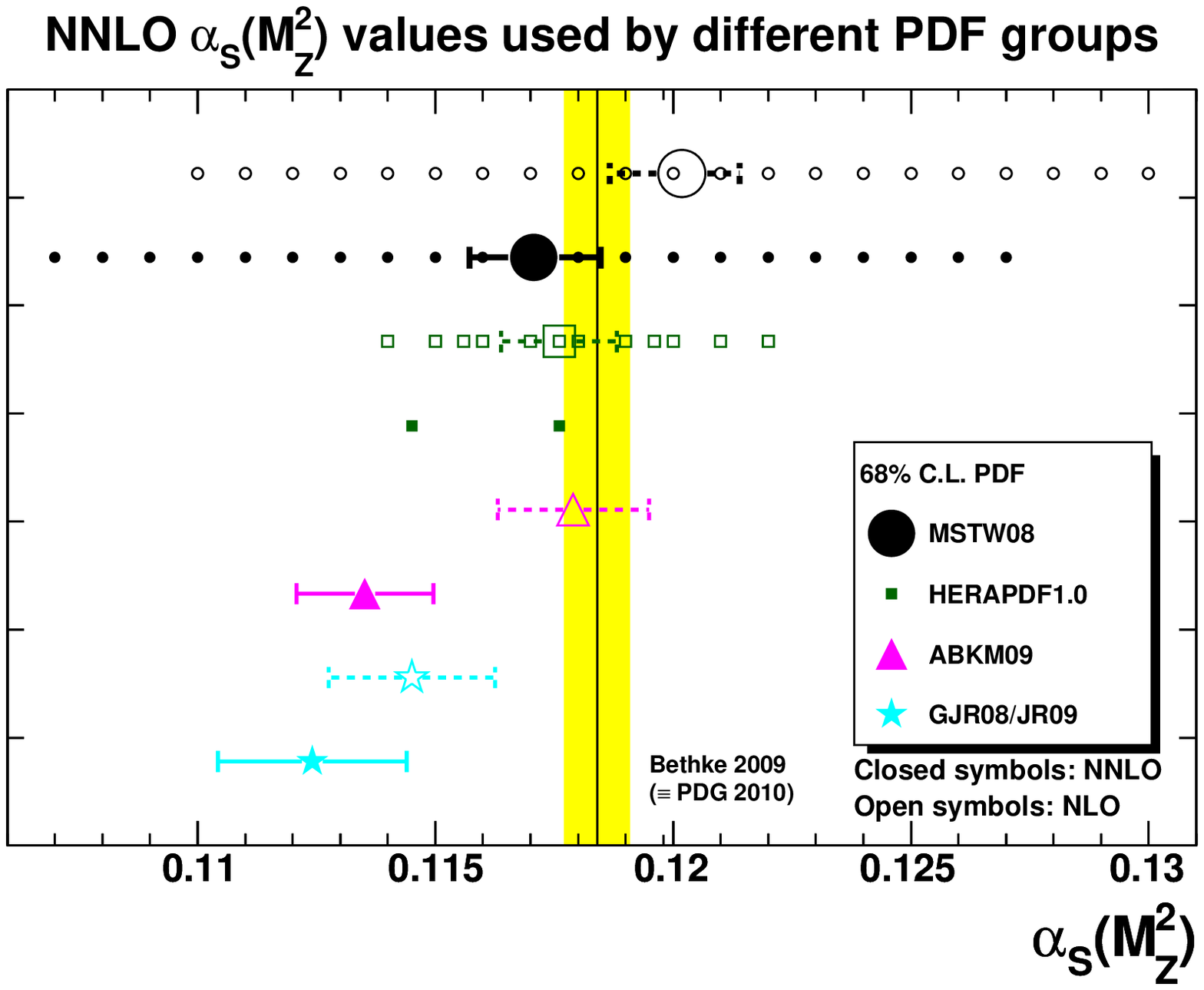}
  \end{minipage}
  \caption{Values of $\alpha_S(M_Z^2)$, and their 1-$\sigma$ uncertainties, used by different PDF fitting groups at (a)~NLO and (b)~NNLO.  The smaller symbols indicate the PDF sets with alternative values of $\alpha_S(M_Z^2)$ provided by each fitting group.  The shaded band indicates the world average $\alpha_S(M_Z^2)$~\cite{Bethke:2009jm}.}
  \label{fig:asmzvalues}
\end{figure}
The various PDF fitting groups take different approaches to the values of the strong coupling $\alpha_S$ and, for consistency, the same value as used in the fit should be used in subsequent cross-section calculations.  The values of $\alpha_S(M_Z^2)$, and the corresponding uncertainties, for MSTW08, ABKM09 and GJR08/JR09 are obtained from a simultaneous fit with the PDF parameters.  Other groups choose a fixed value, generally close to the world average~\cite{Bethke:2009jm}, and for those groups we assume a 1-$\sigma$ uncertainty of $\pm0.0012$~\cite{Alekhin:2011sk}, very similar to the MSTW08 uncertainty.  The central values and 1-$\sigma$ uncertainties are depicted in figure~\ref{fig:asmzvalues} as the larger symbols and error bars, while the smaller symbols indicate the PDF sets with alternative values of $\alpha_S(M_Z^2)$ provided by each fitting group.  The fitted NLO $\alpha_S(M_Z^2)$ value is always larger than the corresponding NNLO $\alpha_S(M_Z^2)$ value in an attempt by the fit to mimic the missing higher-order corrections, which are generally positive.  The world average $\alpha_S(M_Z^2)$~\cite{Bethke:2009jm}, shown in figure~\ref{fig:asmzvalues}, combines determinations made at a variety of perturbative orders, but in most cases an increase in the order corresponds to a decrease in the value of $\alpha_S(M_Z^2)$ obtained.

\begin{figure}
  \centering
  \begin{minipage}{0.5\textwidth}
    (a)\\
    \includegraphics[width=\textwidth]{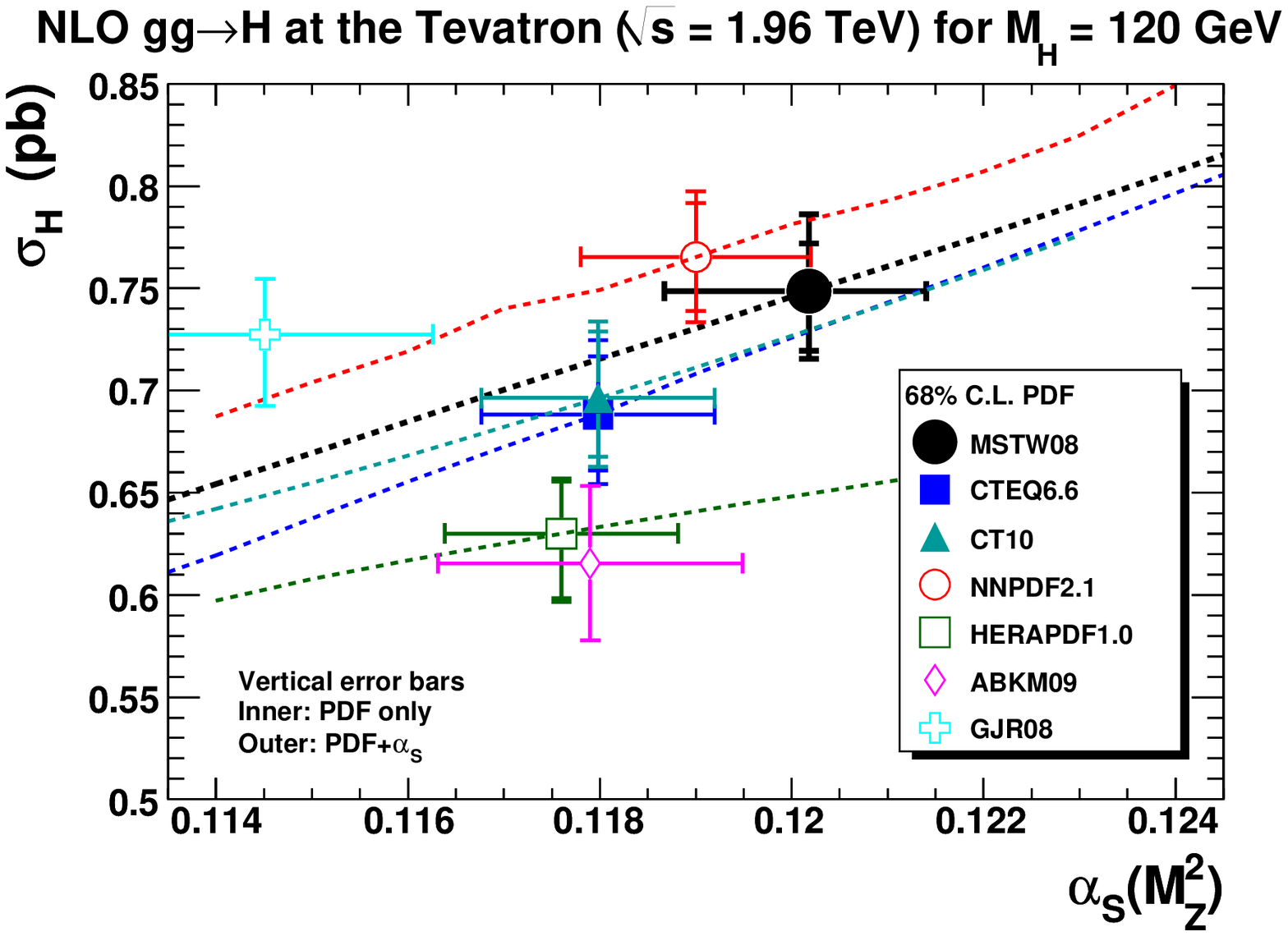}
  \end{minipage}%
  \begin{minipage}{0.5\textwidth}
    (b)\\
    \includegraphics[width=\textwidth]{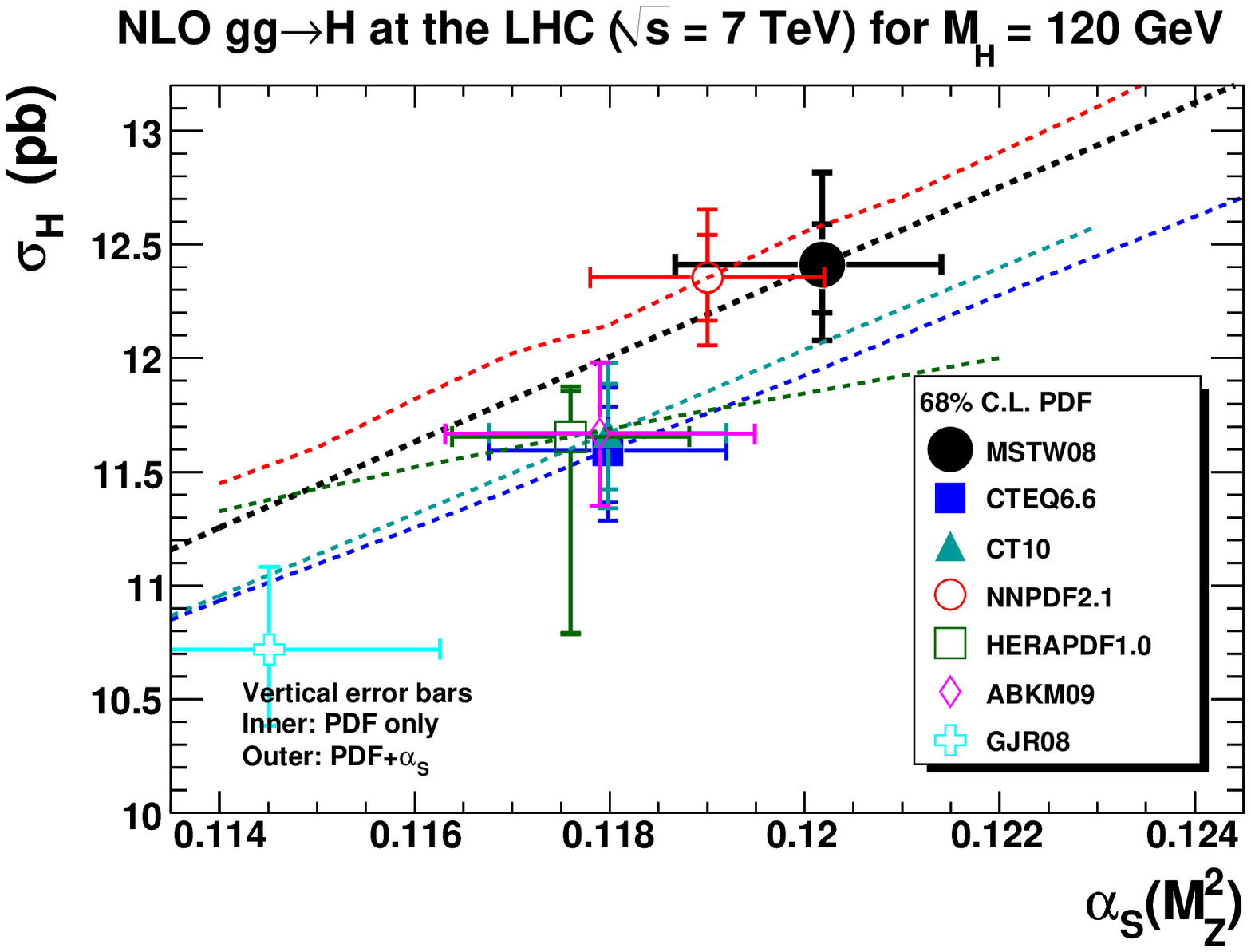}
  \end{minipage}
  \begin{minipage}{0.5\textwidth}
    (c)\\
    \includegraphics[width=\textwidth]{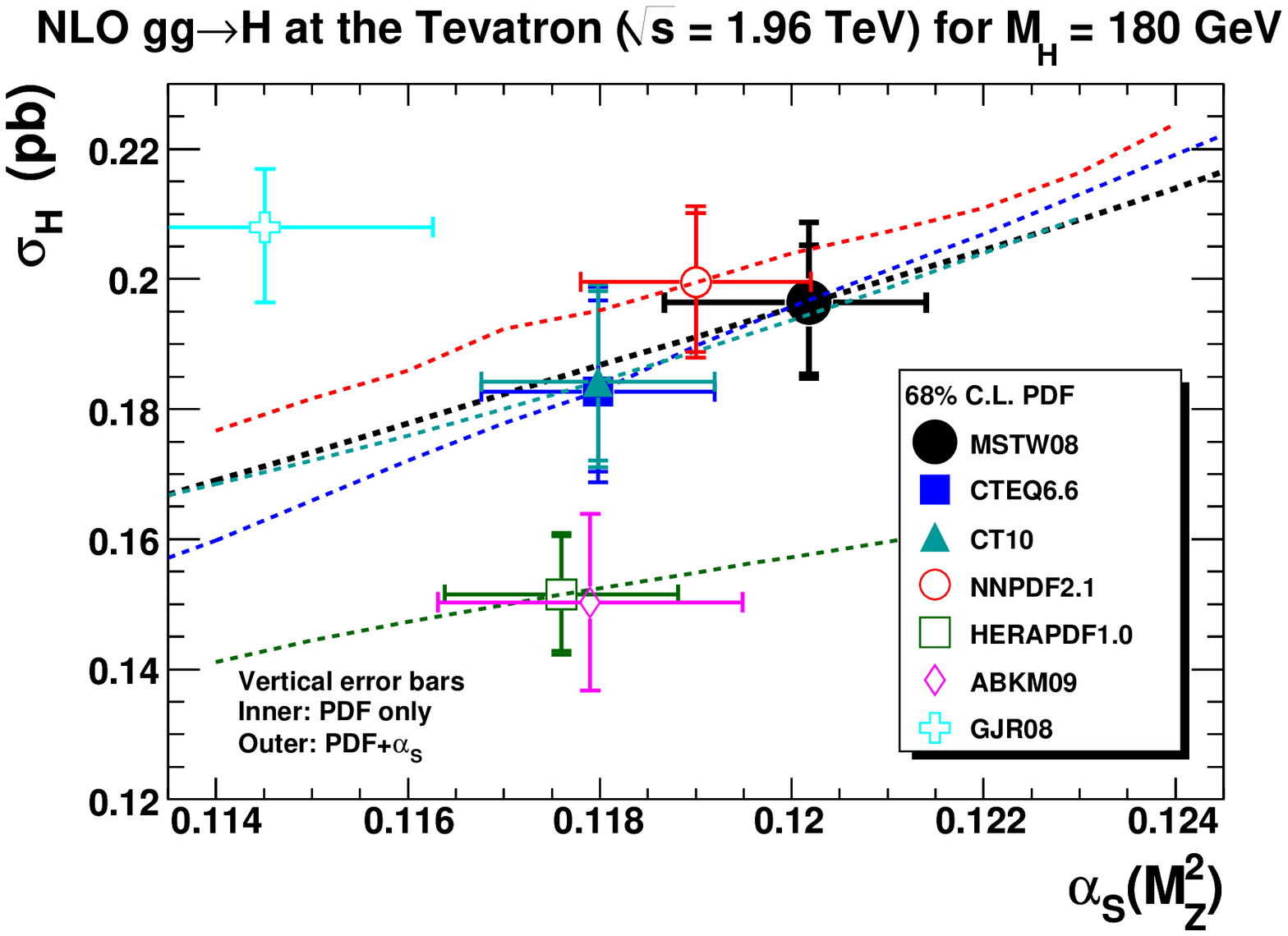}
  \end{minipage}%
  \begin{minipage}{0.5\textwidth}
    (d)\\
    \includegraphics[width=\textwidth]{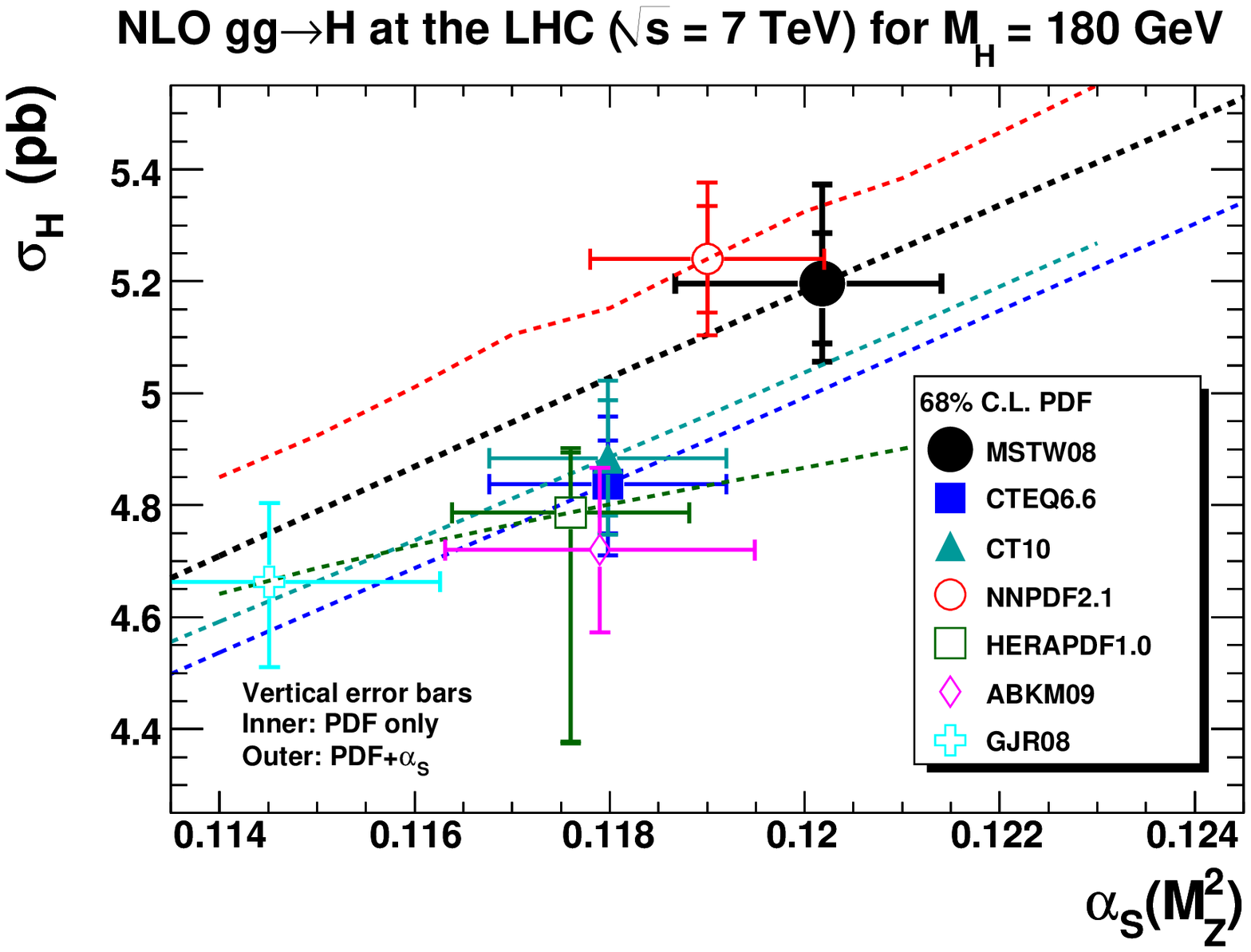}
  \end{minipage}
  \begin{minipage}{0.5\textwidth}
    (e)\\
    \includegraphics[width=\textwidth]{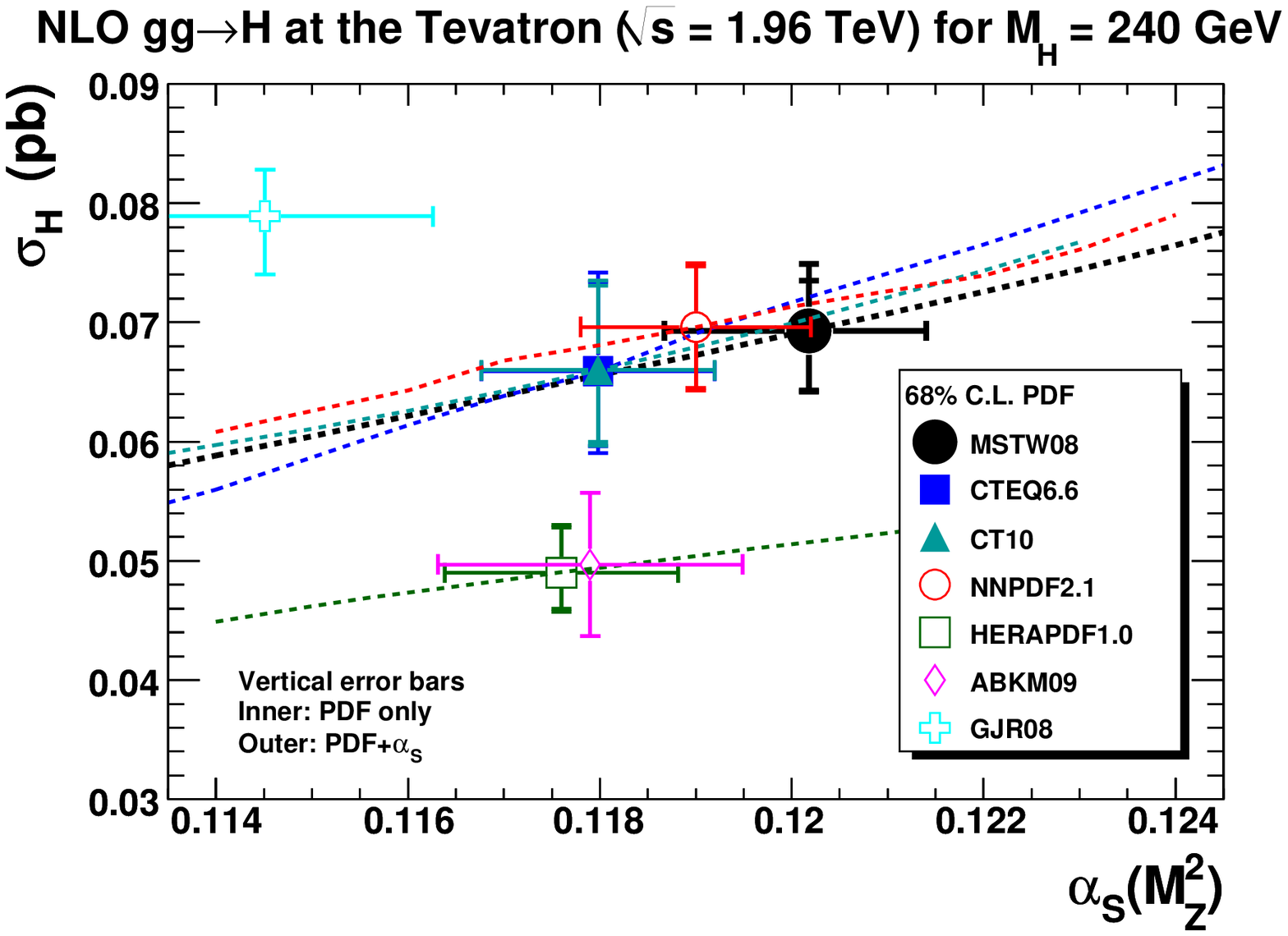}
  \end{minipage}%
  \begin{minipage}{0.5\textwidth}
    (f)\\
    \includegraphics[width=\textwidth]{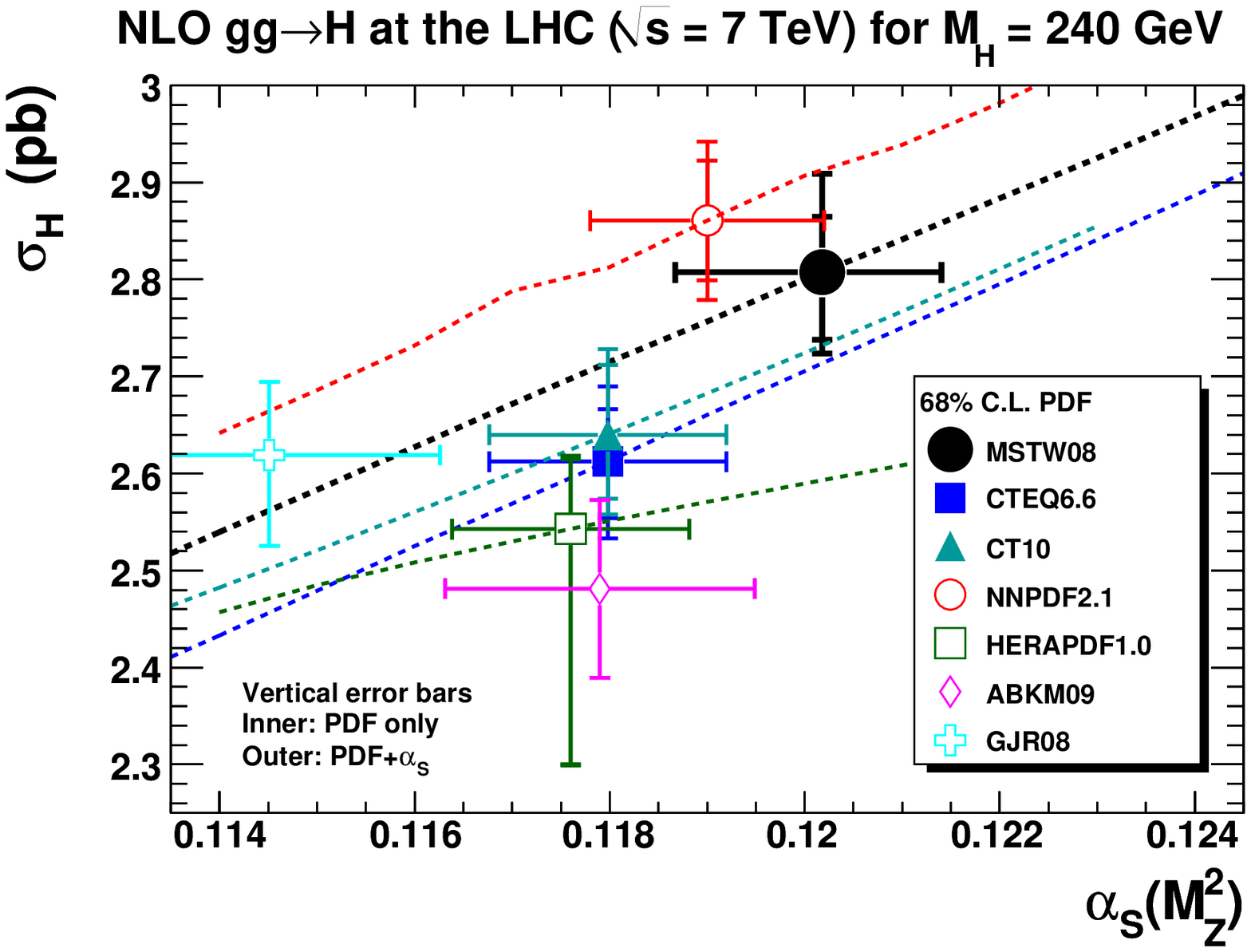}
  \end{minipage}
  \caption{$gg\to H$ total cross sections, plotted as a function of $\alpha_S(M_Z^2)$, at NLO.}
  \label{fig:gghvsasmznlo}
\end{figure}
\begin{figure}
  \centering
  \begin{minipage}{0.5\textwidth}
    (a)\\
    \includegraphics[width=\textwidth]{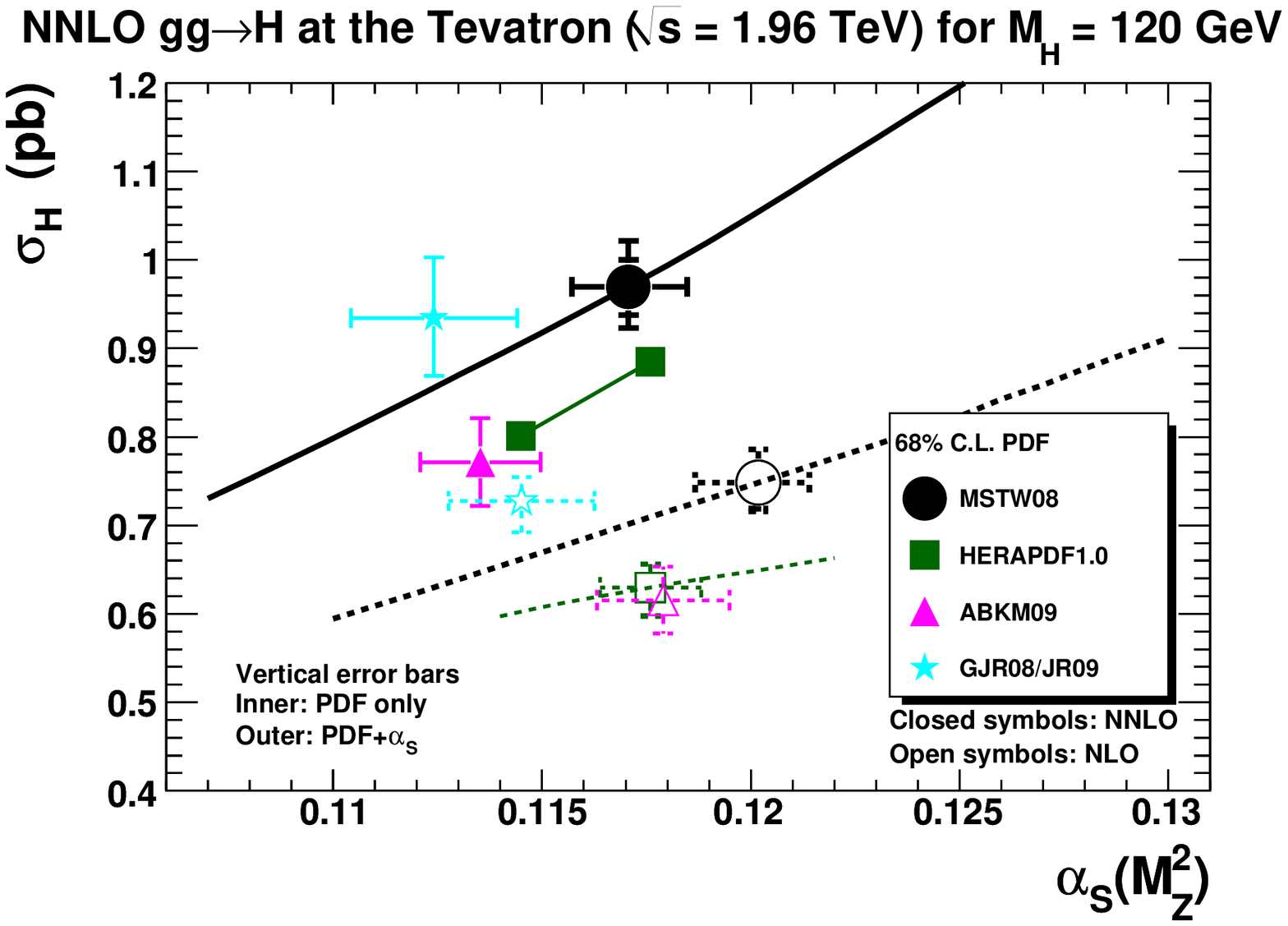}
  \end{minipage}%
  \begin{minipage}{0.5\textwidth}
    (b)\\
    \includegraphics[width=\textwidth]{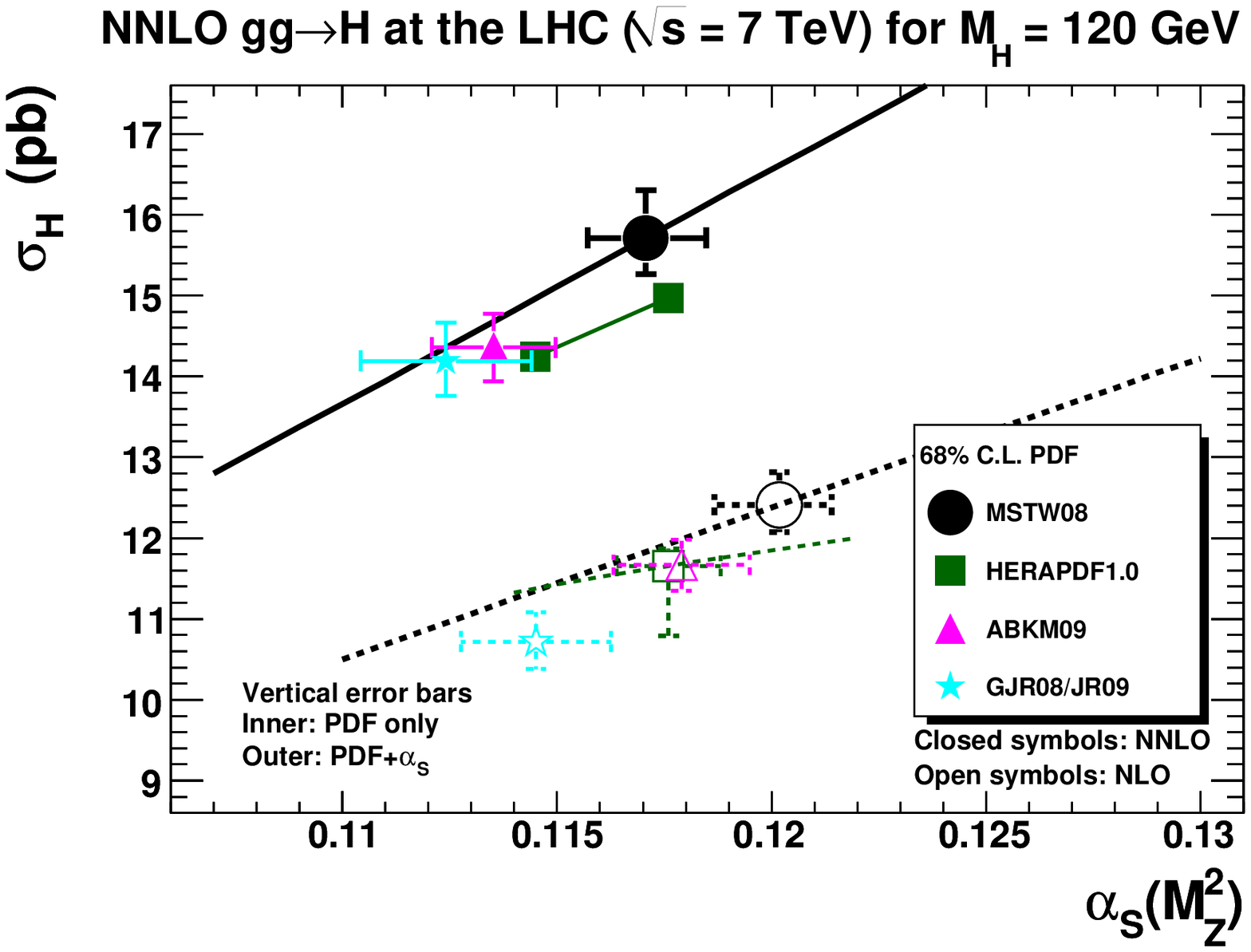}
  \end{minipage}
  \begin{minipage}{0.5\textwidth}
    (c)\\
    \includegraphics[width=\textwidth]{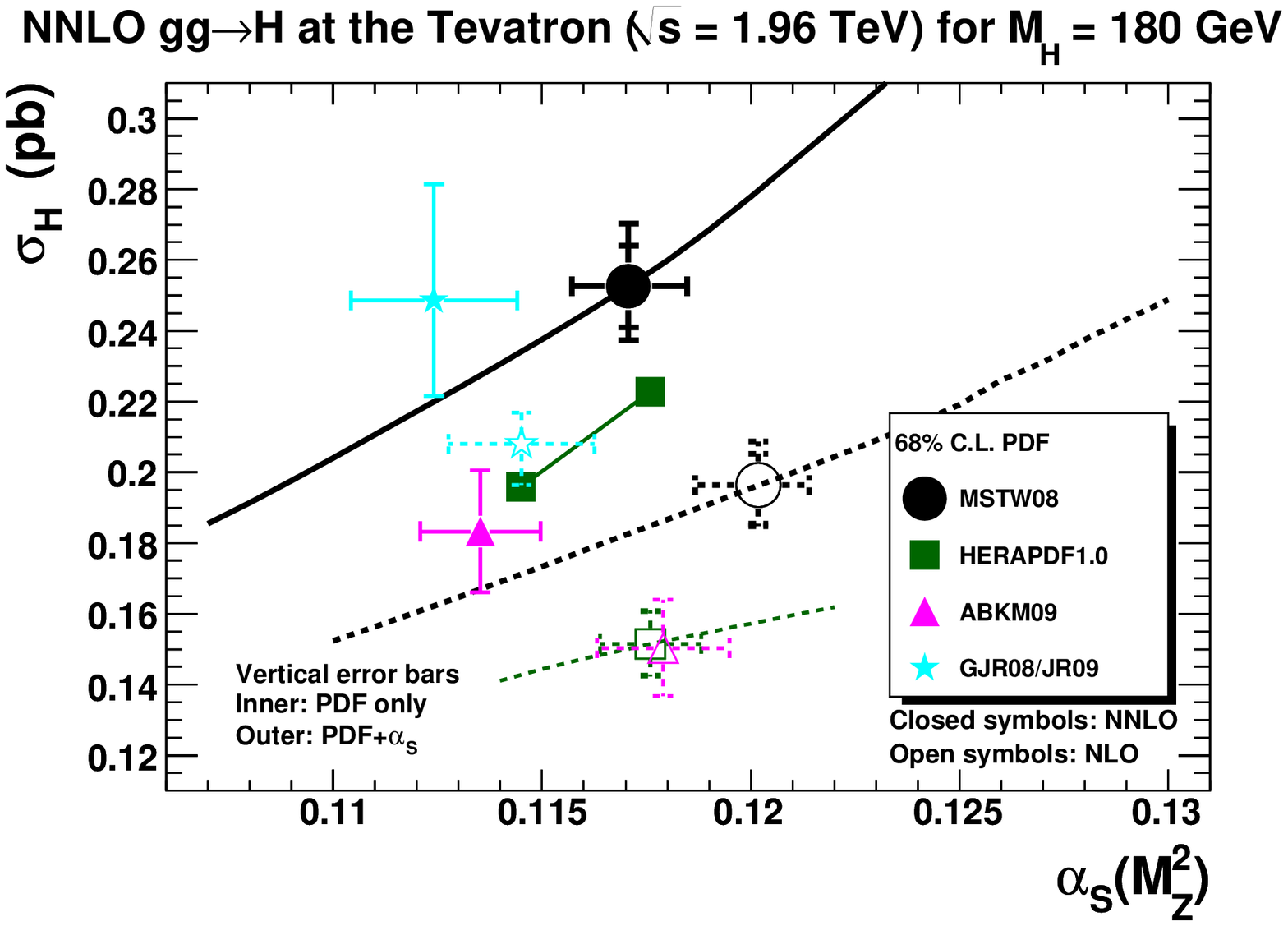}
  \end{minipage}%
  \begin{minipage}{0.5\textwidth}
    (d)\\
    \includegraphics[width=\textwidth]{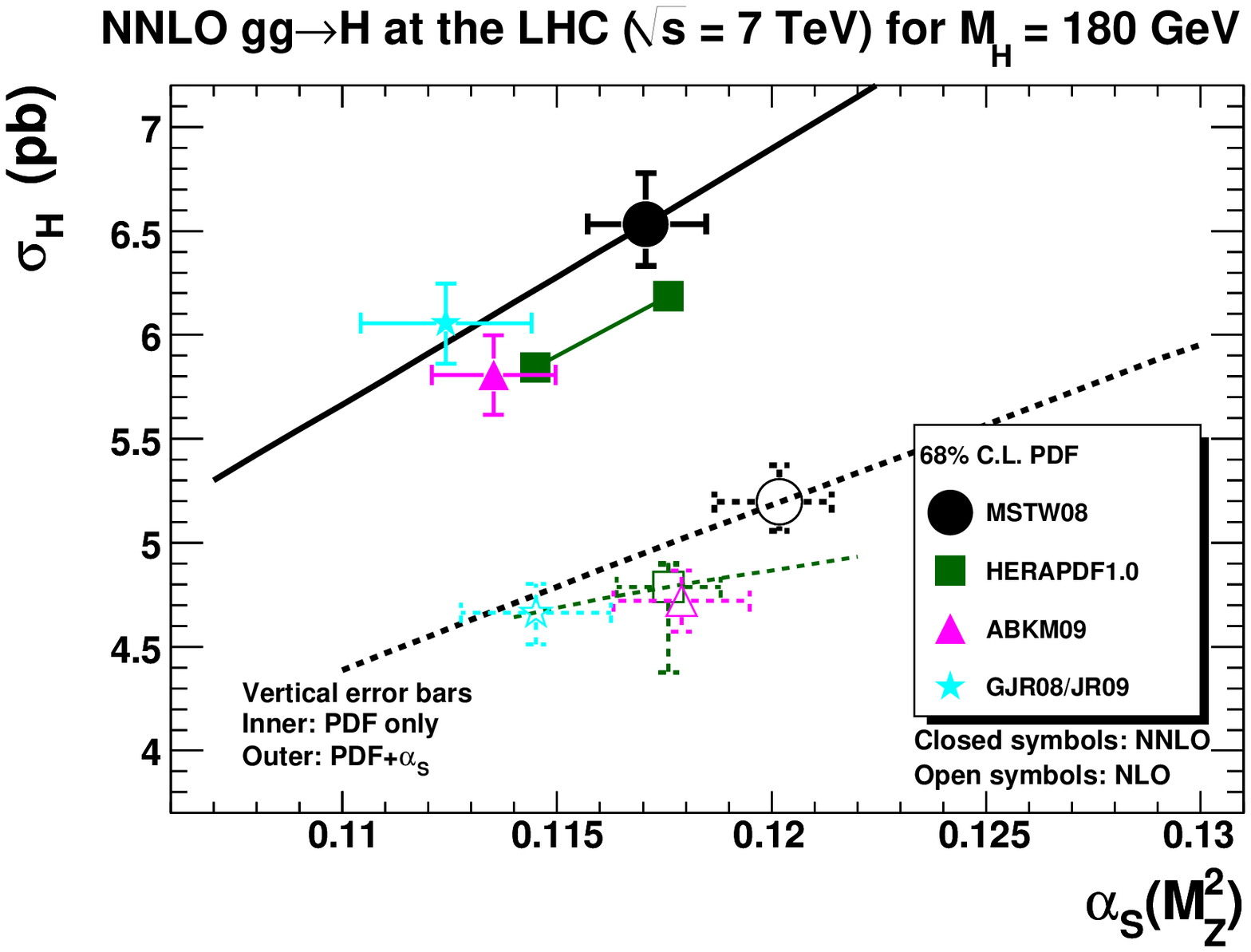}
  \end{minipage}
  \begin{minipage}{0.5\textwidth}
    (e)\\
    \includegraphics[width=\textwidth]{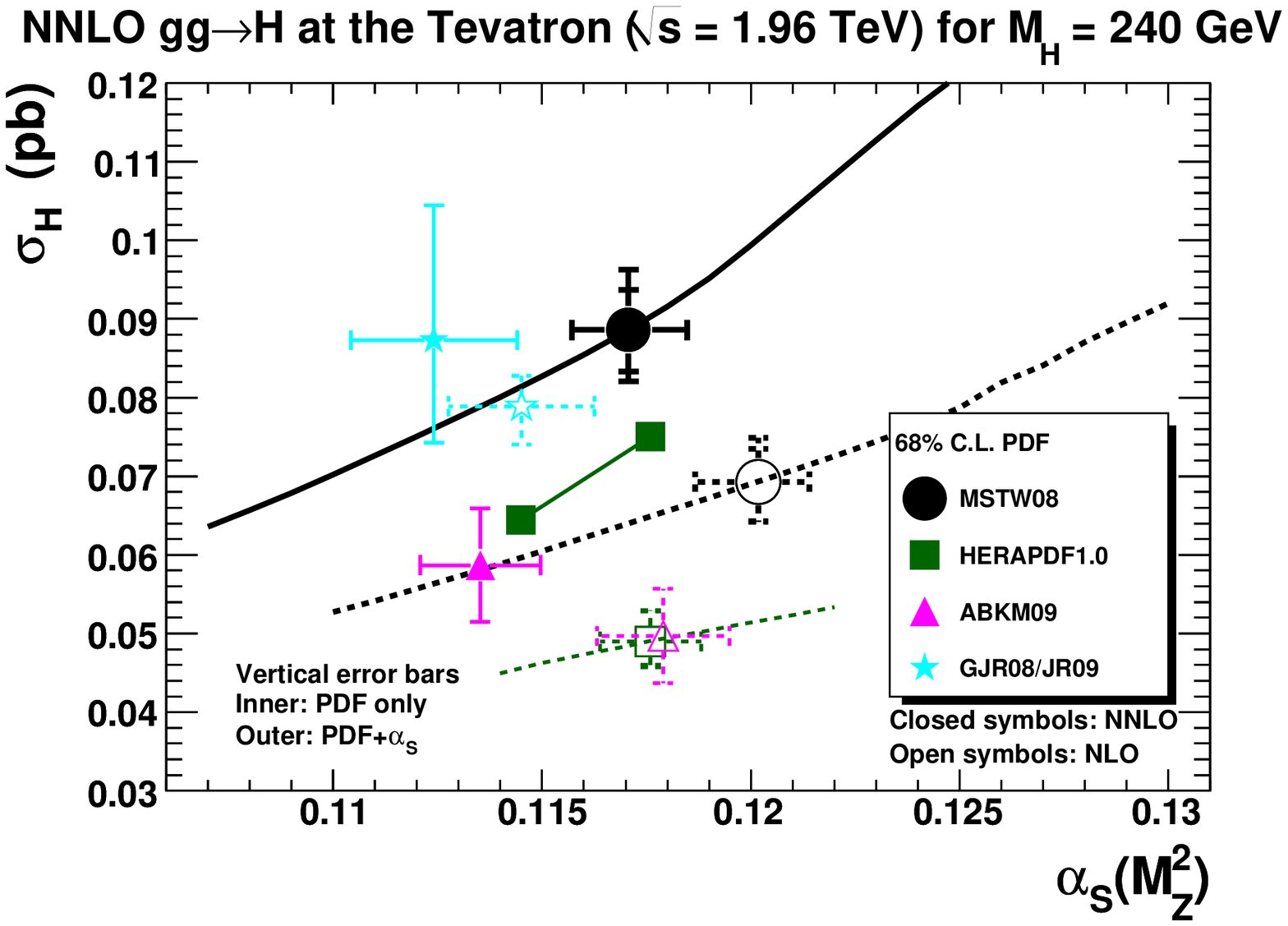}
  \end{minipage}%
  \begin{minipage}{0.5\textwidth}
    (f)\\
    \includegraphics[width=\textwidth]{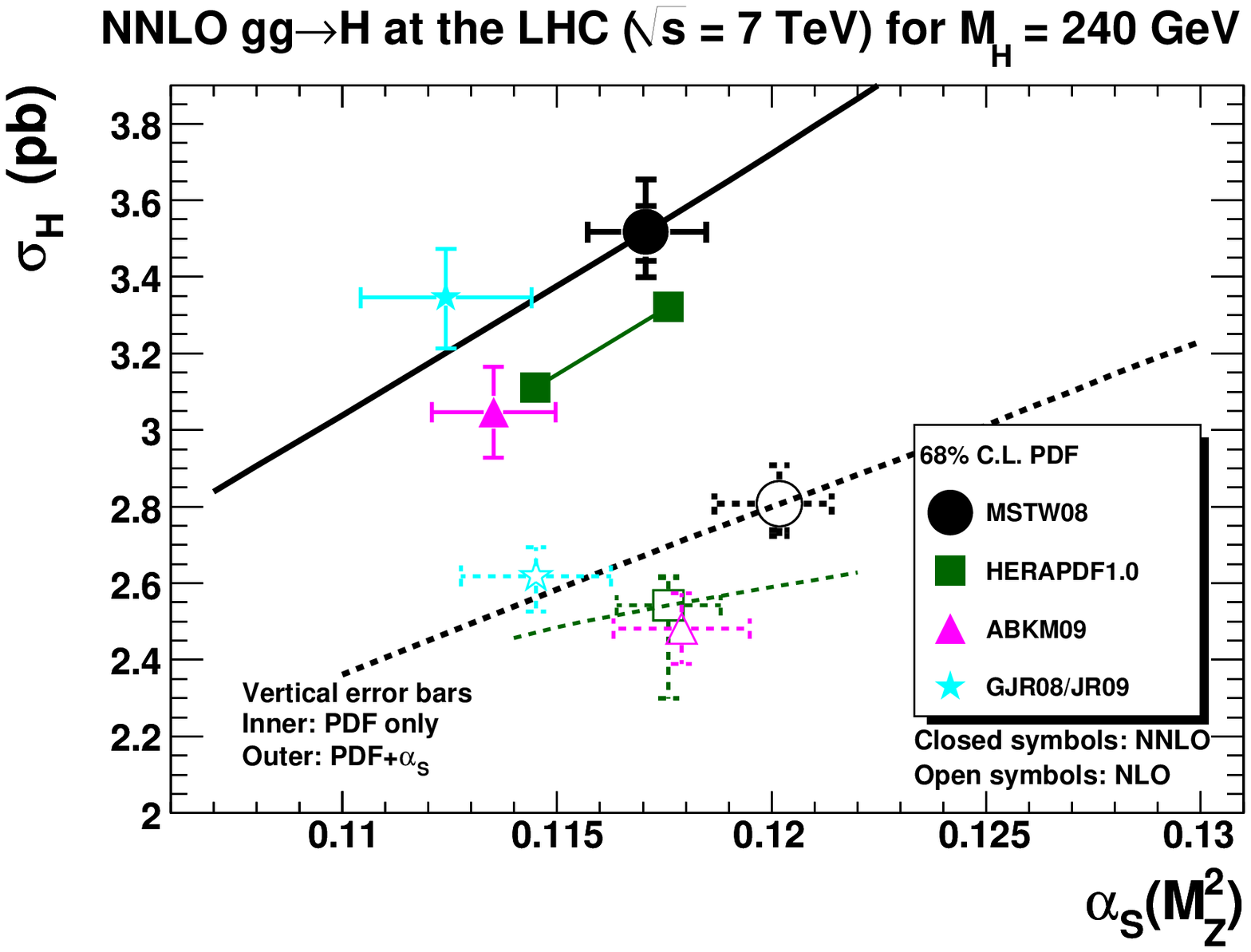}
  \end{minipage}
  \caption{$gg\to H$ total cross sections, plotted as a function of $\alpha_S(M_Z^2)$, at NNLO.}
  \label{fig:gghvsasmznnlo}
\end{figure}
The $gg\to H$ cross sections at the Tevatron and LHC start at $\mathcal{O}(\alpha_S^2)$ at LO, with anomalously large higher-order corrections, therefore they are directly sensitive to the value of $\alpha_S(M_Z^2)$.  Moreover, there is a known correlation between the value of $\alpha_S$ and the gluon distribution, which additionally affects the $gg\to H$ cross sections.  In figures~\ref{fig:gghvsasmznlo} and \ref{fig:gghvsasmznnlo} we show this sensitivity by plotting the Higgs cross sections versus $\alpha_S(M_Z^2)$ at the Tevatron and LHC for Higgs masses $M_H=\{120,180,240\}$~GeV.  We plot both NLO and NNLO predictions for a fixed scale choice $\mu_R=\mu_F=M_H$.  The format of the plots is that the markers are centred on the default $\alpha_S(M_Z^2)$ value and the corresponding predicted cross-section of each group.  The horizontal error bars span the $\alpha_S(M_Z^2)$ uncertainty, the inner vertical error bars span the ``PDF only'' uncertainty where possible (i.e.~not for ABKM09 or GJR08/JR09, where $\alpha_S$ is mixed with the input PDF parameters in the error matrix), and the outer vertical error bars span the PDF+$\alpha_S$ uncertainty.  The effect of the additional $\alpha_S$ uncertainty is sizeable.  The dashed lines at NLO or the solid lines at NNLO interpolate the cross-section predictions calculated with the alternative PDF sets provided by each group, represented by the smaller symbols in figure~\ref{fig:asmzvalues}.  The NNLO plots in figure~\ref{fig:gghvsasmznnlo} also show the NLO predictions (open symbols and dashed lines) together with the corresponding NNLO predictions (closed symbols and solid lines) to explicitly demonstrate how the size of the NNLO corrections depends on both the $\alpha_S(M_Z^2)$ choice and the PDF choice.  It is apparent from the plots that at least part of the MSTW08/ABKM09 discrepancy for Higgs cross sections is due to using quite different values of $\alpha_S(M_Z^2)$ at NNLO, specifically $\alpha_S(M_Z^2) = 0.1135\pm 0.0014$ for ABKM09~\cite{Alekhin:2009ni} compared to $\alpha_S(M_Z^2) = 0.1171\pm 0.0014$ for MSTW08~\cite{Martin:2009iq,Martin:2009bu}.  Comparing cross-section predictions at the same value of $\alpha_S(M_Z^2)$ would reduce the MSTW08/ABKM09 discrepancy at the LHC, but there would still be a significant discrepancy at the Tevatron (see also the later table~\ref{tab:nmc} in section~\ref{sec:nmcdata}).

\subsection{Theoretical uncertainties\texorpdfstring{ on $\alpha_S$}{}}

In ref.~\cite{Martin:2009bu} we gave a prescription for calculating the ``PDF+$\alpha_S$'' uncertainty on an observable such as a hadronic cross section, due to only \emph{experimental} errors on the data fitted.  An estimate of the \emph{theoretical} uncertainty on $\alpha_S$ was given as $\pm0.003$ at NLO and at most $\pm0.002$ at NNLO, where these values should be interpreted as roughly 1-$\sigma$ (68\%~C.L.).  However, this additional uncertainty was \emph{not} recommended to be propagated to the ``PDF+$\alpha_S$'' uncertainty on cross sections, in the same way that theoretical errors on PDFs are not generally provided and propagated to uncertainties on cross sections.  It was intended simply to be an estimate of how much the value of $\alpha_S(M_Z^2)$ might change if extracted at even higher orders.  It has subsequently been proposed (by Baglio and Djouadi) to include the theoretical uncertainty on $\alpha_S$ in the cross-section calculation for the $gg\to H$ process at the Tevatron~\cite{Baglio:2010um} and LHC~\cite{Baglio:2010ae}, which somewhat reduces the apparent inconsistency between MSTW08 and ABKM09 seen in figures~\ref{fig:ratioTEV} and \ref{fig:ratioLHC}.
\begin{figure}
  \centering
  \begin{minipage}{0.5\textwidth}
    (a)\\
    \includegraphics[width=\textwidth]{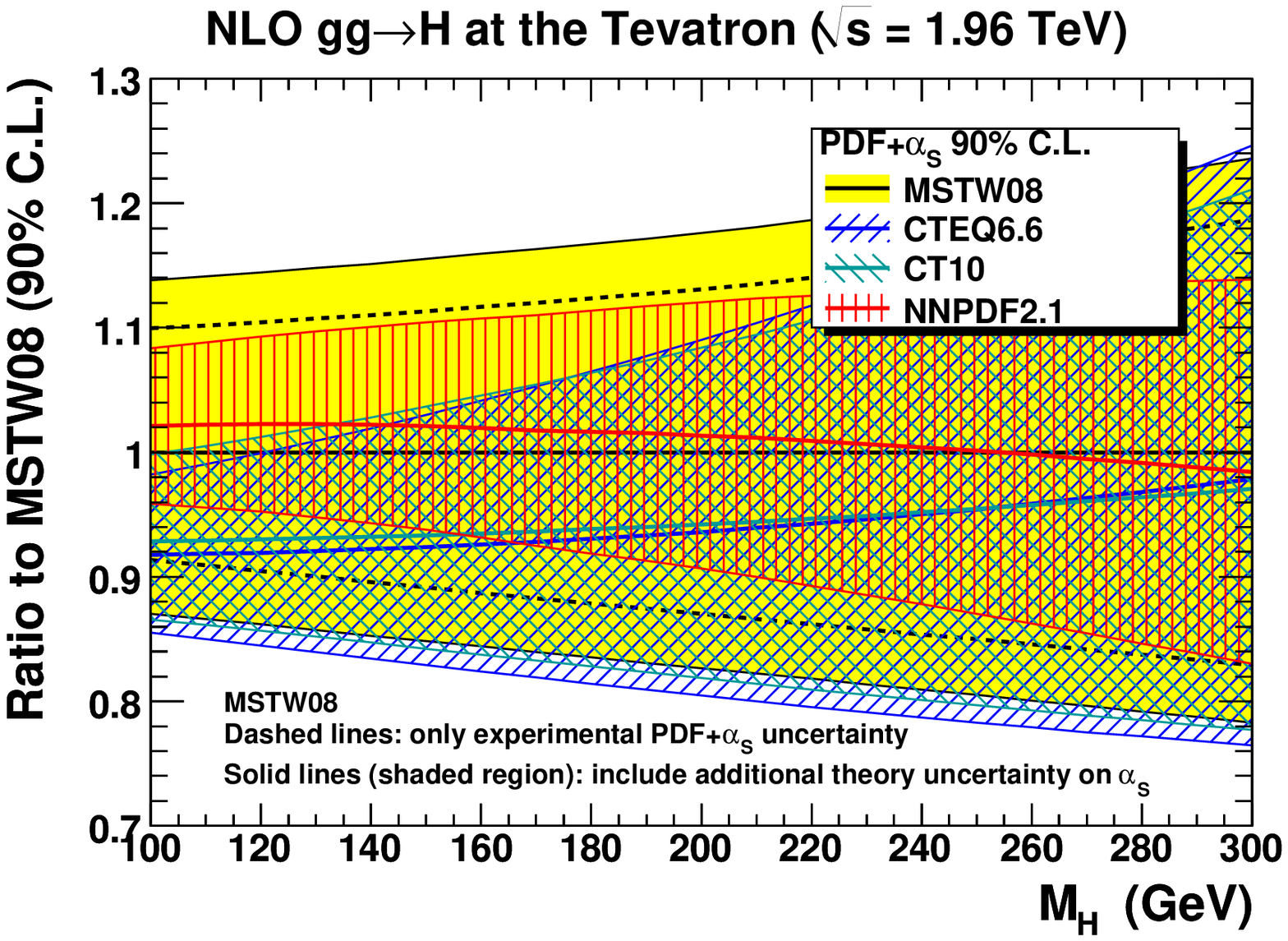}
  \end{minipage}%
  \begin{minipage}{0.5\textwidth}
    (b)\\
    \includegraphics[width=\textwidth]{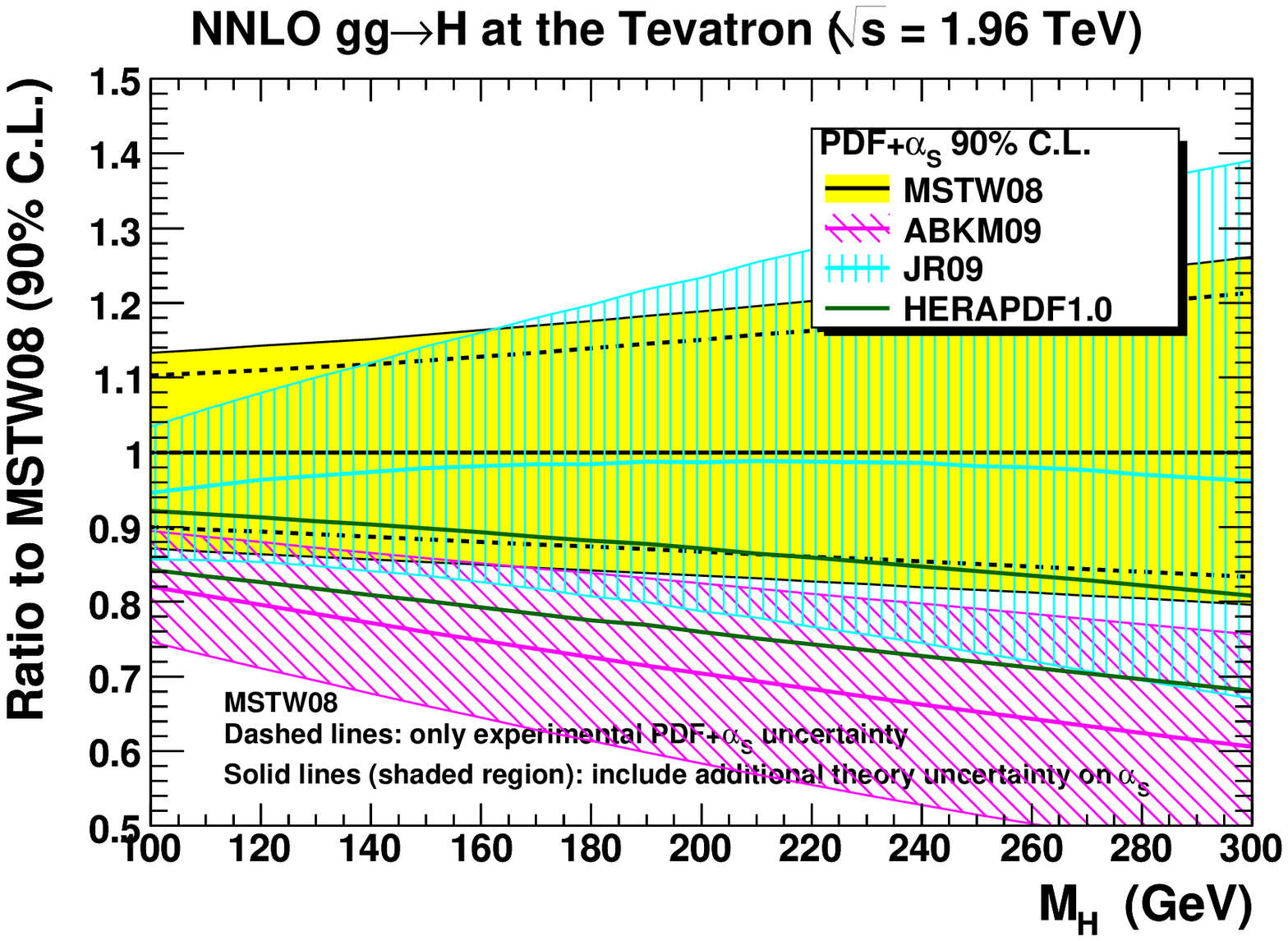}
  \end{minipage}
  \begin{minipage}{0.5\textwidth}
    (c)\\
    \includegraphics[width=\textwidth]{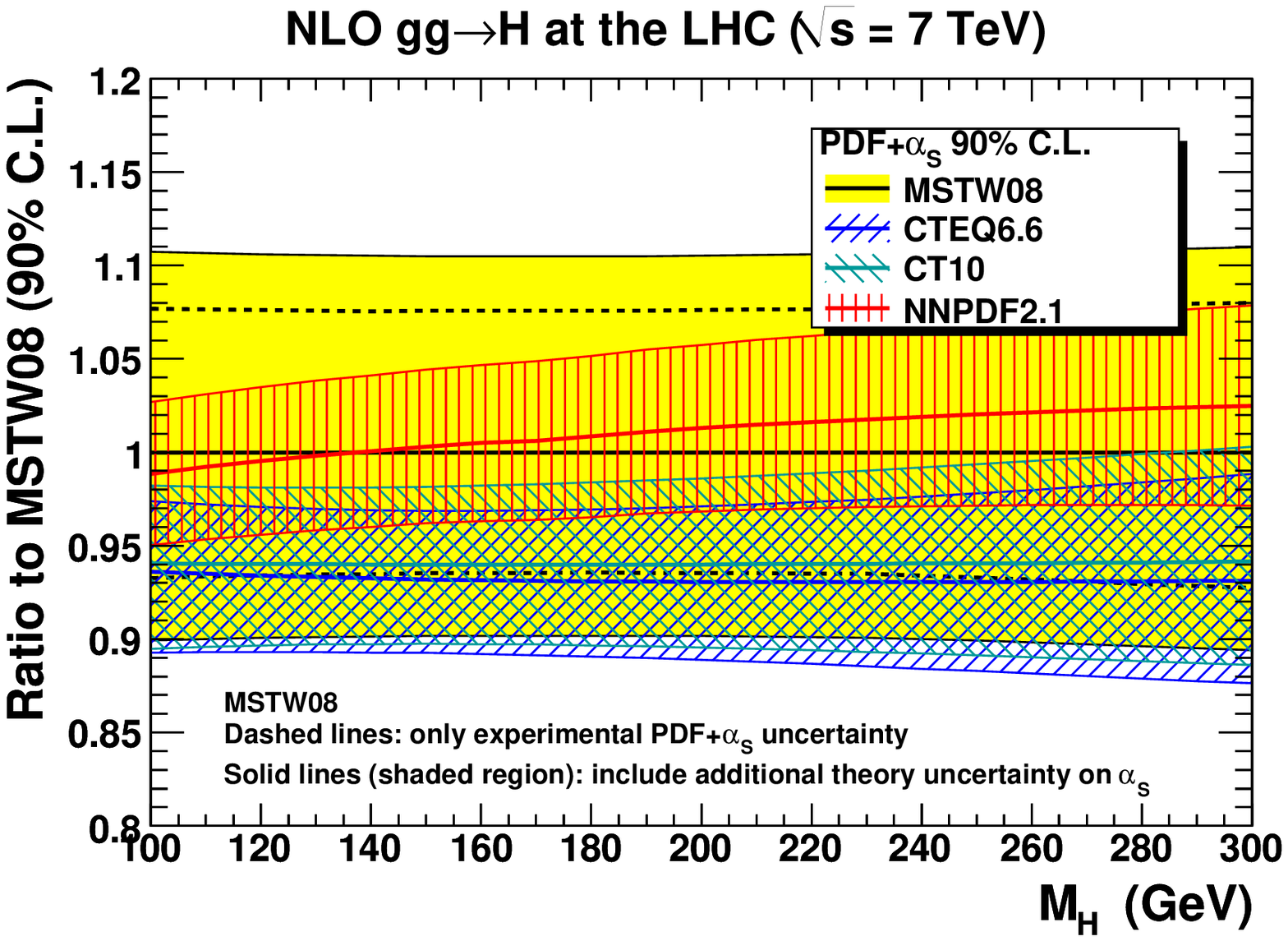}
  \end{minipage}%
  \begin{minipage}{0.5\textwidth}
    (d)\\
    \includegraphics[width=\textwidth]{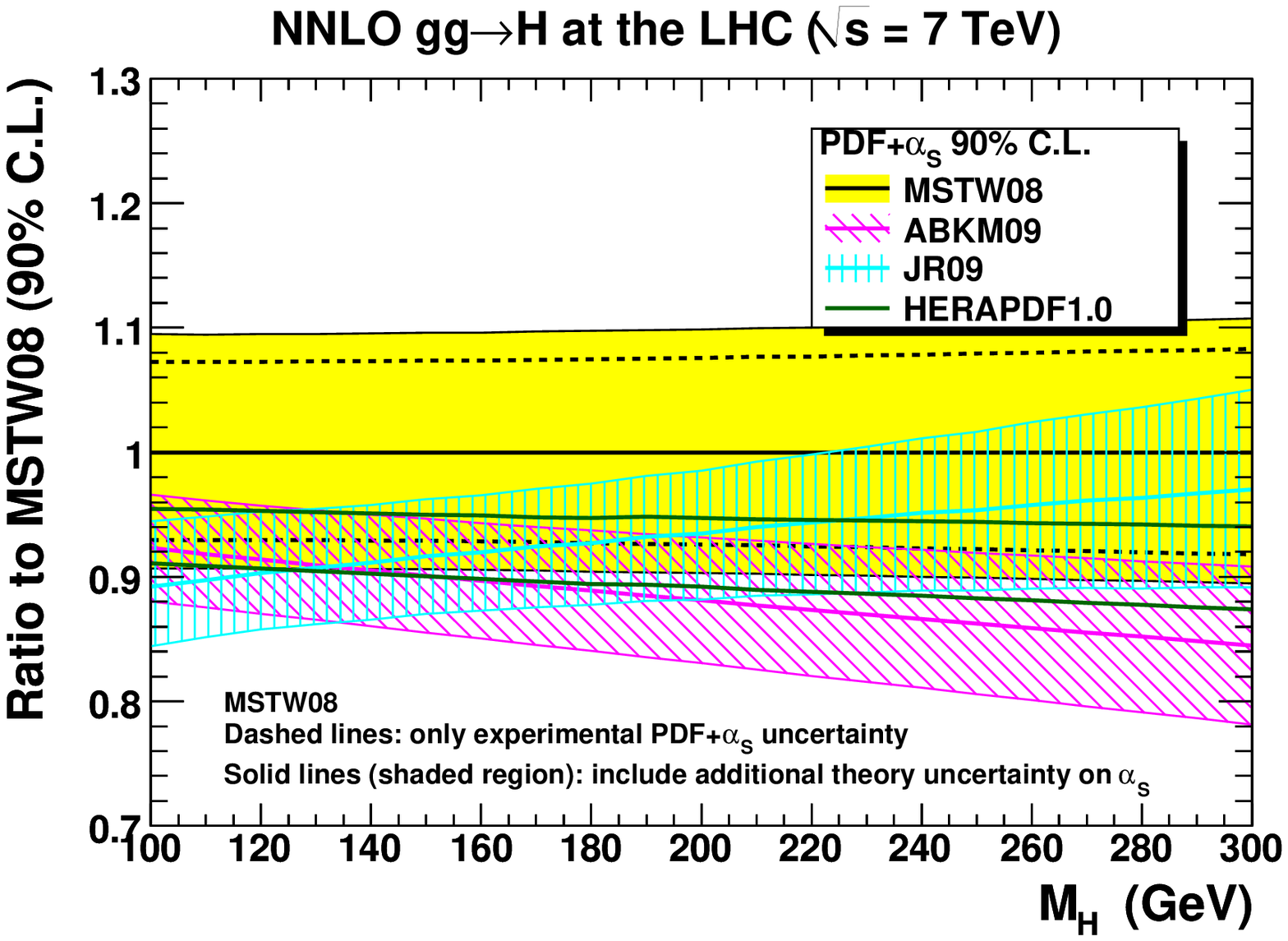}
  \end{minipage}
  \caption{Effect of including an additional theoretical uncertainty on $\alpha_S$ on the 90\%~C.L.~PDF+$\alpha_S$ uncertainty for $gg\to H$ at (a)~NLO at the Tevatron, (b)~NNLO at the Tevatron, (c)~NLO at the LHC, and (d)~NNLO at the LHC.}
  \label{fig:ratioTHas}
\end{figure}
In figure~\ref{fig:ratioTHas} we show the effect of adding in quadrature an additional theoretical uncertainty on $\alpha_S$ to the 90\%~C.L.~MSTW08 PDF+$\alpha_S$ uncertainty for the $gg\to H$ cross sections at both the Tevatron and LHC, at both NLO and NNLO, plotted as a function of the Higgs mass $M_H$.\footnote{We calculate the cross sections evaluated with $\alpha_S(M_Z^2)=0.120\pm0.003$ at NLO and $\alpha_S(M_Z^2)=0.117\pm0.002$ at NNLO, to determine the variation due to the additional theoretical uncertainty on $\alpha_S$ at 68\%~C.L., then we scale this uncertainty by 1.64485 to get the 90\%~C.L.~theoretical uncertainty.}  If a similar theoretical uncertainty on $\alpha_S$ was also added to the ABKM09 uncertainty band, which includes only experimental uncertainties, then the MSTW09 and ABKM09 uncertainty bands would overlap at the Tevatron, at least in the $M_H$ range shown here.  However, even if the additional $\alpha_S$ uncertainty is applied in this manner, it is misleading to claim that it leads to more of an agreement in the predictions obtained using the two PDF sets, since variations of cross sections with $\alpha_S$ are very highly correlated between different PDF sets.  We will see in the rest of this paper that differences between groups in $\alpha_S$ values, gluon distributions and Higgs cross sections are largely due to the selection of data fitted, and it is not the case that the discrepancies should be attributed to unaccounted theoretical uncertainties.

\section{\texorpdfstring{Constraints from jet production}{Jets} at the Tevatron} \label{sec:tevjets}

Here we present a quantitative study of the description of the Tevatron Run II inclusive jet data~\cite{Abulencia:2007ez,Aaltonen:2008eq,Abazov:2008hua} and dijet data~\cite{Abazov:2010fr} by different PDF sets.  The goal is to compare the description of Tevatron jet data in a similar manner to the benchmark cross-section study of ref.~\cite{bench7TeV}, i.e.~we use the \emph{same} code and settings for all NLO and NNLO PDF sets (with the correct $\alpha_S$ value for each set) to ensure that observed differences are only due to the PDF choice rather than any other factor.  We do not consider the less reliable Tevatron Run I data, which prefer a much harder high-$x$ gluon distribution~\cite{Martin:2009iq}, and are obtained using less sophisticated jet algorithms.  The three data sets on inclusive jet production from the Tevatron Run II~\cite{Abulencia:2007ez,Aaltonen:2008eq,Abazov:2008hua} were all found to be compatible~\cite{Martin:2009iq}.  The MSTW 2008 analysis~\cite{Martin:2009iq} included the CDF Run II inclusive jet data using the $k_T$ jet algorithm~\cite{Abulencia:2007ez} and the D{\O} Run II inclusive jet data using a cone jet algorithm~\cite{Abazov:2008hua}.  Consistency was checked with the CDF Run II inclusive jet data using the cone-based Midpoint jet algorithm~\cite{Aaltonen:2008eq}, but this data set was not included in the final MSTW08 fit, since it is essentially the same measurement (using 1.13~fb$^{-1}$) as ref.~\cite{Abulencia:2007ez} (using 1.0~fb$^{-1}$), differing mainly by the choice of jet algorithm.  The $k_T$ jet algorithm is theoretically preferred due to its property of infrared safety, and the corresponding CDF Run II data~\cite{Abulencia:2007ez} was already published and implemented in the MSTW08 analysis by the time the CDF Run II Midpoint data~\cite{Aaltonen:2008eq} appeared.  The D{\O} Run II inclusive jet data~\cite{Abazov:2008hua} and dijet data~\cite{Abazov:2010fr}, both defined using a cone jet algorithm, are also measured from essentially the same 0.7~fb$^{-1}$ of data, differing mainly by the kinematic binning, so as with the two CDF data sets it would be double-counting to include both in the same PDF extraction.  We will concentrate on the inclusive jet data (section~\ref{sec:incljet}), but we will also make a first quantitative comparison to the more recent D{\O} dijet data (section~\ref{sec:dijet}).  However, first in section~\ref{sec:chisq} we precisely define the goodness-of-fit measure used for the comparison of data and theory.

One obvious problem is that the complete NNLO partonic cross section ($\hat{\sigma}$) for inclusive jet production is currently unknown, and needs to be approximated with the NLO $\hat{\sigma}$ supplemented by 2-loop threshold corrections~\cite{Kidonakis:2000gi}, while even these 2-loop threshold corrections are unavailable for the dijet cross section.  We calculate jet cross sections using \textsc{fastnlo}~\cite{Kluge:2006xs} (based on \textsc{nlojet++}~\cite{Nagy:2001fj,Nagy:2003tz}), which includes these 2-loop threshold corrections.  Following the usual way of estimating theoretical uncertainties due to unknown higher-order corrections, we take different scale choices $\mu_R=\mu_F=\mu=\{p_T/2,p_T,2p_T\}$ as some indication of the theoretical uncertainty.  Smaller scale choices raise the partonic cross section, so favour softer high-$x$ gluon distributions~\cite{Martin:2009iq}, and the central $\mu=p_T$ was chosen for the final MSTW08 fit~\cite{Martin:2009iq}.  We comment on the scale dependence in section~\ref{sec:scaledep}, we present distributions of pulls and systematic shifts in section~\ref{sec:pulls}, we briefly discuss other collider data on jet cross sections in section~\ref{sec:otherjets}, then finally we summarise our findings in section~\ref{sec:summary}.

\subsection{Definition of goodness-of-fit\texorpdfstring{, $\chi^2$}{}} \label{sec:chisq}

It is important to account for \emph{correlated} systematic uncertainties of the experimental data points.  The full correlated error information is accounted for by using a goodness-of-fit ($\chi^2$) definition given by~\cite{Stump:2001gu,Pumplin:2002vw}
\begin{equation} \label{eq:chisqcorr}
  \chi^2 \;=\; \sum_{i=1}^{N_{\rm pts.}} \left(\frac{\hat{D}_{i}-T_{i}}{\sigma_{i}^{\rm uncorr.}}\right)^2 \;+\; \sum_{k=1}^{N_{\rm corr.}}r_{k}^2,
\end{equation}
where $T_{i}$ are the theory predictions and
\begin{equation} \label{eq:datashift}
  \hat{D}_{i} \equiv D_{i} - \sum_{k=1}^{N_{\rm corr.}}r_{k}\,\sigma_{k,i}^{\rm corr.}
\end{equation}
are the data points allowed to shift by the systematic errors in order to give the best fit.  Here, $i=1,\ldots,N_{\rm pts.}$ labels the individual data points and $k=1,\ldots,N_{\rm corr.}$ labels the individual correlated systematic errors.  The data points $D_{i}$ have uncorrelated (statistical and systematic) errors $\sigma_{i}^{\rm uncorr.}$ and correlated systematic errors $\sigma_{k,i}^{\rm corr.}$.  Minimising the $\chi^2$ in eq.~\eqref{eq:chisqcorr} with respect to the systematic shifts $r_{k}$ gives the analytic result that~\cite{Stump:2001gu,Pumplin:2002vw}
\begin{equation} \label{eq:rksolution}
  r_{k} = \sum_{k^\prime=1}^{N_{\rm corr.}} (A^{-1})_{kk^\prime} B_{k^\prime},
\end{equation}
where
\begin{equation}
  A_{kk^\prime} = \delta_{kk^\prime} + \sum_{i=1}^{N_{\rm pts.}}\frac{\sigma_{k,i}^{\rm corr.}\;\sigma_{k^\prime,i}^{\rm corr.}}{(\sigma_{i}^{\rm uncorr.})^2}, \quad \label{eq:Akkprime}
  B_{k} = \sum_{i=1}^{N_{\rm pts.}}\frac{\sigma_{k,i}^{\rm corr.}\left(D_{i}-T_{i}\right)}{(\sigma_{i}^{\rm uncorr.})^2},
\end{equation}
and $\delta_{kk^\prime}$ is the Kronecker delta.  Therefore, the optimal shifts of the data points by the systematic errors, eq.~\eqref{eq:datashift}, are solved for analytically.  Here we use the same notation\footnote{We note a typo, already pointed out in ref.~\cite{Martin:2010db}, in the formula for $A_{kk^\prime}$ in eq.~(40) of ref.~\cite{Martin:2009iq} where $\sigma_{i}^{\rm uncorr.}$ should appear squared.  This typo is corrected in eq.~\eqref{eq:Akkprime} above.} as in the MSTW08 paper~\cite{Martin:2009iq}.  We treat the luminosity uncertainty as any other correlated systematic.  However, we find that the relevant systematic shift $r_{\rm lumi.}\sim 3$--$5$ for some PDF sets with soft high-$x$ gluon distributions (e.g.~ABKM09 and HERAPDF1.0), which is clearly completely unreasonable, as it means that the data points are normalised downwards by 3--5 times the nominal luminosity uncertainty (around 6\% for both CDF and D{\O}).  The penalty term $r_{\rm lumi.}^2$ will contribute only 9--25 units to the total $\chi^2$ given by eq.~\eqref{eq:chisqcorr}, which can therefore still lead to reasonably low overall $\chi^2$ values (see appendix~\ref{sec:freenorm} for details).

It is the usual situation at collider experiments that the luminosity determination is common to all cross sections measured from a given data set (see, for example, refs.~\cite{Papadimitriou:2008zza,Burkhardt:1347440}), so the requirement of a single common luminosity is mandatory when fitting multiple measurements taken during a single running period.
\begin{figure}
  \centering
  \begin{minipage}{0.5\textwidth}
    (a)\\
    \includegraphics[width=\textwidth]{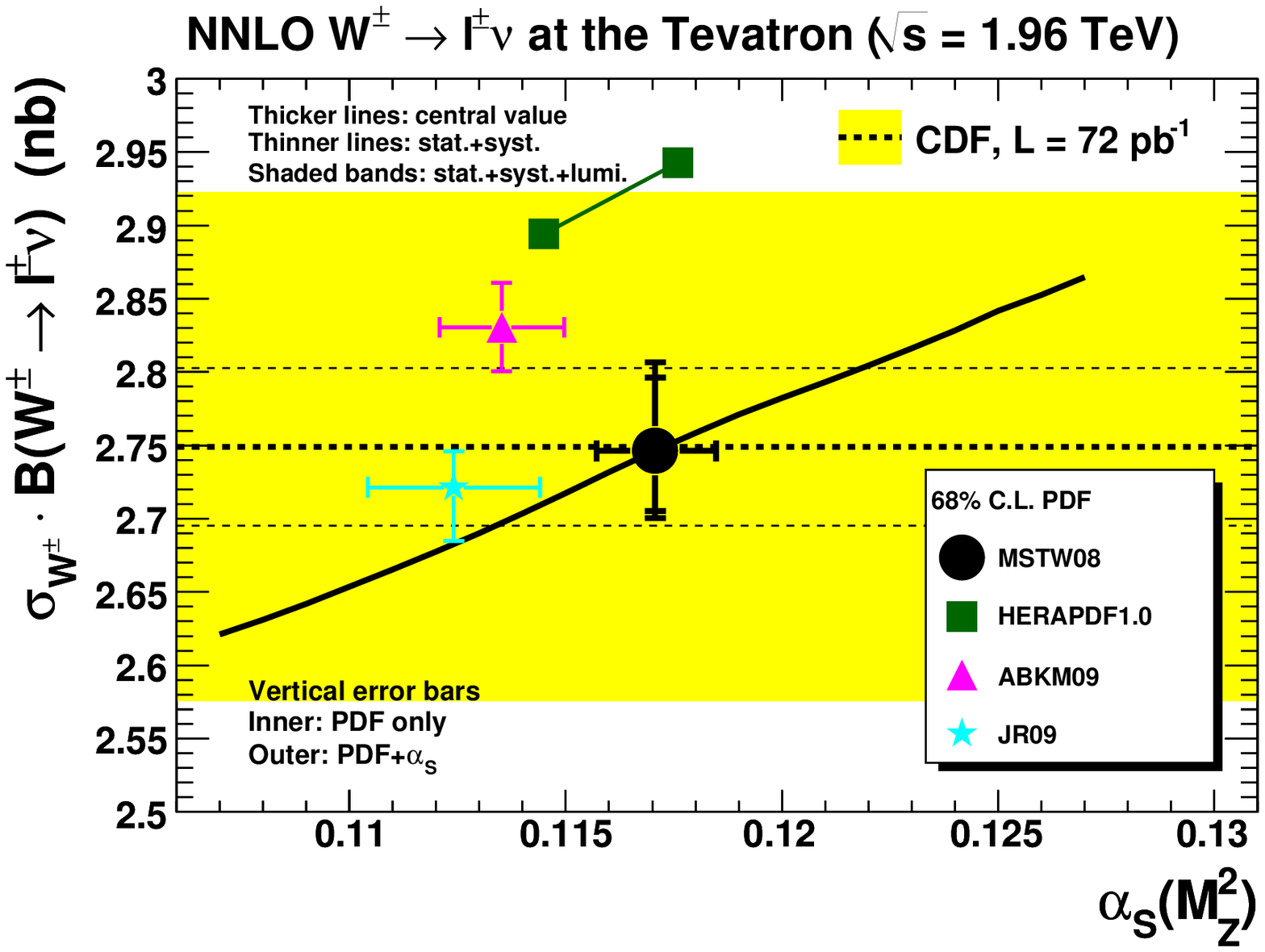}
  \end{minipage}%
  \begin{minipage}{0.5\textwidth}
    (b)\\
    \includegraphics[width=\textwidth]{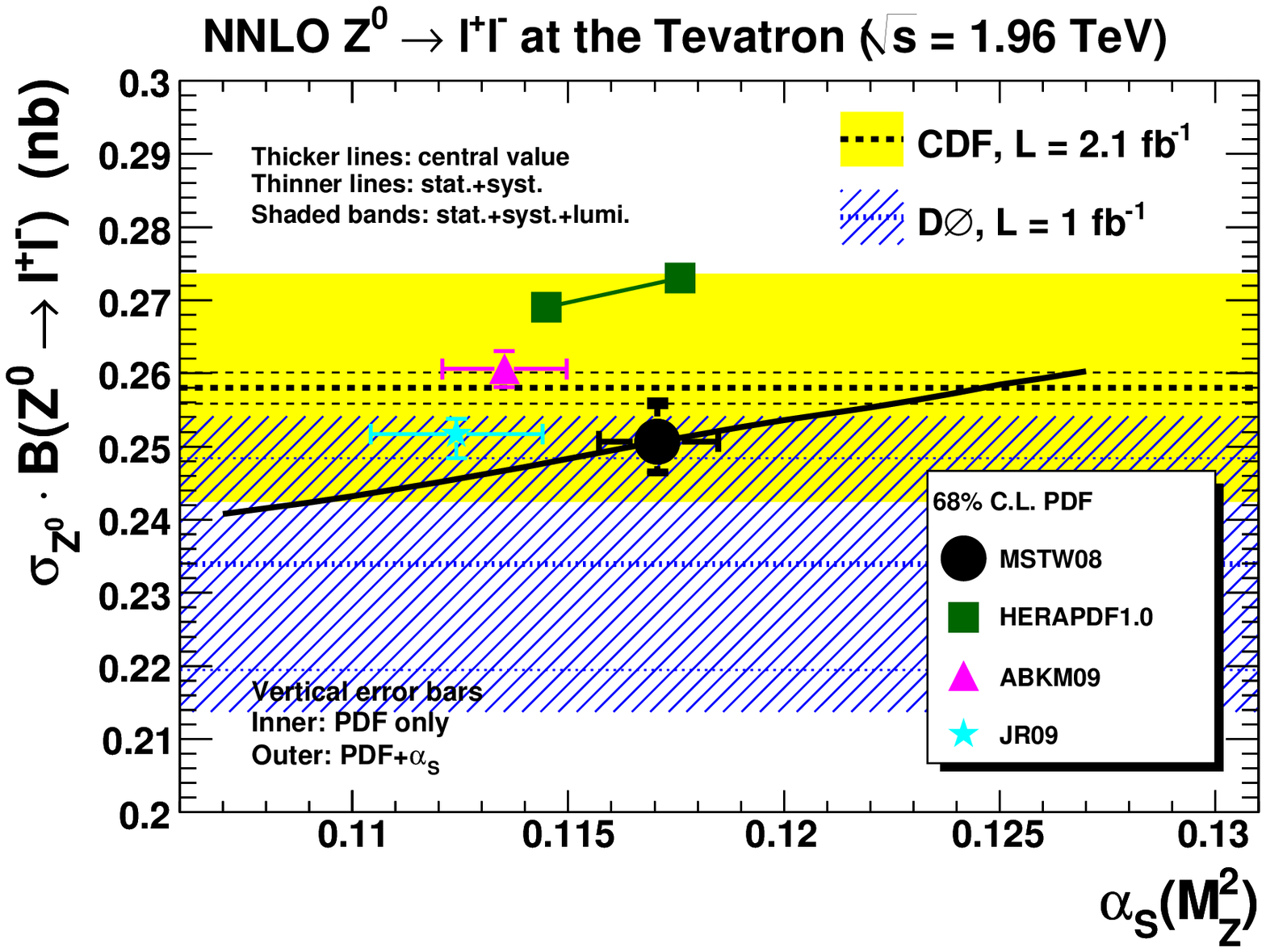}
  \end{minipage}
  \caption{NNLO predictions for (a)~$W$ and (b)~$Z$ total cross sections at the Tevatron Run II, plotted as a function of $\alpha_S(M_Z^2)$, compared to CDF $W$~\cite{Abulencia:2005ix}, CDF $Z$~\cite{Aaltonen:2010zza} and D{\O} $Z$~\cite{Abazov:2008ez} data.}
  \label{fig:wzTEV}
\end{figure}
In figure~\ref{fig:wzTEV} we compare NNLO predictions for the $W$ and $Z$ total cross sections at the Tevatron Run II, calculated in the zero-width approximation with settings described in ref.~\cite{bench7TeV}; see also similar comparisons in ref.~\cite{Alekhin:2010dd}.  The format of the plots in figure~\ref{fig:wzTEV} is the same as for the $gg\to H$ cross sections in section~\ref{sec:ggHalphaS}, i.e.~we show the cross-section predictions plotted against $\alpha_S(M_Z^2)$.  We compare to CDF Run II data on $W$~\cite{Abulencia:2005ix} and $Z$~\cite{Aaltonen:2010zza} total cross sections, and to D{\O} Run II data on the $Z$ total cross section~\cite{Abazov:2008ez}.  The thicker horizontal lines in figure~\ref{fig:wzTEV} indicate the central value of each experimental measurement, the thinner horizontal lines indicate the statistical and systematic (excluding luminosity) uncertainties added in quadrature, while the shaded regions indicate the total uncertainty obtained by also adding the luminosity uncertainty in quadrature.  The plotted CDF $Z$ measurement with 2.1~fb$^{-1}$~\cite{Aaltonen:2010zza} supersedes the earlier $Z$ measurement with 72~pb$^{-1}$~\cite{Abulencia:2005ix}, but both measurements are dominated by the (common) luminosity uncertainty.  The D{\O} experiment has not published any dedicated $W$ and $Z$ total cross-section measurements from Run II at the Tevatron.  The D{\O} $Z$ total cross section shown in figure~\ref{fig:wzTEV}(b) was obtained as part of the $Z$+jet measurement~\cite{Abazov:2008ez}.  The CDF measurement~\cite{Aaltonen:2010zza} is defined as the $Z/\gamma^*\to ee$ cross section in an invariant mass range $M_{ee}\in[66,116]$~GeV, while the D{\O} measurement~\cite{Abazov:2008ez} is defined as the $Z/\gamma^*\to \mu\mu$ cross section in an invariant mass range $M_{\mu\mu}\in[65,115]$~GeV.  We have therefore multiplied the CDF and D{\O} data by factors of 1.006 and 1.004, respectively, derived using the \textsc{vrap} code~\cite{Anastasiou:2003ds} at NNLO with MSTW08 PDFs, to correct to the $Z$-only cross section with $M_{\ell\ell}=M_Z$.  We note from figure~\ref{fig:wzTEV} that the MSTW08, ABKM09 and JR09 NNLO predictions for the $W$ and $Z$ total cross sections at the Tevatron are in good agreement with the CDF data~\cite{Abulencia:2005ix,Aaltonen:2010zza}, and lie around 1-$\sigma$ above the D{\O} data~\cite{Abazov:2008ez}.  In the MSTW08 fit~\cite{Martin:2009iq}, the luminosity shift for the CDF jet data was correctly tied to be the same as for the more-constraining CDF $Z$ rapidity distribution, ${\rm d}\sigma_Z/{\rm d}y$~\cite{Aaltonen:2010zza}, which therefore effectively acted as a luminosity monitor.  The optimal CDF normalisation in the MSTW08 NNLO fit~\cite{Martin:2009iq} was found to be very close to the nominal value, therefore it is not surprising that the CDF $Z$ total cross section is well described in figure~\ref{fig:wzTEV}(b).  The D{\O} experiment instead measured the $Z$ rapidity \emph{shape} distribution, $(1/\sigma_Z){\rm d}\sigma_Z/{\rm d}y$~\cite{Abazov:2007jy}, also included in the MSTW08 fit, which is one reason why the D{\O} jet data were found to be less constraining than the CDF jet data; see ref.~\cite{Martin:2009bu}.  The optimal D{\O} normalisation in the MSTW08 NNLO fit~\cite{Martin:2009iq}, determined only from jet data, was around 1-$\sigma$ above the nominal value, consistent with the D{\O} $Z$ total cross section shown in figure~\ref{fig:wzTEV}(b).  If the Tevatron jet data were normalised downwards by 20--30\% (i.e.~3--5 times the luminosity uncertainty), the Tevatron $W$ and $Z$ total cross sections would need to normalised downwards by the same amount, resulting in complete disagreement with all theory predictions shown in figure~\ref{fig:wzTEV}.  This example illustrates the utility of simultaneously fitting $W$ and $Z$ cross sections together with jet cross sections at the Tevatron (and LHC).  The luminosity shifts, common to both data sets, are effectively determined by the more precise $W$ and $Z$ cross sections.  The luminosity uncertainty is then effectively removed from the jet cross sections, thereby allowing the jet data to provide a tighter constraint on the gluon distribution (and $\alpha_S$).

To avoid these completely unrealistic luminosity shifts, $r_{\rm lumi.}\sim 3$--$5$, without going into the complication of simultaneously including $W$ and $Z$ cross sections in the $\chi^2$ computation, we will calculate the $\chi^2$ values for the Tevatron jet data using eq.~\eqref{eq:chisqcorr}, but with the simple restriction that the relevant systematic shift $|r_{\rm lumi.}|\le 1$.  More practically, this means that if $|r_{\rm lumi.}|>1$ for any particular PDF set, we fix $r_{\rm lumi.}$ at $\pm1$ and reevaluate eq.~\eqref{eq:chisqcorr} with the luminosity removed from the list of correlated systematics.  However, we note from figure~\ref{fig:wzTEV} that the ABKM09 predictions are slightly above the central value of the CDF $W$ and $Z$ data, and the HERAPDF1.0 predictions are higher by around 1-$\sigma$, while both ABKM09 and HERAPDF1.0 lie above the 1-$\sigma$ limit of the D{\O} $Z$ data.  Allowing luminosity shifts downwards by even 1-$\sigma$ is therefore distinctly generous, particularly for HERAPDF1.0, and upwards luminosity shifts would bring the ABKM09 and HERAPDF1.0 predictions into better agreement with the CDF $W$ and $Z$ data, and especially the D{\O} $Z$ data.  Therefore, it should be understood that the $\chi^2$ values quoted in the tables we will present in section~\ref{sec:incljet} and \ref{sec:dijet} are rather optimistic for ABKM09 and HERAPDF1.0, and more realistic constraints in the luminosity shifts would result in even worse $\chi^2$ values.

The form of eq.~\eqref{eq:chisqcorr} is slightly different from the treatment of normalisation uncertainties adopted in eq.~(38) of the MSTW08 paper~\cite{Martin:2009iq}, but is the form used, for example, in the CT10 analysis~\cite{Lai:2010vv}.  Rescaling only the central value of the data in eq.~\eqref{eq:chisqcorr}, but not the uncertainties, leads to so-called ``d'Agostini bias''~\cite{D'Agostini:1993uj,Ball:2009qv}.  However, since we are only comparing and not fitting PDFs, we use the simpler form of eq.~\eqref{eq:chisqcorr} which has the major advantage that all shifts $r_{k}$ can be solved for analytically.  A more sophisticated approach to the treatment of normalisation uncertainties may somewhat lessen the preference of some PDF sets for large downwards luminosity shifts, but should not affect our main conclusions.  The normalisation uncertainties were treated as multiplicative rather than additive in the MSTW08 fit~\cite{Martin:2009iq}, i.e.~the uncertainties were correctly rescaled to reduce bias.  Moreover, large normalisation shifts for any experiment were discouraged through use of a quartic penalty term rather than the usual quadratic penalty term in eq.~\eqref{eq:chisqcorr}.  These small differences in $\chi^2$ definition mean that the MSTW08 $\chi^2$ values we quote here will be slightly different from the values quoted in ref.~\cite{Martin:2009iq}.

Even considering the constraint on the CDF and D{\O} luminosities from the comparison to the weak boson cross sections (see figure~\ref{fig:wzTEV}), it might be considered that imposing $|r_{\rm lumi.}|<1$ is too restrictive if the luminosity uncertainty is assumed to be Gaussian.  However, as another reason for limiting the luminosity shifts to some extent, we note that it has been claimed (see section 6.7.4 on ``Normalizations'', pg.~170" in \cite{Devenish:2004pb}) that, for many experiments, quoted normalisation uncertainties represent the limits of a box-shaped distribution rather than the standard deviation of a Gaussian distribution.  This was one motivation for the more severe quartic penalty term for normalisation uncertainties in the MSTW08 analysis; see discussion in section~5.2.1 of ref.~\cite{Martin:2009iq}.  Nevertheless, if we instead impose $|r_{\rm lumi.}|<2$ rather than $|r_{\rm lumi.}|<1$, then the change in the $\chi^2/N_{\rm pts.}$ values for the most relevant ABKM09 NNLO PDF set with $\mu=p_T$ is $\{2.76\to2.10, 1.94\to1.81, 1.55\to1.55, 1.49\to1.41\}$ for the \{CDF $k_T$~\cite{Abulencia:2007ez}, CDF Midpoint~\cite{Aaltonen:2008eq}, D{\O} inclusive~\cite{Abazov:2008hua}, D{\O} dijet~\cite{Abazov:2010fr}\} data, respectively, so there is not a significant improvement in the $\chi^2$ values.  However, as discussed above, our main argument does not rely on the precise form of the uncertainty on the luminosity determination, but that we can use the $W$ and $Z$ cross sections as a \emph{luminosity monitor}, where the predictions have small theoretical uncertainties, effectively providing an accurate luminosity determination independently of the CDF and D{\O} values.  Combining these arguments, we consider allowing luminosity shifts downwards by more than 1-$\sigma$ to be excessively generous.

There is a clear trade-off between the systematic shifts $r_k$ and the parameters of the gluon distribution.  Deficiencies in the theory calculation can be masked to some extent by large systematic shifts, therefore it is important to check that the optimal $r_k$ values are not unreasonable.  This is straightforward when using a $\chi^2$ definition like eq.~\eqref{eq:chisqcorr}, but is more difficult using an equivalent form written in terms of the experimental covariance matrix,
\begin{equation} \label{eq:covmat}
  V_{ii^\prime} \;=\; \delta_{ii^\prime}\,(\sigma_{i}^{\rm uncorr.})^2\;+\;\sum_{k=1}^{N_{\rm corr.}}\sigma_{k,i}^{\rm corr.}\,\sigma_{k,i^\prime}^{\rm corr.}.
\end{equation}
Then eq.~\eqref{eq:chisqcorr} is equivalent~\cite{Stump:2001gu} to the more traditional $\chi^2$ form written in terms of the inverse of the experimental covariance matrix:
\begin{equation} \label{eq:chisqcov}
  \chi^2 \;=\; \sum_{i=1}^{N_{\rm pts.}}\sum_{i^\prime=1}^{N_{\rm pts.}}(D_i-T_i)\left(V^{-1}\right)_{ii^\prime}(D_{i^\prime}-T_{i^\prime}),
\end{equation}
as used by the ABKM and NNPDF fitting groups.  More precisely, NNPDF use a refinement to treat normalisation errors as multiplicative~\cite{Ball:2009qv}, while Alekhin (ABKM) treats all correlated systematic errors as multiplicative~\cite{Alekhin:1996za,Alekhin:2000ch}.

It can easily be seen from eqs.~\eqref{eq:covmat} and \eqref{eq:chisqcov} that treating the correlated errors as uncorrelated ($V_{ii^\prime}\propto \delta_{ii^\prime}$) leads to the familiar form of
\begin{equation} \label{eq:chisqtot}
  \chi^2 \;=\; \sum_{i=1}^{N_{\rm pts.}} \left(\frac{D_{i}-T_{i}}{\sigma_{i}^{\rm tot.}}\right)^2,
\end{equation}
where the total error is simply obtained by adding all errors in quadrature,
\begin{equation}
  \left(\sigma_{i}^{\rm tot.}\right)^2 = \left(\sigma_{i}^{\rm uncorr.}\right)^2 + \sum_{k=1}^{N_{\rm corr.}}\left(\sigma_{k,i}^{\rm corr.}\right)^2.
\end{equation}

\subsection{Inclusive jet production} \label{sec:incljet}

\begin{table}
  \centering
  \begin{tabular}{|l|r|r|r|}
    \hline
    NLO PDF (with NLO $\hat{\sigma}$) & $\mu=p_T/2$ & $\mu=p_T$ & $\mu=2p_T$ \\
    \hline
    MRST04 & 1.06 (0.59) & 0.94 (0.31) & 0.84 (0.31) \\
    MSTW08 & {\bf 0.75} (0.30) & {\bf 0.68} (0.28) & 0.91 (0.84) \\
    CTEQ6.6 & 1.25 (0.14) & 1.66 (0.20) & 2.38 (0.84) \\
    CT10 & 1.03 (0.13) & 1.20 (0.19) & 1.81 (0.84) \\
    NNPDF2.1 & {\bf 0.74} (0.29) & {\bf 0.82} (0.25) & 1.23 (0.69) \\
    HERAPDF1.0 & 2.43 (0.39) & 3.26 (0.66) & 4.03 (1.67) \\
    HERAPDF1.5 & 2.26 (0.40) & 3.05 (0.66) & 3.80 (1.66) \\
    ABKM09 & 1.62 (0.52) & 2.21 (0.85) & 3.26 (2.10) \\
    GJR08 & 1.36 (0.23) & 0.94 (0.13) & {\bf 0.79} (0.36) \\
    \hline \multicolumn{4}{c}{} \\ \hline
    NNLO PDF (with NLO+2-loop $\hat{\sigma}$) & $\mu=p_T/2$ & $\mu=p_T$ & $\mu=2p_T$ \\
    \hline
    MRST06 & 2.96 (1.24) & 1.21 (1.18) & 1.03 (0.84) \\
    MSTW08 & 1.39 (0.42) & {\bf 0.69} (0.44) & 0.97 (0.48) \\
    HERAPDF1.0, $\alpha_S(M_Z^2)=0.1145$ & 2.64 (0.36) & 2.15 (0.36) & 2.20 (0.46) \\
    HERAPDF1.0, $\alpha_S(M_Z^2)=0.1176$ & 2.24 (0.35) & 1.17 (0.32) & 1.23 (0.31) \\
    ABKM09 & 2.55 (0.82) & 2.76 (0.89) & 3.41 (1.17) \\
    JR09 & {\bf 0.75} (0.37) & 1.26 (0.41) & 2.21 (0.49) \\
    \hline
  \end{tabular}
  \caption{Values of $\chi^2/N_{\rm pts.}$ for the CDF Run II inclusive jet data using the $k_T$ jet algorithm~\cite{Abulencia:2007ez} with $N_{\rm pts.}=76$ and $N_{\rm corr.}=17$, for different PDF sets and different scale choices $\mu_R=\mu_F=\mu=\{p_T/2,p_T,2p_T\}$.  The $\chi^2$ values are calculated accounting for all 17 sources of correlated systematic uncertainty, using eq.~\eqref{eq:chisqcorr}, including the 5.8\% normalisation uncertainty due to the luminosity determination.  At most a 1-$\sigma$ shift in normalisation is allowed.  We highlight in bold those values lying inside the 90\% C.L.~region, defined by eq.~\eqref{eq:90percentCL}, which gives $\chi^2/N_{\rm pts.} < 0.83$.  The values of $\chi^2/N_{\rm pts.}$ computed using eq.~\eqref{eq:chisqtot}, simply adding all experimental uncertainties in quadrature (including luminosity), are shown in brackets in the table.  If the theory prediction was identically zero, the $\chi^2/N_{\rm pts.}$ values would be 25.0 (37.5) with (without) accounting for correlations between systematic uncertainties.\label{tab:cdfkt}}
\end{table}
\begin{table}
  \centering
  \begin{tabular}{|l|r|r|r|}
    \hline
    NLO PDF (with NLO $\hat{\sigma}$) & $\mu=p_T/2$ & $\mu=p_T$ & $\mu=2p_T$ \\
    \hline
    MRST04 & 2.14 (1.42) & 2.01 (0.54) & 1.57 (0.26) \\
    MSTW08 & 1.52 (0.61) & {\bf 1.40} (0.27) & {\bf 1.16} (0.73) \\
    CTEQ6.6 & 1.93 (0.41) & 1.98 (0.21) & 1.78 (0.78) \\
    CT10 & 1.75 (0.38) & 1.69 (0.19) & 1.50 (0.76) \\
    NNPDF2.1 & 1.69 (0.60) & 1.56 (0.25) & 1.44 (0.60) \\
    HERAPDF1.0 & 2.61 (0.23) & 2.73 (0.49) & 2.53 (1.58) \\
    HERAPDF1.5 & 2.48 (0.24) & 2.60 (0.49) & 2.44 (1.57) \\
    ABKM09 & 1.56 (0.26) & 1.68 (0.65) & 1.69 (2.01) \\
    GJR08 & 2.11 (0.71) & 1.75 (0.24) & 1.52 (0.31) \\
    \hline \multicolumn{4}{c}{} \\ \hline
    NNLO PDF (with NLO+2-loop $\hat{\sigma}$) & $\mu=p_T/2$ & $\mu=p_T$ & $\mu=2p_T$ \\
    \hline
    MRST06 & 2.83 (2.25) & 2.08 (1.56) & 2.11 (0.86) \\
    MSTW08 & 1.67 (0.62) & {\bf 1.39} (0.43) & 1.62 (0.37) \\
    HERAPDF1.0, $\alpha_S(M_Z^2)=0.1145$ & 2.20 (0.25) & 2.06 (0.27) & 2.19 (0.40) \\
    HERAPDF1.0, $\alpha_S(M_Z^2)=0.1176$ & 2.08 (0.55) & 1.76 (0.33) & 1.99 (0.23) \\
    ABKM09 & 1.70 (0.50) & 1.94 (0.71) & 2.26 (1.12) \\
    JR09 & 1.57 (0.41) & 2.05 (0.36) & 2.82 (0.39) \\
    \hline
  \end{tabular}
  \caption{Values of $\chi^2/N_{\rm pts.}$ for the CDF Run II inclusive jet data using the cone-based Midpoint jet algorithm~\cite{Aaltonen:2008eq} with $N_{\rm pts.}=72$ and $N_{\rm corr.}=25$, for different PDF sets and different scale choices $\mu_R=\mu_F=\mu=\{p_T/2,p_T,2p_T\}$.  The $\chi^2$ values are calculated accounting for all 25 sources of correlated systematic uncertainty, using eq.~\eqref{eq:chisqcorr}, including the 5.8\% normalisation uncertainty due to the luminosity determination.  At most a 1-$\sigma$ shift in normalisation is allowed.  We highlight in bold those values lying inside the 90\% C.L.~region, defined by eq.~\eqref{eq:90percentCL}, which gives $\chi^2/N_{\rm pts.} < 1.43$.  The values of $\chi^2/N_{\rm pts.}$ computed using eq.~\eqref{eq:chisqtot}, simply adding all experimental uncertainties in quadrature (including luminosity), are shown in brackets in the table.  If the theory prediction was identically zero, the $\chi^2/N_{\rm pts.}$ values would be 5.30 (38.8) with (without) accounting for correlations between systematic uncertainties.\label{tab:cdfmid}}
\end{table}
\begin{table}
  \centering
  \begin{tabular}{|l|r|r|r|}
    \hline
    NLO PDF (with NLO $\hat{\sigma}$) & $\mu=p_T/2$ & $\mu=p_T$ & $\mu=2p_T$ \\
    \hline
    MRST04 & 1.86 (2.89) & 1.34 (0.96) & {\bf 1.11} (0.30) \\
    MSTW08 & 1.45 (0.89) & {\bf 1.08} (0.20) & {\bf 1.05} (1.22) \\
    CTEQ6.6 & 1.62 (1.15) & 1.56 (0.59) & 1.61 (1.35) \\
    CT10 & 1.39 (0.88) & 1.26 (0.37) & 1.32 (1.29) \\
    NNPDF2.1 & 1.41 (0.87) & 1.29 (0.20) & 1.22 (0.96) \\
    HERAPDF1.0 & 1.73 (0.27) & 1.84 (0.74) & 1.83 (2.79) \\
    HERAPDF1.5 & 1.78 (0.29) & 1.87 (0.75) & 1.84 (2.81) \\
    ABKM09 & 1.39 (0.35) & 1.43 (1.07) & 1.63 (3.66) \\
    GJR08 & 1.90 (1.46) & 1.34 (0.45) & {\bf 1.03} (0.51) \\
    \hline \multicolumn{4}{c}{} \\ \hline
    NNLO PDF (with NLO+2-loop $\hat{\sigma}$) & $\mu=p_T/2$ & $\mu=p_T$ & $\mu=2p_T$ \\
    \hline
    MRST06 & 3.19 (5.00) & 1.77 (3.22) &  1.25 (1.50) \\
    MSTW08 & 1.95 (0.90) & 1.23 (0.44) & {\bf 1.08} (0.35) \\
    HERAPDF1.0, $\alpha_S(M_Z^2)=0.1145$ & 2.11 (0.37) & 1.68 (0.35) & 1.41 (0.63) \\
    HERAPDF1.0, $\alpha_S(M_Z^2)=0.1176$ & 2.28 (0.95) & 1.50 (0.40) & {\bf 1.17} (0.21) \\
    ABKM09 & 1.68 (0.79) & 1.55 (1.21) & 1.63 (2.04) \\
    JR09 & 1.84 (0.47) & 1.61 (0.36) & 1.58 (0.50) \\
    \hline
  \end{tabular}
  \caption{Values of $\chi^2/N_{\rm pts.}$ for the D{\O} Run II inclusive jet data using a cone jet algorithm~\cite{Abazov:2008hua} with $N_{\rm pts.}=110$ and $N_{\rm corr.}=23$, for different PDF sets and different scale choices $\mu_R=\mu_F=\mu=\{p_T/2,p_T,2p_T\}$.  The $\chi^2$ values are calculated accounting for all 23 sources of correlated systematic uncertainty, using eq.~\eqref{eq:chisqcorr}, including the 6.1\% normalisation uncertainty due to the luminosity determination.  At most a 1-$\sigma$ shift in normalisation is allowed.  We highlight in bold those values lying inside the 90\% C.L.~region, defined by eq.~\eqref{eq:90percentCL}, which gives $\chi^2/N_{\rm pts.} < 1.22$.  The values of $\chi^2/N_{\rm pts.}$ computed using eq.~\eqref{eq:chisqtot}, simply adding all experimental uncertainties in quadrature (including luminosity), are shown in brackets in the table.  If the theory prediction was identically zero, the $\chi^2/N_{\rm pts.}$ values would be 7.46 (65.7) with (without) accounting for correlations between systematic uncertainties.\label{tab:d0incl}}
\end{table}
In tables~\ref{tab:cdfkt}, \ref{tab:cdfmid} and \ref{tab:d0incl} we give the $\chi^2$ per data point, calculated using eq.~\eqref{eq:chisqcorr} with the restriction $|r_{\rm lumi.}|<1$, for the Tevatron Run II data on inclusive jet production~\cite{Abulencia:2007ez,Aaltonen:2008eq,Abazov:2008hua}, for different PDF sets and different scale choices $\mu_R=\mu_F=\mu=\{p_T/2,p_T,2p_T\}$, where $p_T$ is the jet transverse momentum.  For NNPDF2.1 the jet cross sections are averaged over 100 replica sets.  We give the $\chi^2/N_{\rm pts.}$ values defined by simply adding all uncertainties in quadrature, eq.~\eqref{eq:chisqtot}, in brackets in the tables.  In this case many PDF sets and scale choices give a $\chi^2/N_{\rm pts.}\ll 1$, so the consistent treatment of correlated uncertainties is vital for the jet data to discriminate.  In the table captions we give the $\chi^2$ values with an identically zero theory prediction, $T_i\equiv 0$, just to illustrate how the correlated systematic shifts can partially accommodate a clearly inadequate theory prediction.  We highlight in bold the $\chi^2$ values lying inside the 90\% C.L.~region defined as
\begin{equation} \label{eq:90percentCL}
  \chi^2 < \left(\frac{\chi_{0}^2}{\xi_{50}}\right)\xi_{90},
\end{equation}
where $\xi_{50}$ and $\xi_{90}$ are the 50th and 90th percentiles of the $\chi^2$-distribution with $N_{\rm pts.}$ degrees of freedom.  (These quantities are defined in detail in section~6.2 of ref.~\cite{Martin:2009iq}.)  Here, $\chi_{0}^2$ is defined as the lowest $\chi^2$ value of all theory predictions in each table, i.e.~assumed to be close to the best possible fit, so that the rescaling factor $\chi_{0}^2/\xi_{50}$ in eq.~\eqref{eq:90percentCL} empirically accounts for any unusual fluctuations preventing the best possible fit having $\chi^2 \simeq \xi_{50} \simeq N_{\rm pts.}$~\cite{Stump:2001gu}.  The 90\% C.L.~region given in this way is used to determine the PDF uncertainties according to the ``dynamical tolerance'' prescription introduced in ref.~\cite{Martin:2009iq}, so PDF sets with $\chi^2$ values far outside this region cannot be considered to give an acceptable description of the data.  We consider NLO PDFs from MRST04~\cite{Martin:2004ir}, MSTW08~\cite{Martin:2009iq}, CTEQ6.6~\cite{Nadolsky:2008zw}, CT10~\cite{Lai:2010vv}, NNPDF2.1~\cite{Ball:2011mu}, HERAPDF1.0~\cite{HERA:2009wt}, HERAPDF1.5 (preliminary)~\cite{HERA:2010}, ABKM09~\cite{Alekhin:2009ni} and GJR08~\cite{Gluck:2007ck,Gluck:2008gs}.  We consider NNLO PDFs from MRST06~\cite{Martin:2007bv}, MSTW08~\cite{Martin:2009iq}, HERAPDF1.0~\cite{HERA:2009wt}, ABKM09~\cite{Alekhin:2009ni} and JR09~\cite{JimenezDelgado:2008hf,JimenezDelgado:2009tv}.  The MRST04 and MRST06 fits only included Tevatron Run I data~\cite{Affolder:2001fa,Abbott:2000ew}, and were superseded by the MSTW08 fits, but we show the $\chi^2$ values here just to demonstrate that these older fits do not give a good description of the newer Tevatron Run II data due to their harder high-$x$ gluon distribution.  The CTEQ6.6 fit includes only the Tevatron Run I data~\cite{Affolder:2001fa,Abbott:2000ew}, while the CT10 fit includes Run II data~\cite{Aaltonen:2008eq,Abazov:2008hua} in \emph{addition} to the Run I data~\cite{Affolder:2001fa,Abbott:2000ew}, contrary to the MSTW08 and NNPDF2.1 fits which include \emph{only} Run II data~\cite{Abulencia:2007ez,Abazov:2010fr}.  The GJR08 fit included some Run I~\cite{Abbott:2000ew} and Run II~\cite{Abulencia:2005yg} data, while the JR09, ABKM09 and HERAPDF fits did not include any Tevatron jet data.

The most constraining data set appears to be the CDF Run II inclusive jet data using the $k_T$ jet algorithm~\cite{Abulencia:2007ez} (see table~\ref{tab:cdfkt}) where, other than MSTW08, only NNPDF2.1 gives an acceptable description for $\mu=p_T$, while HERAPDF1.0 and ABKM09 typically give $\chi^2/N_{\rm pts.}\sim 2$--$3$, and CTEQ6.6/CT10 give better values but still much worse than MSTW08 (and NNPDF2.1).  The GJR08/JR09 sets and the HERAPDF1.0 NNLO set with $\alpha_S(M_Z^2) = 0.1176$ give a reasonable description, at a similar level to CT10, and give predictions for $gg\to H$ cross sections at the Tevatron which are much closer to the MSTW08 predictions than those from ABKM09 and the HERAPDF1.0 NNLO set with $\alpha_S(M_Z^2) = 0.1145$.  The same trend is apparent, but to a somewhat lesser extent, for the CDF Run II inclusive jet data using the cone-based Midpoint jet algorithm~\cite{Aaltonen:2008eq} (see table~\ref{tab:cdfmid}) and the D{\O} Run II inclusive jet data using a cone jet algorithm~\cite{Abazov:2008hua} (see table~\ref{tab:d0incl}).

\begin{figure}
  \centering
  \begin{minipage}{0.5\textwidth}
    (a)\\
    \includegraphics[width=\textwidth]{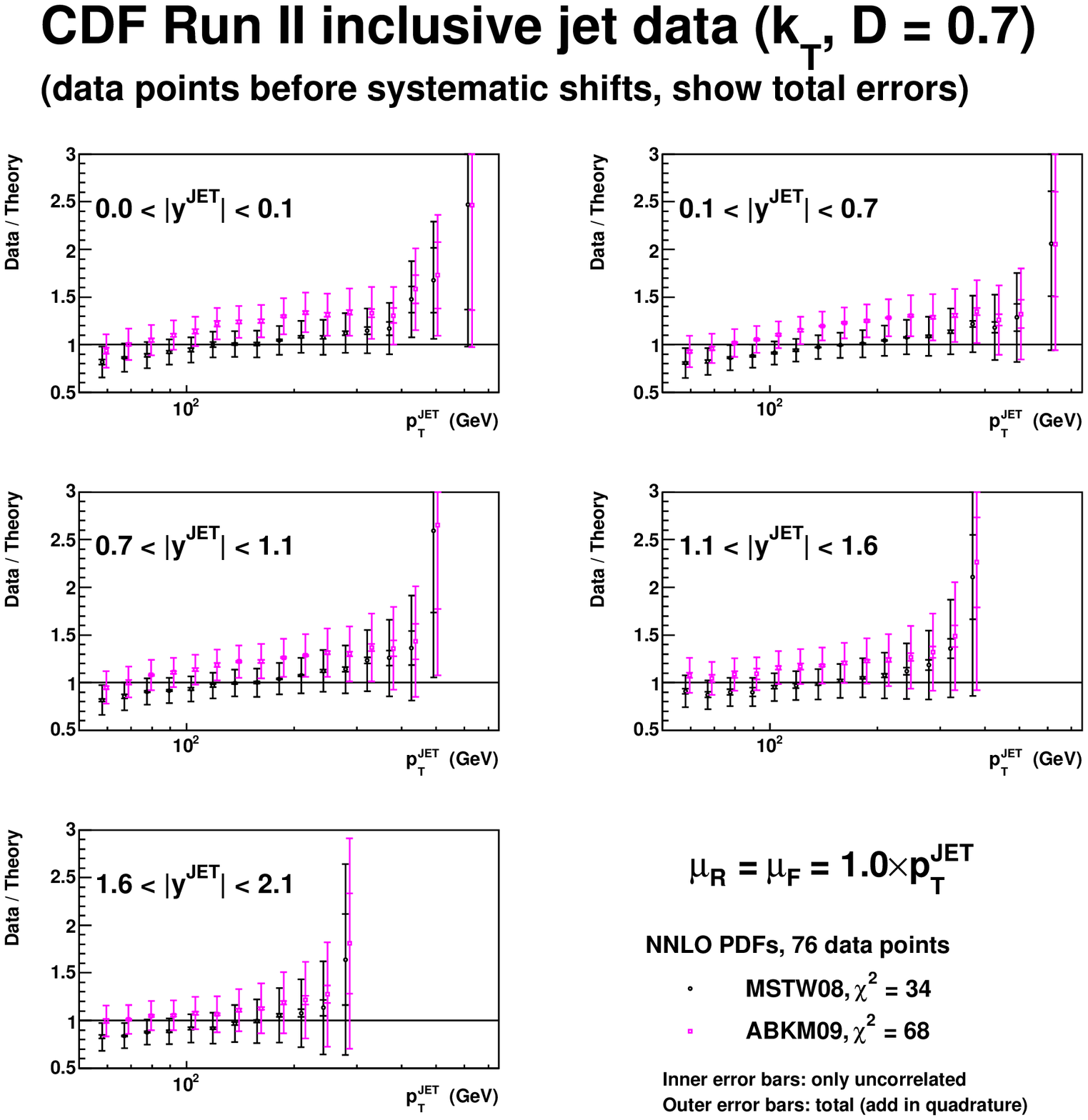}
  \end{minipage}%
  \begin{minipage}{0.5\textwidth}
    (b)\\
    \includegraphics[width=\textwidth]{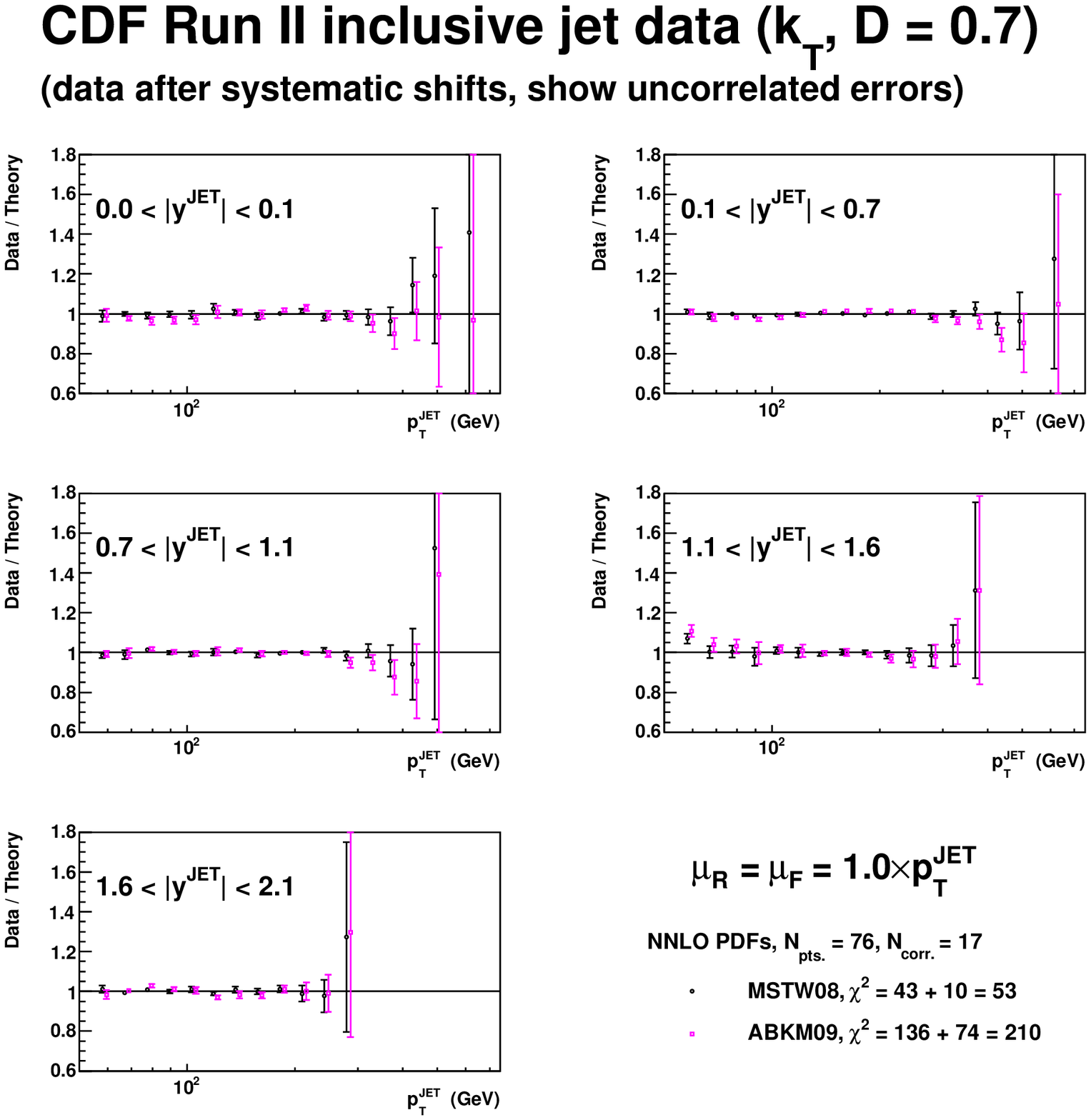}
  \end{minipage}
  \caption{Data/theory ratios for the CDF Run II inclusive jet data using the $k_T$ jet algorithm~\cite{Abulencia:2007ez} with $N_{\rm pts.}=76$ and $N_{\rm corr.}=17$, for MSTW08 and ABKM09 NNLO PDFs with NLO partonic cross sections supplemented by 2-loop threshold corrections, with scale choice $\mu_R=\mu_F=p_T$, and (a)~all experimental errors added in quadrature, then (b)~accounting for correlated systematic uncertainties using eq.~\eqref{eq:chisqcorr} and showing only the uncorrelated experimental errors.}
  \label{fig:cdfkt}
\end{figure}
\begin{figure}
  \centering
  \begin{minipage}{0.5\textwidth}
    (a)\\
    \includegraphics[width=\textwidth]{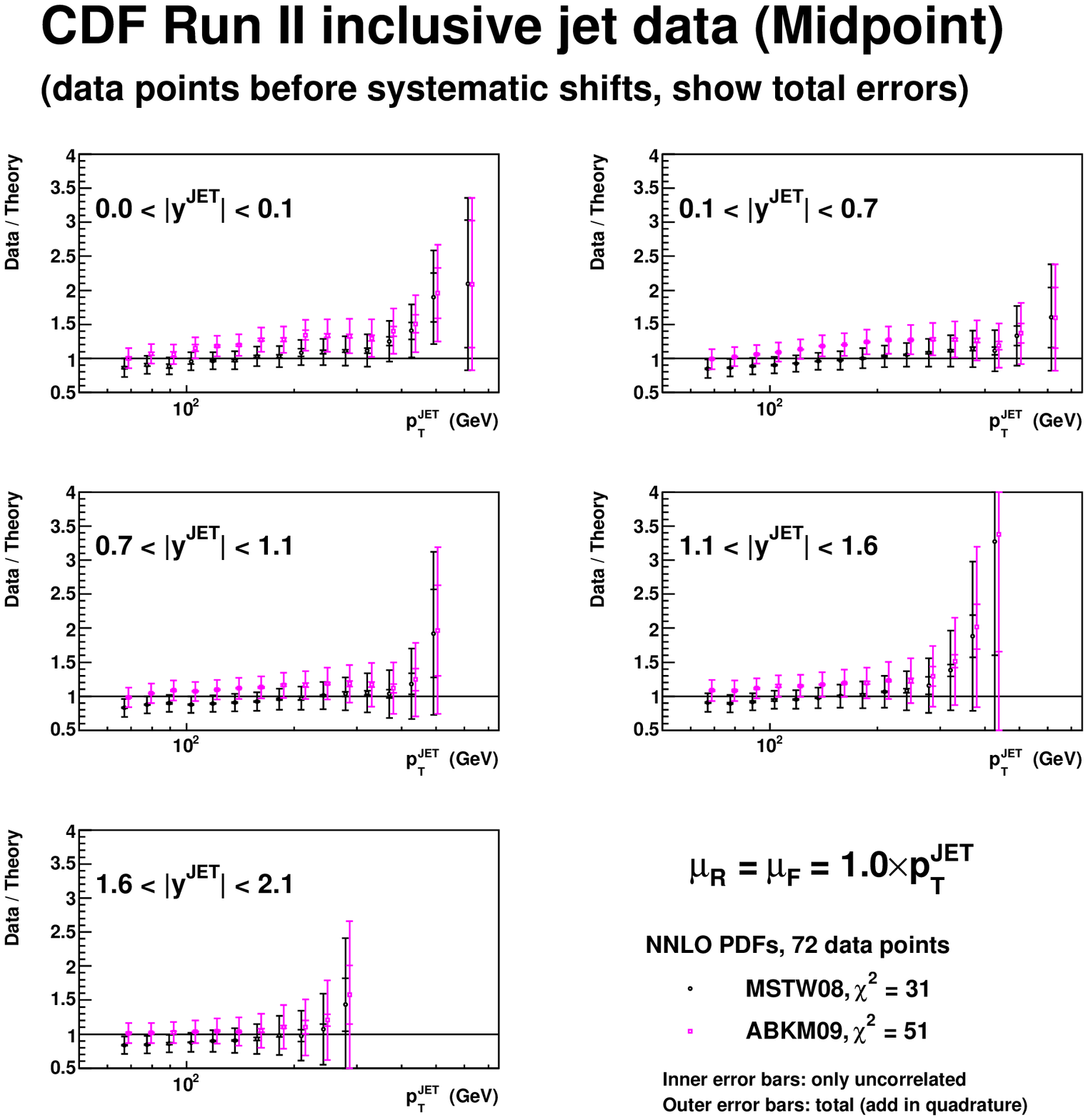}
  \end{minipage}%
  \begin{minipage}{0.5\textwidth}
    (b)\\
    \includegraphics[width=\textwidth]{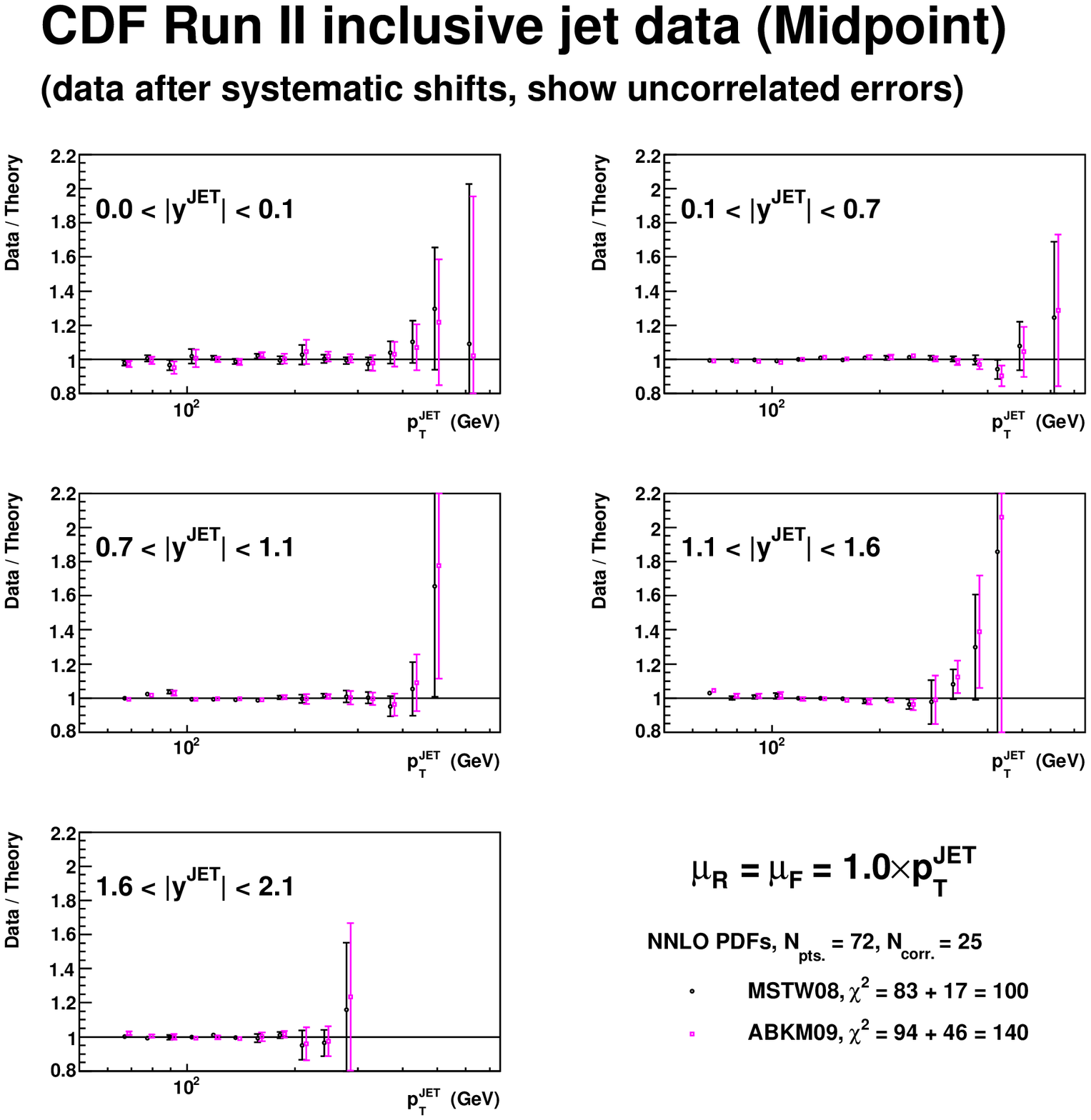}
  \end{minipage}
  \caption{Data/theory ratios for the CDF Run II inclusive jet data using the cone-based Midpoint jet algorithm~\cite{Aaltonen:2008eq} with $N_{\rm pts.}=72$ and $N_{\rm corr.}=25$, for MSTW08 and ABKM09 NNLO PDFs with NLO partonic cross sections supplemented by 2-loop threshold corrections, with scale choice $\mu_R=\mu_F=p_T$, and (a)~all experimental errors added in quadrature, then (b)~accounting for correlated systematic uncertainties using eq.~\eqref{eq:chisqcorr} and showing only the uncorrelated experimental errors.}
  \label{fig:cdfmid}
\end{figure}
\begin{figure}
  \centering
  \begin{minipage}{0.5\textwidth}
    (a)\\
    \includegraphics[width=\textwidth]{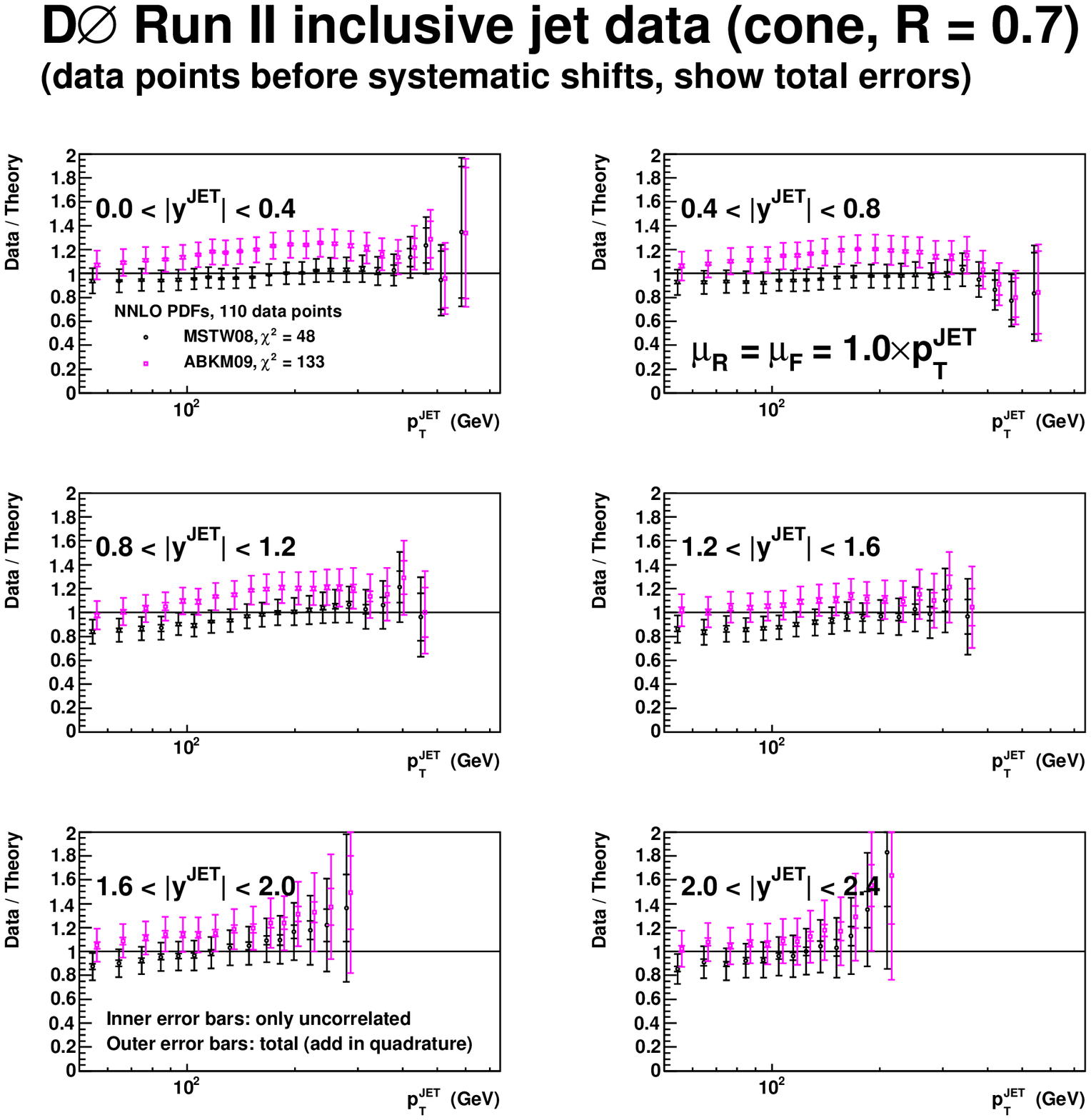}
  \end{minipage}%
  \begin{minipage}{0.5\textwidth}
    (b)\\
    \includegraphics[width=\textwidth]{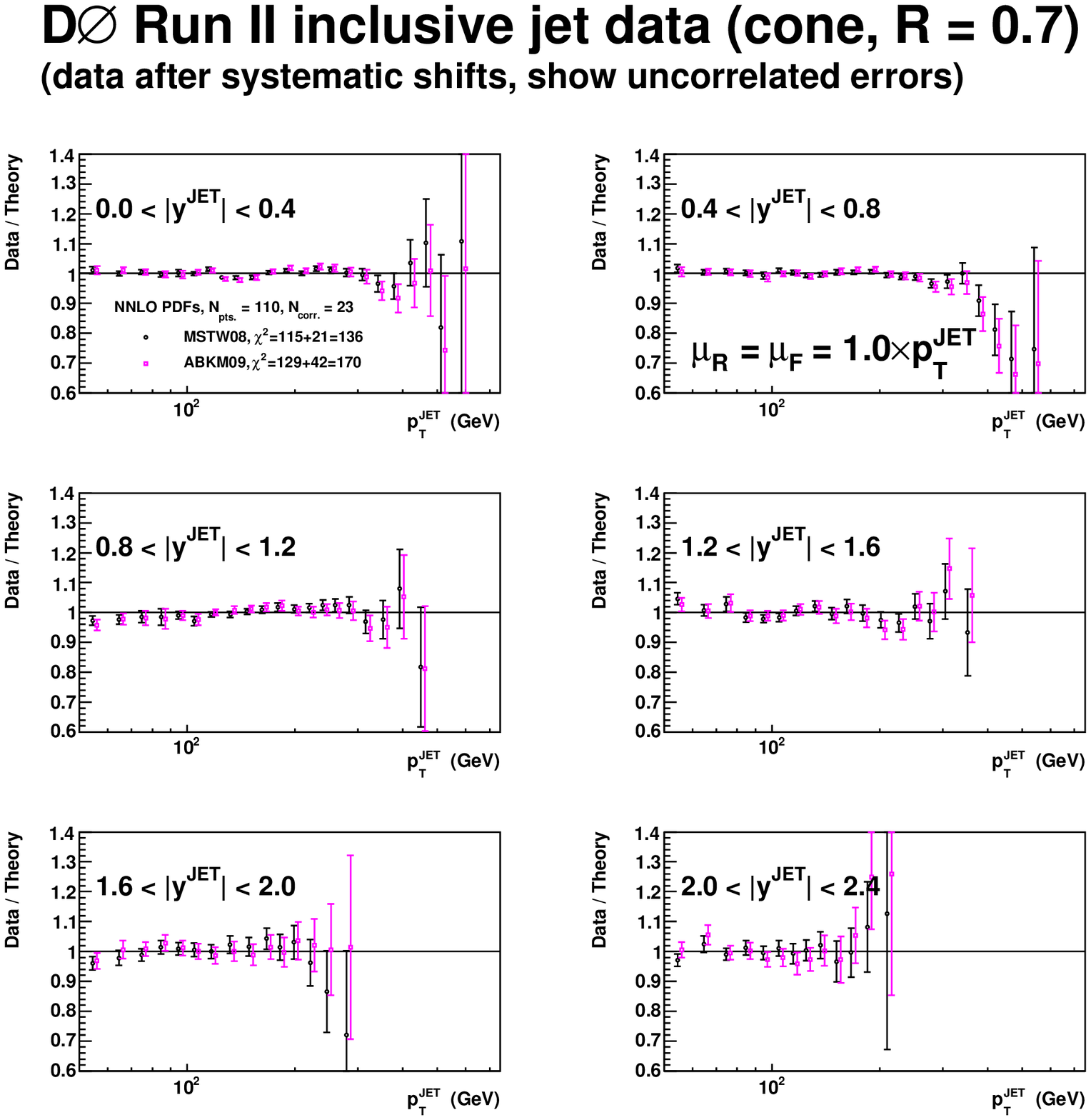}
  \end{minipage}
  \caption{Data/theory ratios for the D{\O} Run II inclusive jet data using a cone jet algorithm~\cite{Abazov:2008hua} with $N_{\rm pts.}=110$ and $N_{\rm corr.}=23$, for MSTW08 and ABKM09 NNLO PDFs with NLO partonic cross sections supplemented by 2-loop threshold corrections, with scale choice $\mu_R=\mu_F=p_T$, and (a)~all experimental errors added in quadrature, then (b)~accounting for correlated systematic uncertainties using eq.~\eqref{eq:chisqcorr} and showing only the uncorrelated experimental errors.}
  \label{fig:d0incl}
\end{figure}
\begin{figure}
  \centering
  \begin{minipage}{0.5\textwidth}
    (a)\\
    \includegraphics[width=\textwidth]{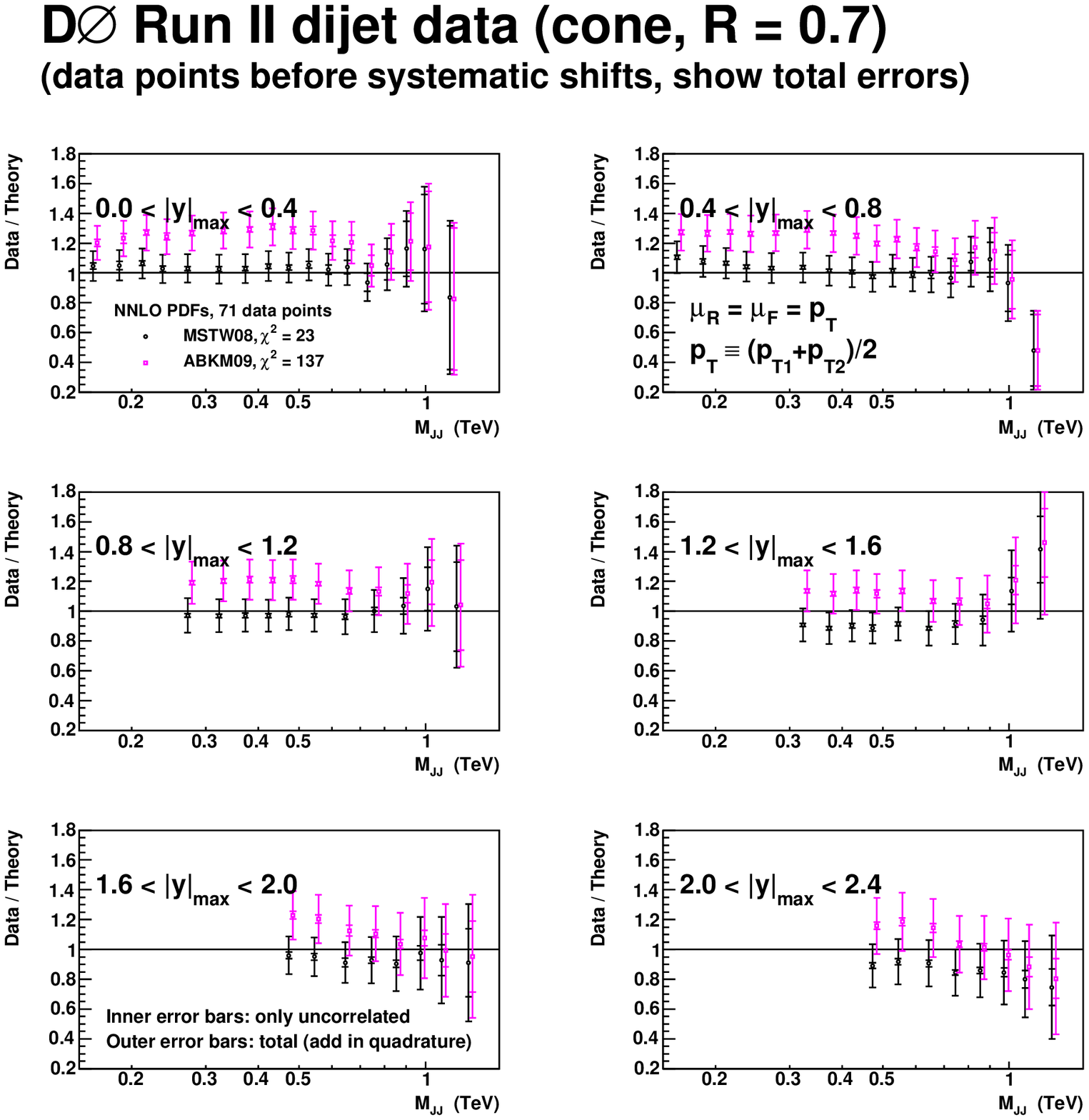}
  \end{minipage}%
  \begin{minipage}{0.5\textwidth}
    (b)\\
    \includegraphics[width=\textwidth]{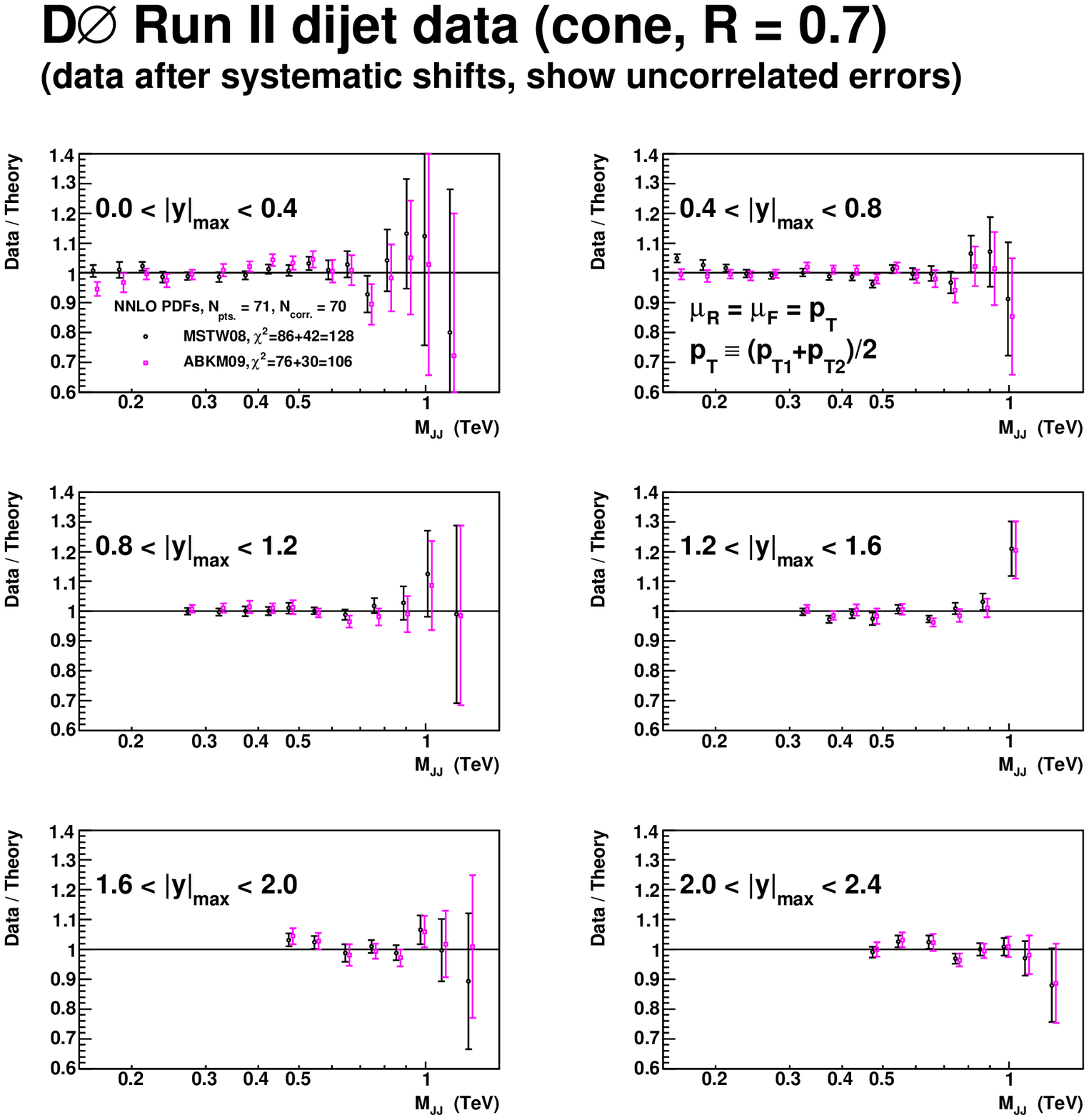}
  \end{minipage}
  \caption{Data/theory ratios for the D{\O} Run II dijet data using a cone jet algorithm~\cite{Abazov:2010fr} with $N_{\rm pts.}=71$ and $N_{\rm corr.}=70$, for MSTW08 and ABKM09 NNLO PDFs and NLO partonic cross sections, with scale choice $\mu_R=\mu_F=p_T$, where $p_T\equiv(p_{T1}+p_{T2})/2$, with (a)~all experimental errors added in quadrature, then (b)~accounting for correlated systematic uncertainties using eq.~\eqref{eq:chisqcorr} and showing only the uncorrelated experimental errors.}
  \label{fig:d0dijet}
\end{figure}

In figures~\ref{fig:cdfkt}, \ref{fig:cdfmid} and \ref{fig:d0incl} we compare the description of the Tevatron inclusive jet data by the MSTW08 and ABKM09 NNLO PDFs (recall that the latter give the lowest predictions for Tevatron Higgs cross sections) by showing the ratio of data to theory defined in two different ways: (a)~first we use the original data points $D_{i}/T_{i}$ with uncertainties given by adding all errors in quadrature (including luminosity), $\sigma_{i}^{\rm tot.}/T_{i}$, with the appropriate $\chi^2$ value in the plot legends obtained using eq.~\eqref{eq:chisqtot},  then (b) we use the shifted data points $\hat{D}_{i}/T_{i}$ with uncertainties given by $\sigma_{i}^{\rm uncorr.}/T_{i}$, with the $\chi^2$ calculated according to eq.~\eqref{eq:chisqcorr} and showing the two terms separately in the plot legends.  The $p_T$ values for ABKM09 are slightly offset for clarity in the plots.  The size of the second penalty term in eq.~\eqref{eq:chisqcorr} is some measure of how much the data points are shifted compared to their systematic errors.  For example, if the penalty term $\sum_{k=1}^{N_{\rm corr.}}r_{k}^2 > N_{\rm corr.}$, then the data points are shifted by, on average, more then 1-$\sigma$ for each systematic source $k$.  In general, a poor description of data before the systematic shifts leads to a large penalty term and a poor description also after the systematic shifts, although this general statement is not universally true.  We note that the shape of the data/theory ratio, both before and after the systematic shifts, looks remarkably similar as a function of both transverse momentum $p_T$ and rapidity $y$ in figures~\ref{fig:cdfkt} and \ref{fig:cdfmid}.  This demonstrates very clearly that the two CDF inclusive jet measurements~\cite{Abulencia:2007ez,Aaltonen:2008eq} each contain the same data, but simply analysed in a different way, and the change in analysis method is accounted for extremely well by the change in the theory.  Hence, it is not at all surprising that the two data sets can be well described by the same PDF set.  Indeed, it was explicitly demonstrated in ref.~\cite{Aaltonen:2008eq} that the ratios of the cross sections measured with the two jet algorithms were in reasonable agreement with theoretical expectations.

\subsection{Dijet production} \label{sec:dijet}

\begin{table}
  \centering
  \begin{tabular}{|l|r|r|r|}
    \hline
    NLO PDF (with NLO $\hat{\sigma}$) & $\mu=p_T/2$ & $\mu=p_T$ & $\mu=2p_T$ \\
    \hline
    MRST04 & 6.04 (4.93) & 4.54 (2.93) & 2.75 (0.72) \\
    MSTW08 & 3.15 (1.63) & 2.25 (0.70) & 1.56 (0.70) \\
    CTEQ6.6 & 5.41 (2.22) & 4.85 (1.79) & 3.36 (1.52) \\
    CT10 & 4.74 (1.87) & 4.06 (1.32) & 2.70 (1.21) \\
    NNPDF2.1 & 2.67 (1.56) & 1.93 (0.66) & 1.47 (0.55) \\
    HERAPDF1.0 & 2.05 (0.38) & 2.21 (0.77) & 2.11 (2.28) \\
    HERAPDF1.5 & 1.90 (0.34) & 2.00 (0.67) & 1.88 (2.16) \\
    ABKM09 & 1.49 (0.33) & 1.41 (0.80) & 1.34 (2.78) \\
    GJR08 & 10.7 (3.92) & 7.91 (2.36) & 5.30 (0.66) \\
    \hline \multicolumn{4}{c}{} \\ \hline
    NNLO PDF (with NLO $\hat{\sigma}$) & $\mu=p_T/2$ & $\mu=p_T$ & $\mu=2p_T$ \\
    \hline
    MRST06 & 8.06 (5.07) & 6.55 (3.21) & 4.07 (0.96) \\
    MSTW08 & 2.38 (0.63) & 1.80 (0.33) & 1.31 (1.24) \\
    HERAPDF1.0, $\alpha_S(M_Z^2)=0.1145$ & 2.61 (0.48) & 2.55 (0.89) & 2.40 (2.40) \\
    HERAPDF1.0, $\alpha_S(M_Z^2)=0.1176$ & 2.72 (0.83) & 2.31 (0.50) & 1.96 (1.08) \\
    ABKM09 & 1.36 (0.98) & 1.49 (1.93) & 1.57 (4.53) \\
    JR09 & 3.29 (0.42) & 2.55 (0.24) & 1.88 (1.26) \\
    \hline
  \end{tabular}
  \caption{Values of $\chi^2/N_{\rm pts.}$ for the D{\O} dijet data using a cone jet algorithm~\cite{Abazov:2010fr} with $N_{\rm pts.}=71$ and $N_{\rm corr.}=70$, for different NLO PDF sets and different scale choices $\mu_R=\mu_F=\mu=\{p_T/2,p_T,2p_T\}$, where $p_T\equiv(p_{T1}+p_{T2})/2$.  Only NLO partonic cross sections are used with the NNLO PDFs, since the 2-loop threshold corrections are only available for the inclusive jet cross section.  The $\chi^2$ values are calculated accounting for all 70 sources of correlated systematic uncertainty, using eq.~\eqref{eq:chisqcorr}, including the 6.1\% normalisation uncertainty due to the luminosity determination.  At most a 1-$\sigma$ shift in normalisation is allowed.  The values of $\chi^2/N_{\rm pts.}$ computed using eq.~\eqref{eq:chisqtot}, simply adding all experimental uncertainties in quadrature (including luminosity), are shown in brackets in the table.  If the theory prediction was identically zero, the $\chi^2/N_{\rm pts.}$ values would be 5.86 (60.5) with (without) accounting for correlations between systematic uncertainties.\label{tab:d0dijet}}
\end{table}

In table~\ref{tab:d0dijet} and figure~\ref{fig:d0dijet} we show similar results for the D{\O} Run II dijet data~\cite{Abazov:2010fr}, measured as a function of the dijet invariant mass, $M_{\rm JJ}$, and the largest absolute rapidity, $|y|_{\rm max}$, of the two jets with the largest transverse momentum.  Again, the $M_{\rm JJ}$ values for ABKM09 are slightly offset for clarity in figure~\ref{fig:d0dijet}.  The \textsc{fastnlo} grids are provided with a scale choice proportional to the mean transverse momentum of these two jets, $p_T\equiv(p_{T1}+p_{T2})/2$, and we show results with $\mu_R=\mu_F=\mu=\{p_T/2,p_T,2p_T\}$ in table~\ref{tab:d0dijet}.  Taking $\mu=p_T/4$ leads to negative cross sections at large $M_{\rm JJ}$ and large $|y|_{\rm max}$.  We multiply the \textsc{fastnlo} predictions by a factor 4 to account for a mismatch in the bin width factors of the provided grids.  There are no 2-loop threshold corrections available, so we are forced to use only the pure NLO partonic cross sections with the NNLO PDFs.  It can be seen that the trend in the $\chi^2$ values for the dijet data shown in table~\ref{tab:d0dijet} appears to be rather different from the inclusive jet data shown in tables~\ref{tab:cdfkt}, \ref{tab:cdfmid} and \ref{tab:d0incl}.  In particular, in contrast to the case for inclusive jets, the ABKM09 set gives the best description for $\mu=p_T$, whereas MSTW08 and NNPDF2.1 have $\chi^2/N_{\rm pts.}\sim 2$ and CTEQ6.6/CT10 has $\chi^2/N_{\rm pts.}\sim 4$--$5$.  For $\mu=2p_T$ there is a significant improvement in $\chi^2$ for MSTW08 and NNPDF2.1, and MSTW08 NNLO for $\mu=2p_T$ gives the best description out of all PDF sets and scale choices, while the CTEQ6.6/CT10 sets still have $\chi^2/N_{\rm pts.}\sim 3$ even for the larger scale choice.  However, it is interesting to note that while figures~\ref{fig:cdfkt}(a) and \ref{fig:cdfmid}(a) show a very similar trend for the data/theory ratios, figures~\ref{fig:d0incl}(a) and \ref{fig:d0dijet}(a) show quite a different trend, implying that the change in theory in using the NLO dijet cross section at the same scale as the inclusive jet cross section does not account for the difference in the data produced by the two methods~\cite{Abazov:2008hua,Abazov:2010fr} of binning and analysis.

At LO we have $M_{\rm JJ}=2p_T\cosh y^*$ where $y^*=|y_1-y_2|/2$, with $y_{1,2}$ the rapidities of the two jets.  It is clear that $p_T$ is a better measure of the ``hardness'' of the process than $M_{\rm JJ}$ and therefore $\mu=p_T$ is the most common scale choice for dijet production.  (Consider, for example, the extreme case of elastic $pp$ scattering where each final-state proton is considered to be a ``jet'', then $M_{\rm JJ}\approx\sqrt{s}$, but $p_T\approx0$.)   More generally, typical scale choices in fixed-order perturbative QCD calculations are usually, for example, the mass or transverse momentum of a produced particle, or a scalar sum of such scales added either linearly or in quadrature.  However, it is clear that choices of scale involving both $p_T$ and $y^*$ are perfectly feasible for dijets, whereas some multiple of $p_T$ seems more obviously the scale choice for inclusive jets. There is no reason that the choice which best mimics the full calculation at fixed order for inclusive jets need be the same as for dijets binned in $M_{\rm JJ}$, i.e.~the structure of higher order corrections is not automatically the same.  Indeed, a hybrid scale choice was proposed in ref.~\cite{Ellis:1992en} to interpolate between a scale choice based on $p_T$ and one based on $M_{\rm JJ}$, namely $\mu=AM_{\rm JJ}/(2\cosh By^*)$, with the two adjustable parameters chosen to be $A=0.5$ and $B=0.7$ so that the difference between the $\mathcal{O}(\alpha_S^3)$ calculation and the Born calculation was small over the angular region of interest~\cite{Ellis:1992en}.  It would be interesting to investigate whether such a scale choice could resolve the somewhat different conclusions reached from the Tevatron Run II inclusive and dijet data.  There is no requirement that the scale choice for dijets be the same as for inclusive jets.  Taking $\mu=p_T$ for inclusive jets and $\mu=2p_T=p_{T1}+p_{T2}$ for dijets, then the MSTW08 (and NNPDF2.1) PDFs would give a good description of all four Tevatron Run II jet data sets~\cite{Abulencia:2007ez,Aaltonen:2008eq,Abazov:2008hua,Abazov:2010fr}.

Another difference, possibly correlated to the issue of scale choice, is that the dijet data may probe higher $x$ values than the inclusive jet data.  If there are two jets labelled ``1'' and ``2'', and jet ``1'' has high $p_T$ in the forward region, then the phase space for the jet ``2'' is integrated over in the inclusive jet cross section, but will typically lie in the central region, creating an imbalance in the $x$ values of the two initial partons.  On the other hand, for the dijet cross section at high $M_{\rm JJ}$ values, if jet ``1'' lies in the forward region, then jet ``2'' will typically lie at the same absolute rapidity in the opposite direction, giving similarly large $x$ values of the two initial partons.  Since high-$x$ PDFs evolve very quickly, probing two high-$x$ PDFs increases sensitivity to (factorisation) scale choices.  This sensitivity will be most extreme when both PDFs are evolving quickly in the same direction (for example, both getting smaller with increasing scale), rather than one PDF getting smaller and one PDF getting larger as would be the case with one high-$x$ parton and one low-$x$ parton.  This effect automatically means that the higher-order corrections must be slightly different in the two cases of inclusive jet and dijet production.

\subsection{Scale dependence\texorpdfstring{ of jet cross sections}{}} \label{sec:scaledep}

\begin{figure}
  \centering
  \begin{minipage}{0.5\textwidth}
    (a)\\
    \includegraphics[width=\textwidth]{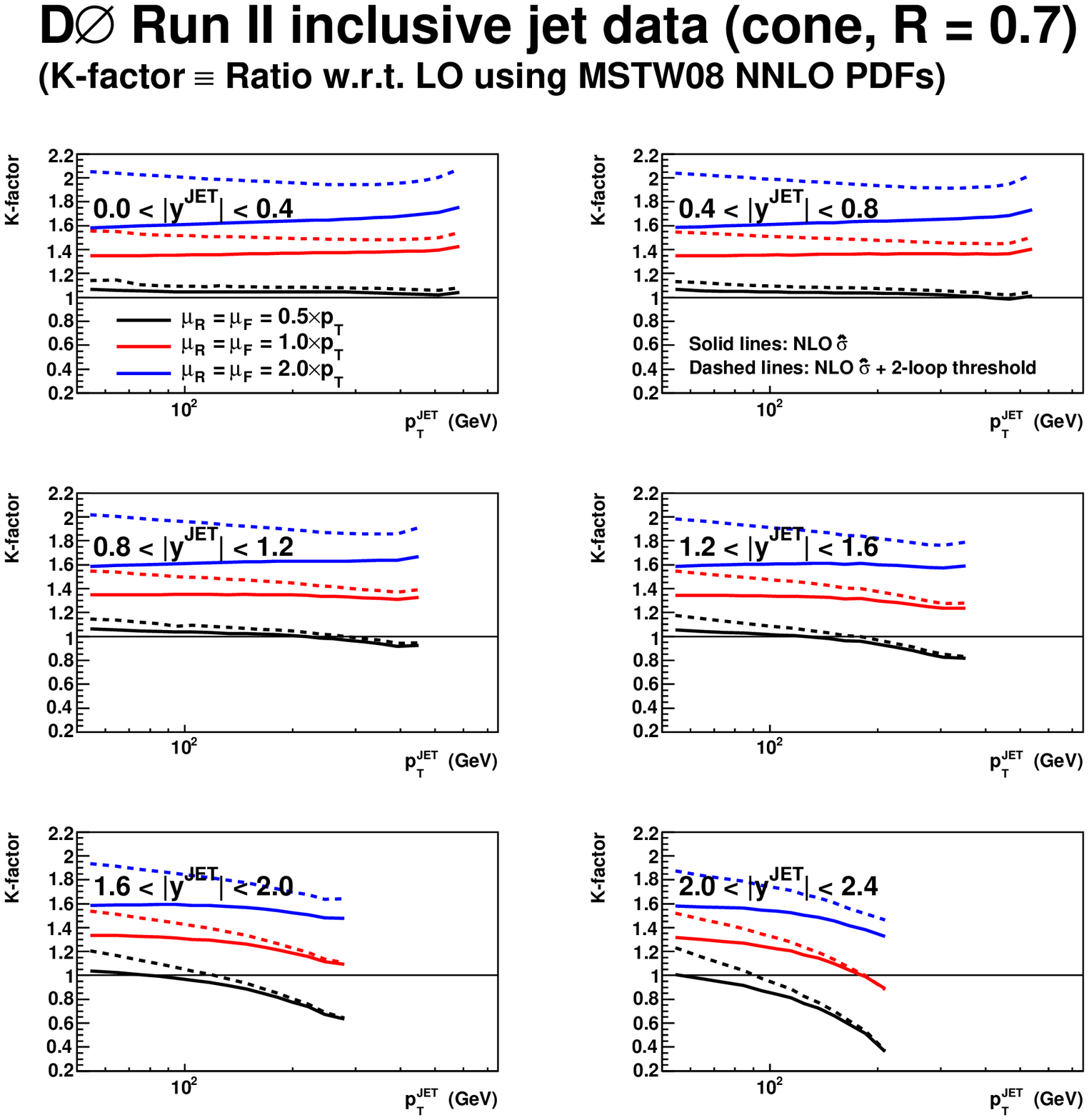}
  \end{minipage}%
  \begin{minipage}{0.5\textwidth}
    (b)\\
    \includegraphics[width=\textwidth]{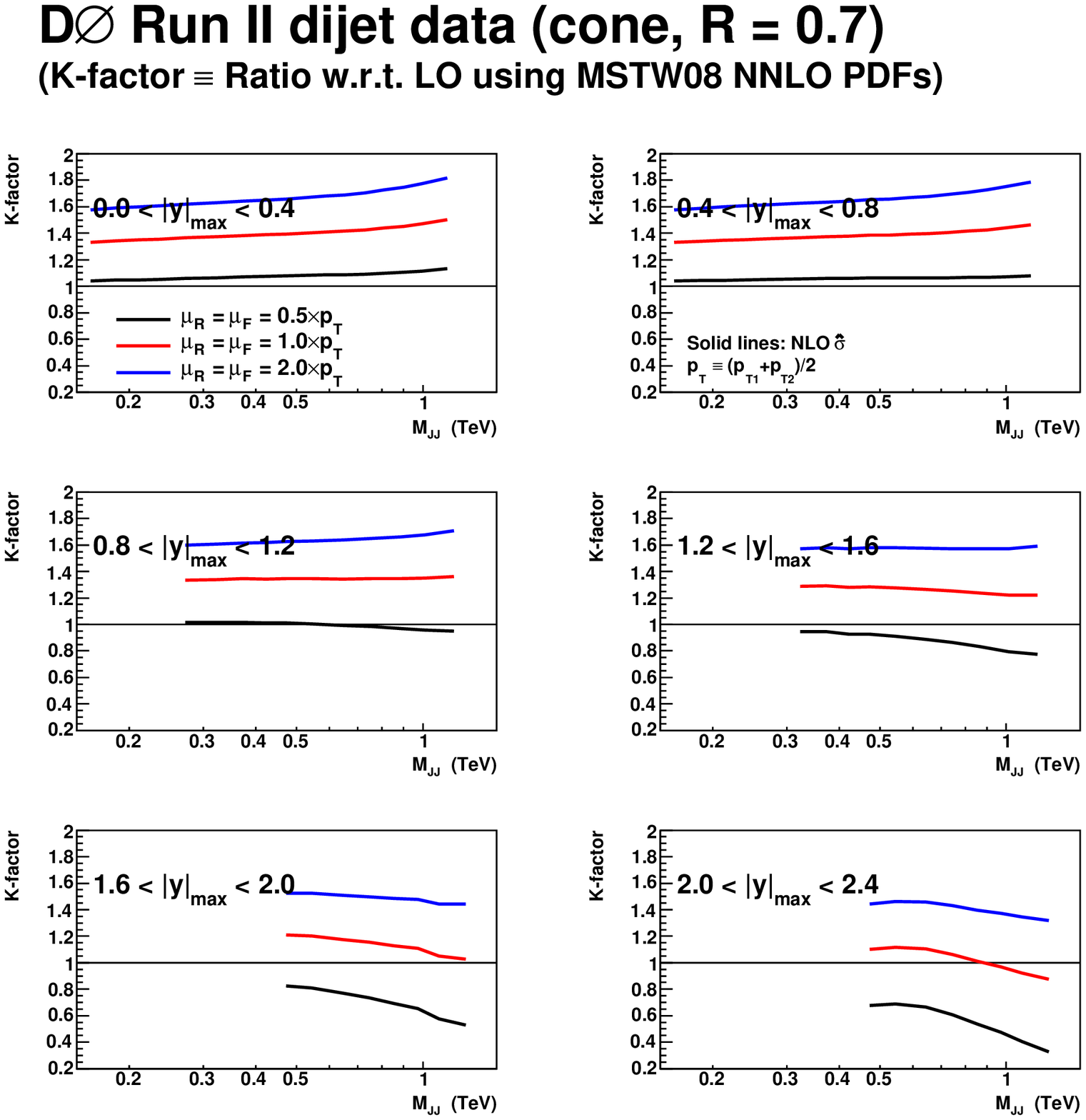}
  \end{minipage}
  \caption{$K$-factors using MSTW08 NNLO PDFs for (a)~inclusive jet and (b)~dijet production.}
  \label{fig:Kfactors}
\end{figure}
\begin{figure}
  \centering
  \begin{minipage}{0.5\textwidth}
    (a)\\
    \includegraphics[width=\textwidth]{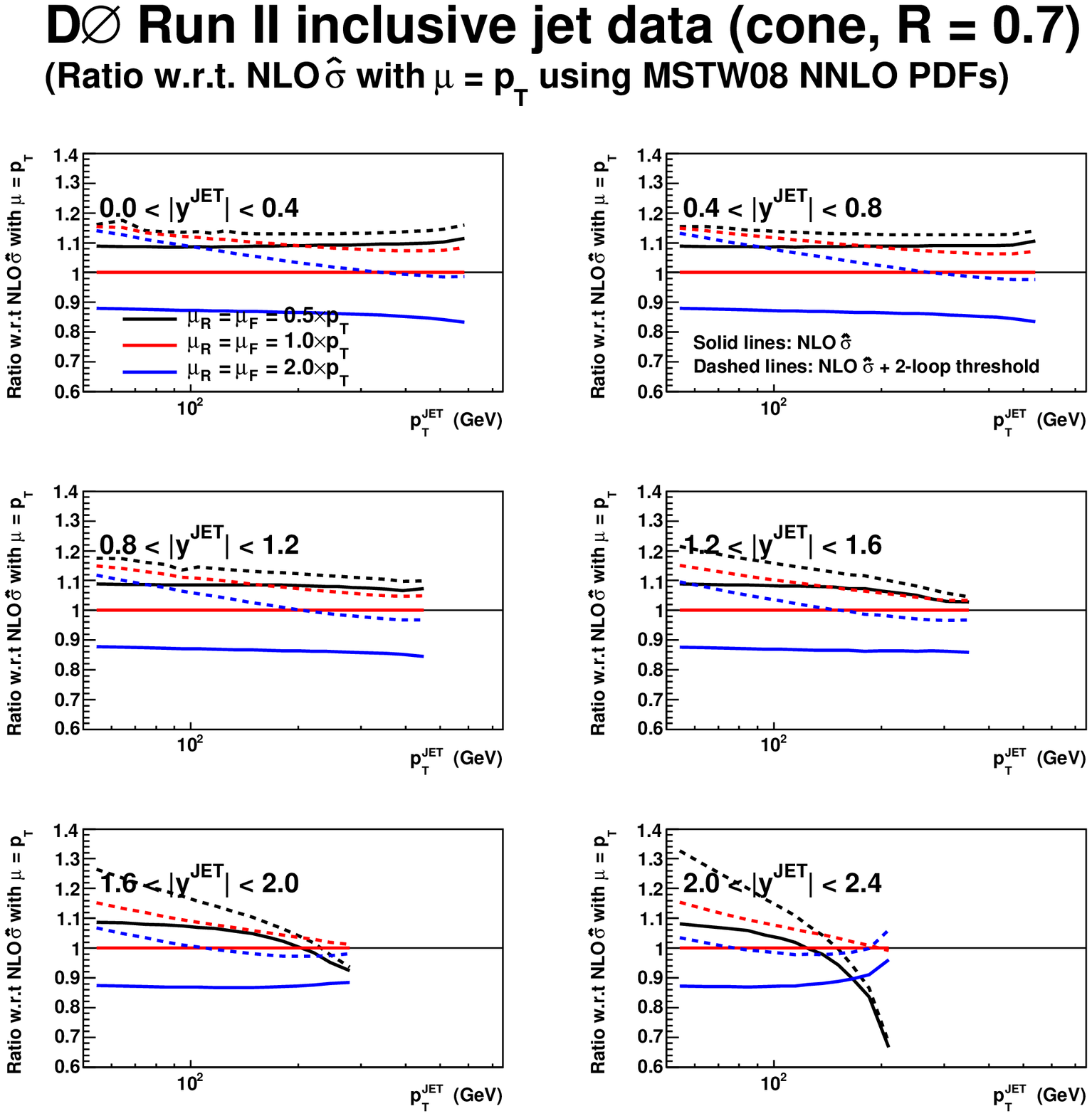}
  \end{minipage}%
  \begin{minipage}{0.5\textwidth}
    (b)\\
    \includegraphics[width=\textwidth]{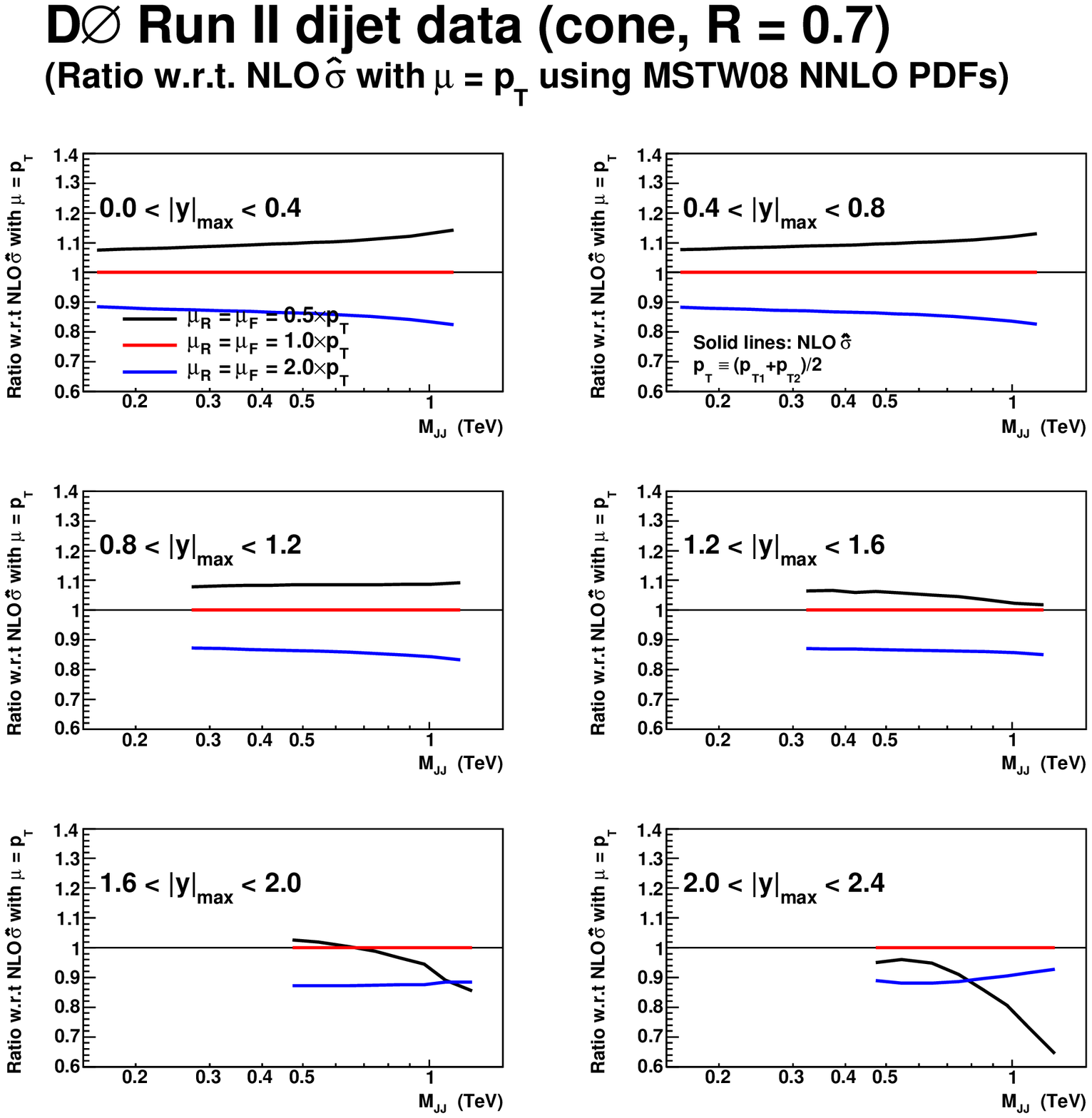}
  \end{minipage}
  \caption{Ratio of jet cross sections to those with NLO $\hat{\sigma}$ and scale choice $\mu_R=\mu_F=p_T$ using MSTW08 NNLO PDFs for (a)~inclusive jet and (b)~dijet production.}
  \label{fig:Rfactors}
\end{figure}
In figure~\ref{fig:Kfactors} we compare the $K$-factors for the D{\O} inclusive and dijet data, defined as the ratio of the NLO (both with/without the 2-loop threshold corrections) jet cross sections to the LO jet cross sections, computed with the same MSTW08 NNLO PDFs (and $\alpha_S$) in the numerator and denominator of the ratio.  Using another PDF choice, such as ABKM09 NNLO, makes little difference to the $K$-factors.  The choice $\mu=p_T/2$ has historically been favoured in MRST/CTEQ fits because the $K$-factor is close to 1 at central rapidity.  However, going to forward rapidities with the choice $\mu=p_T/2$, the $K$-factor decreases substantially with increasing $p_T$.  The $K$-factor with the choice $\mu=p_T$ is more uniform (with moderate size) across all rapidity bins and $p_T$ values, hence $\mu=p_T$ was chosen for the MSTW08 analysis~\cite{Martin:2009iq}.  It is striking, however, that although the NLO corrections are $\sim 60\%$ for $\mu=2p_T$, and a further $20\%$ or more with the 2-loop threshold corrections, the shape of the $K$-factor is rather more stable across all rapidity bins and $p_T$ with this choice.  In figure~\ref{fig:Rfactors} we show the ratio of the NLO (both with/without the 2-loop threshold corrections) jet cross sections with different scale choices to the NLO jet cross section with $\mu_R=\mu_F=p_T$, again computed with the same MSTW08 NNLO PDFs (and $\alpha_S$) in the numerator and denominator of the ratio.  It can be seen that the use of the 2-loop threshold corrections for the inclusive jet cross sections stabilises the scale dependence (except at the very highest rapidity and $p_T$ values where the low scale choice still leads to a large variation).  To some extent, different scale choices will be compensated by different systematic shifts, particularly for the luminosity (see appendix~\ref{sec:freenorm}).  The predictions for $\mu=p_T$ are generally in the middle of the other two choices, but this breaks down at high rapidity and $p_T$ values.  Indeed, for dijets $\mu=p_T$ ceases to be the central prediction at nearly all $p_T$ in the two highest rapidity bins, and is progressively less so in the middle rapidity bins than for the case of inclusive jets.  This supports the idea that the optimal choice for dijets might be $p_T$ multiplied by a function $f(y^*)$, growing with increasing $y^*$, so that $\mu = p_T\cdot f(y^*)$ would be the central prediction over all $|y|_{\rm max}$ bins.  (The $y^*$ variable is closely related to the $|y|_{\rm max}$ variable used by the D{O} dijet data~\cite{Abazov:2010fr}.)  In the absence of readily-available theory predictions for such a scale choice, the best description of dijet data by PDFs obtained from fitting to inclusive jet data seems to be given, as a compromise, by a scale of $\mu=c\,p_T$ with $c>1$, with our specific example being $c=2$.

\subsection{\texorpdfstring{Distributions of pulls}{Pulls} and systematic shifts} \label{sec:pulls}

\begin{figure}
  \centering
  \begin{minipage}{0.5\textwidth}
    (a)\\
    \includegraphics[width=\textwidth]{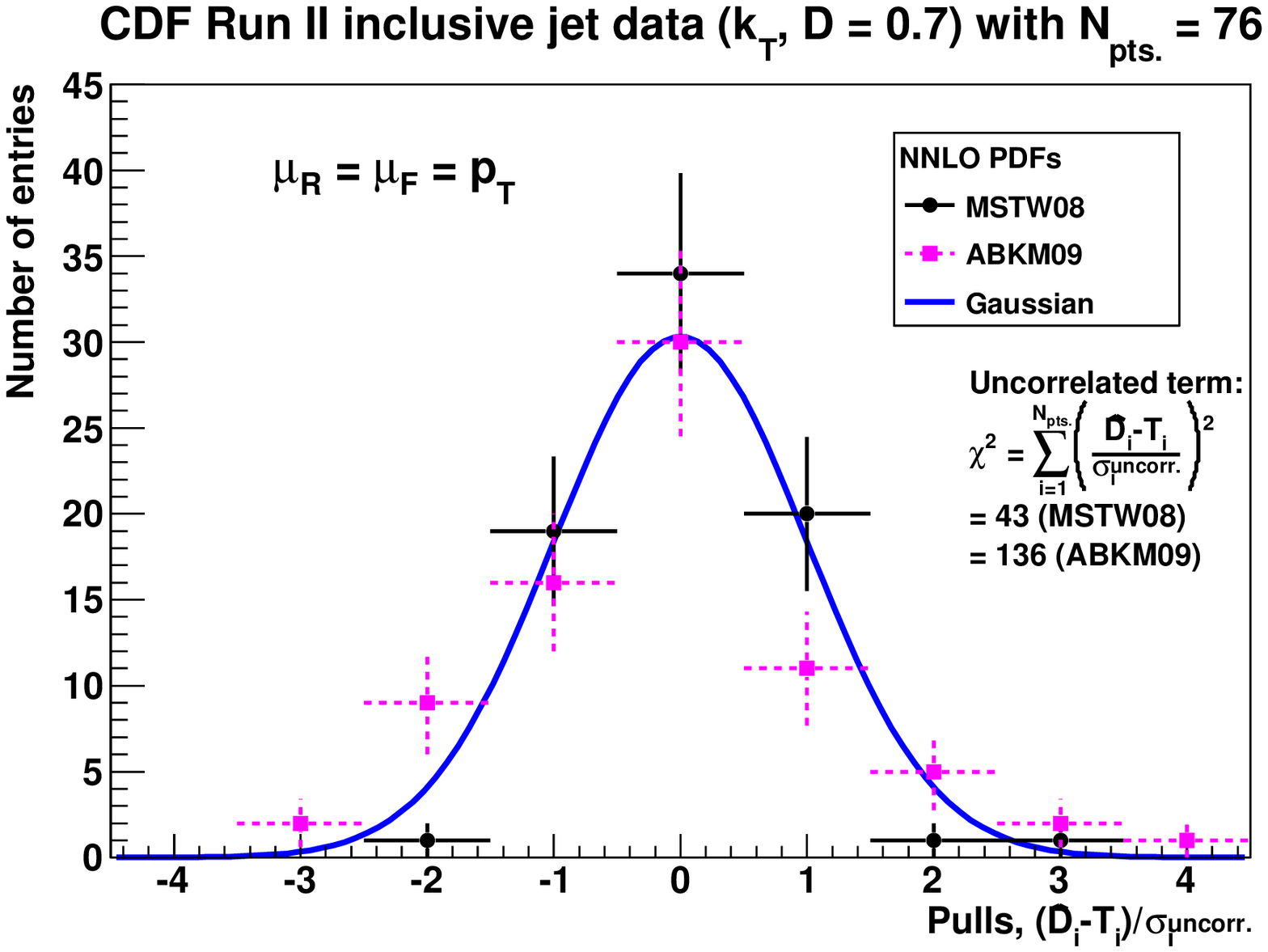}
  \end{minipage}%
  \begin{minipage}{0.5\textwidth}
    (b)\\
    \includegraphics[width=\textwidth]{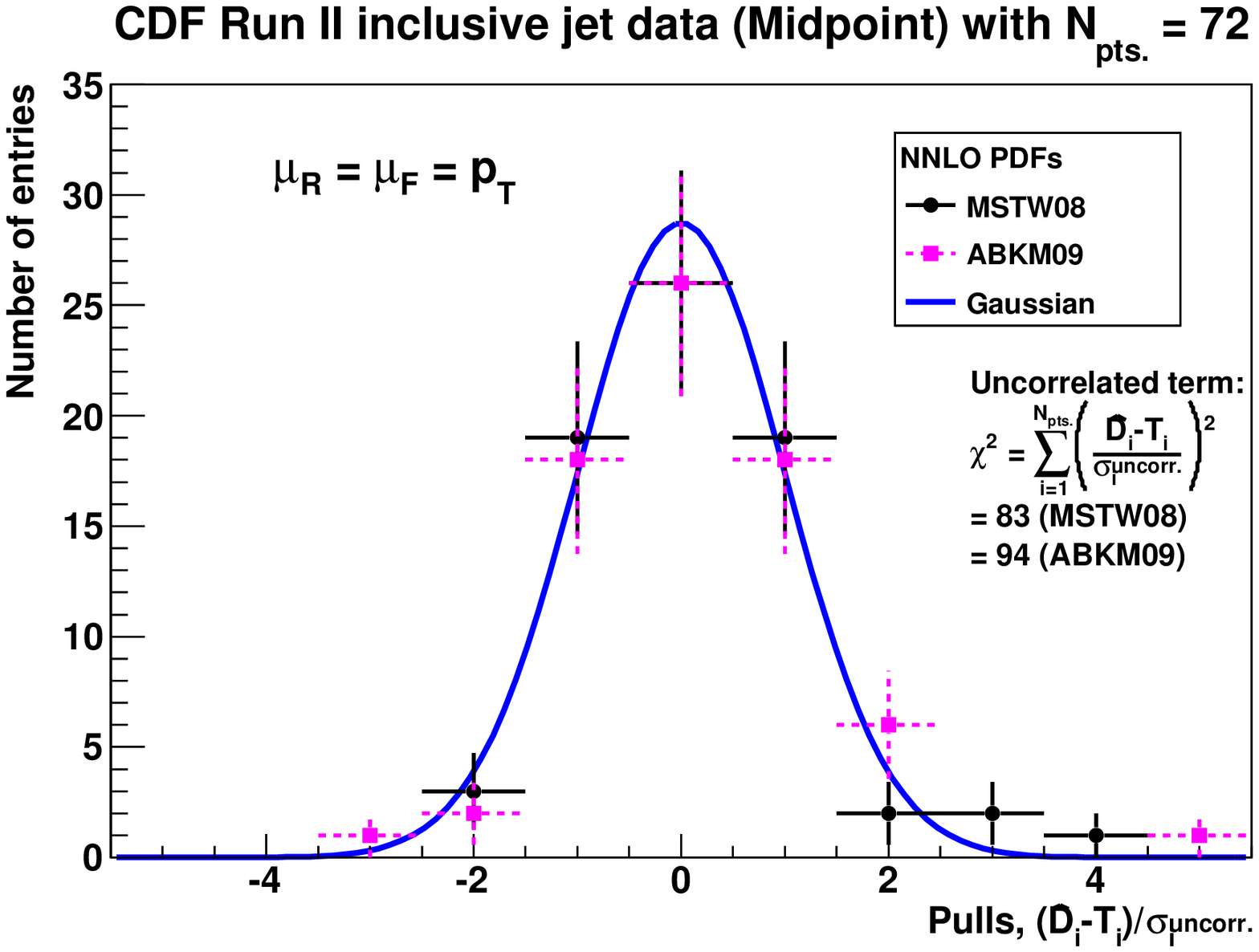}
  \end{minipage}
  \begin{minipage}{0.5\textwidth}
    (c)\\
    \includegraphics[width=\textwidth]{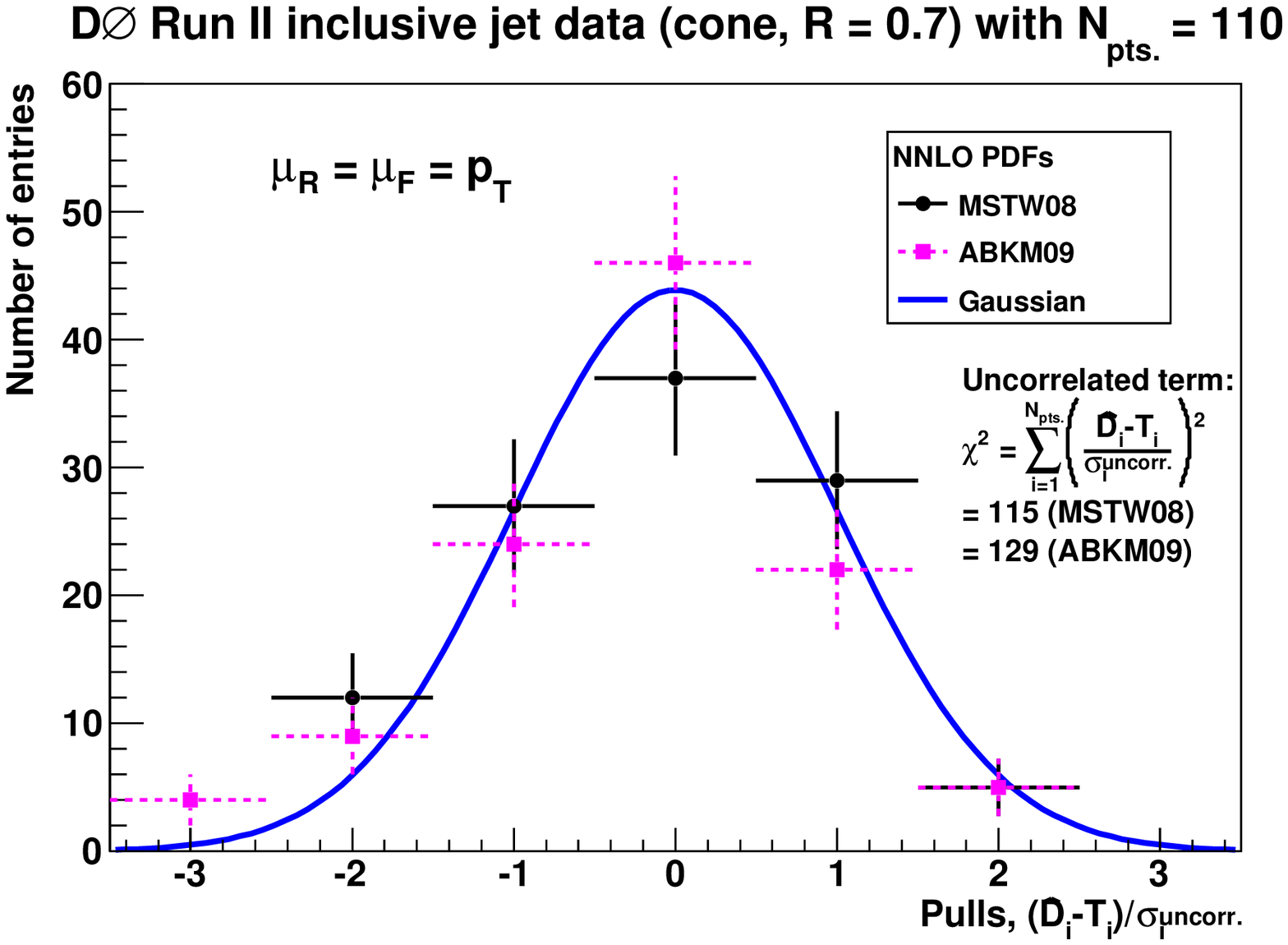}
  \end{minipage}%
  \begin{minipage}{0.5\textwidth}
    (d)\\
    \includegraphics[width=\textwidth]{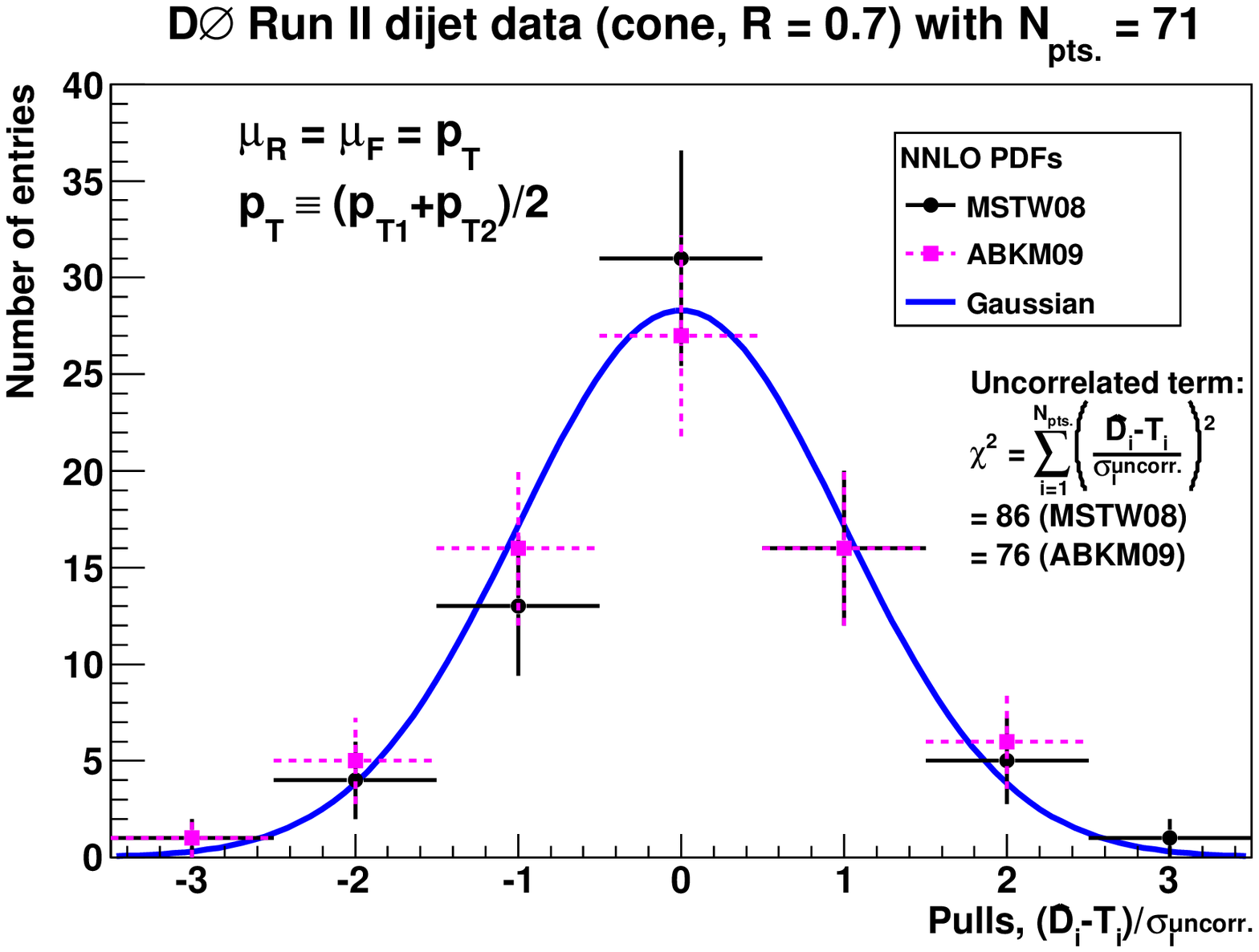}
  \end{minipage}
  \caption{Distributions of the pulls, $(\hat{D}_i-T_i)/\sigma_{i}^{\rm uncorr.}$, for each of the four Tevatron data sets on jet production, with theory predictions calculated using either MSTW08 or ABKM09 NNLO PDFs, compared to the expectation of a Gaussian distribution with unit width.}
  \label{fig:plotpulls}
\end{figure}
In figure~\ref{fig:plotpulls} we show the distributions of pulls, $(\hat{D}_{i}-T_{i})/\sigma_{i}^{\rm uncorr.}$, for all four Tevatron Run II data sets on jet production, with theory predictions calculated using either MSTW08 or ABKM09 NNLO PDFs and a scale choice $\mu_R=\mu_F=\mu=p_T$.  We show the expected behaviour of a Gaussian distribution with unit width, and the first $\chi^2$ term in eq.~\eqref{eq:chisqcorr} given simply by the sum of pulls over all data points.  The histogram error bars are simply given by the square root of the number of entries.  We see that the distribution of pulls is fairly close to the expected Gaussian behaviour for all four data sets, although the tails for the inclusive jet data with ABKM09 are somewhat broader than expected, leading to larger $\chi^2$ contributions than for MSTW08, particularly for the CDF data using the $k_T$ jet algorithm~\cite{Abulencia:2007ez} shown in figure~\ref{fig:plotpulls}(a).  However, it is clear that this source does not account for the complete differences in $\chi^2$ seen previously.
\begin{figure}
  \centering
  \begin{minipage}{0.5\textwidth}
    (a)\\
    \includegraphics[width=\textwidth]{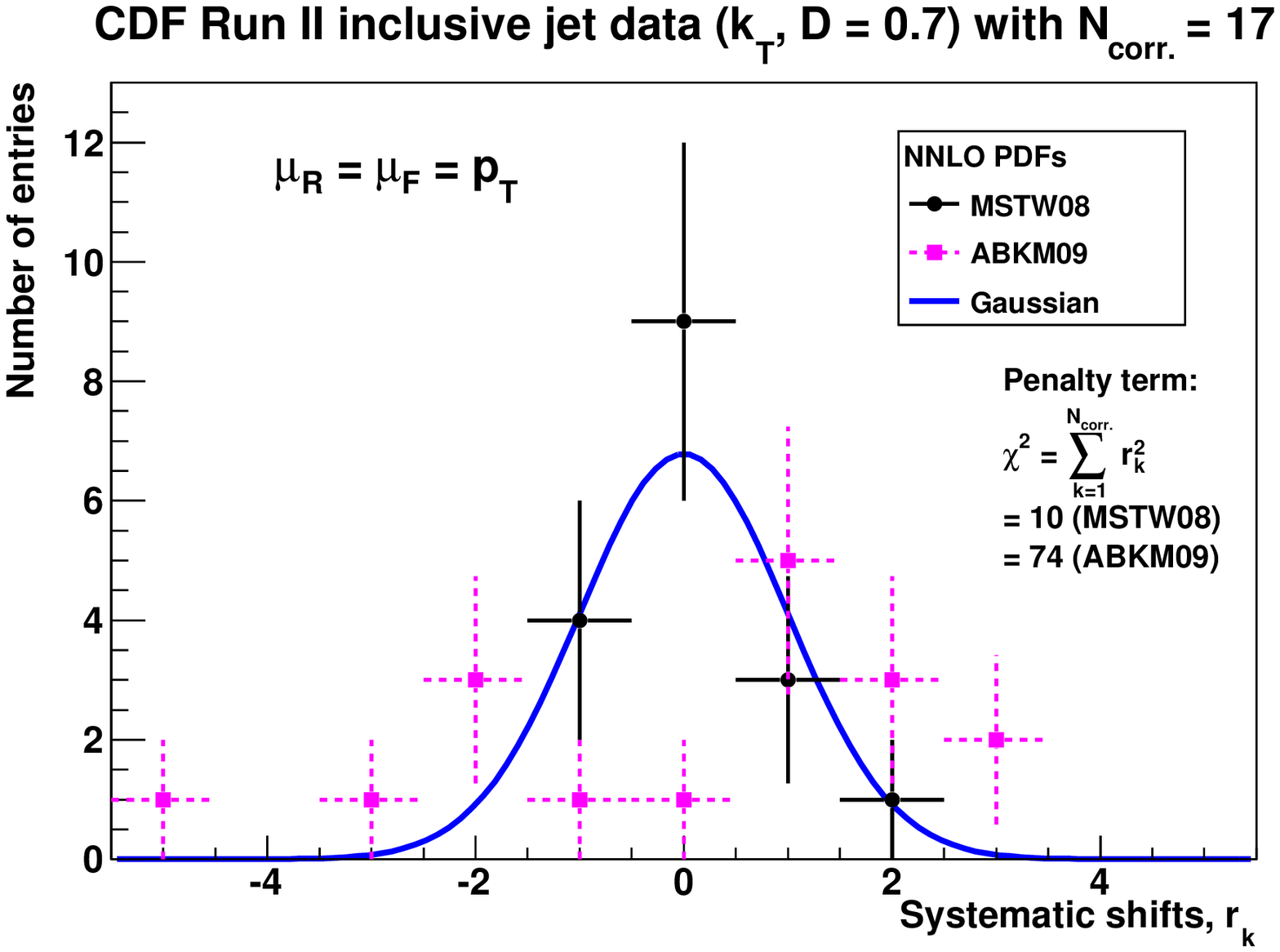}
  \end{minipage}%
  \begin{minipage}{0.5\textwidth}
    (b)\\
    \includegraphics[width=\textwidth]{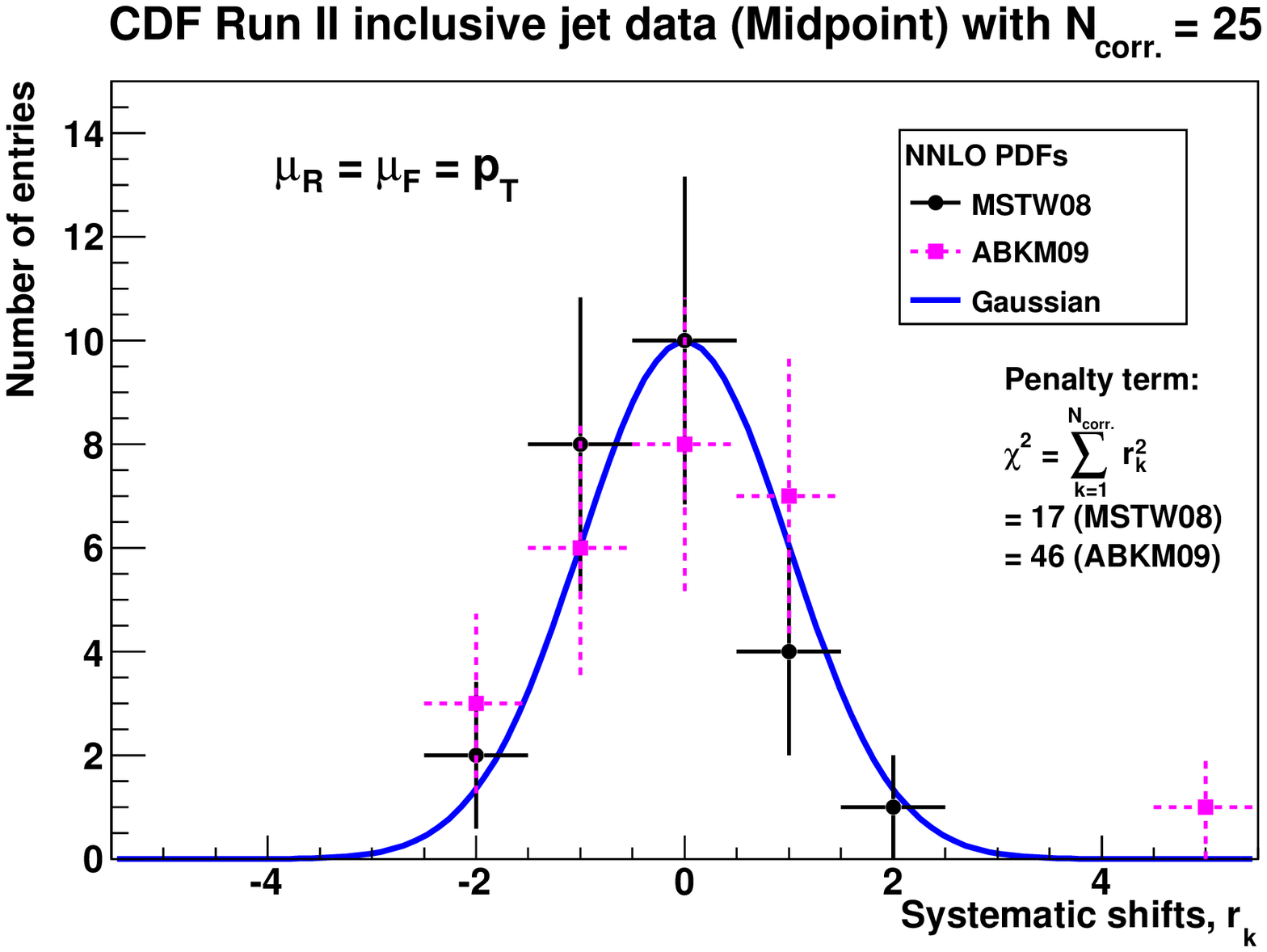}
  \end{minipage}
  \begin{minipage}{0.5\textwidth}
    (c)\\
    \includegraphics[width=\textwidth]{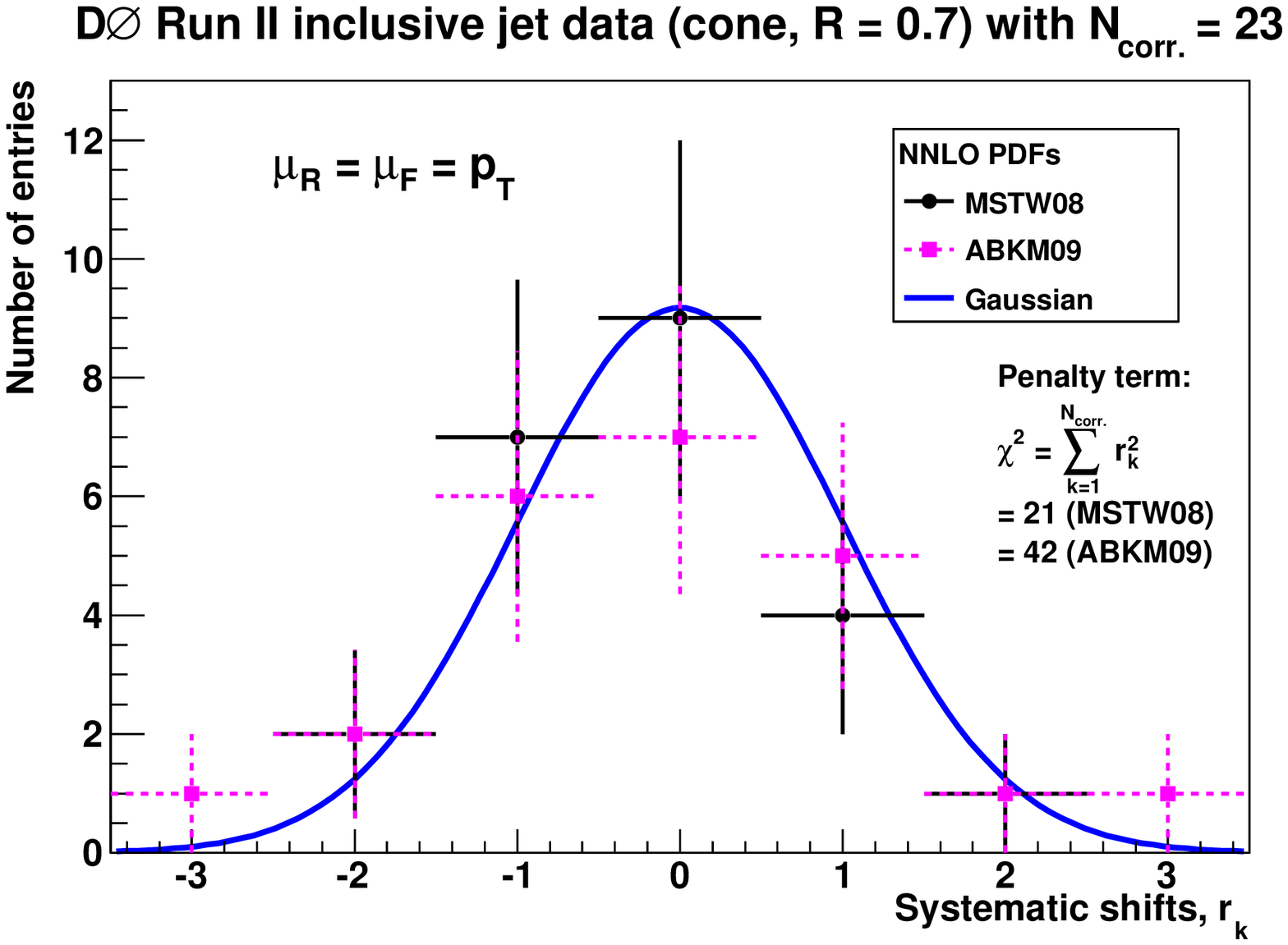}
  \end{minipage}%
  \begin{minipage}{0.5\textwidth}
    (d)\\
    \includegraphics[width=\textwidth]{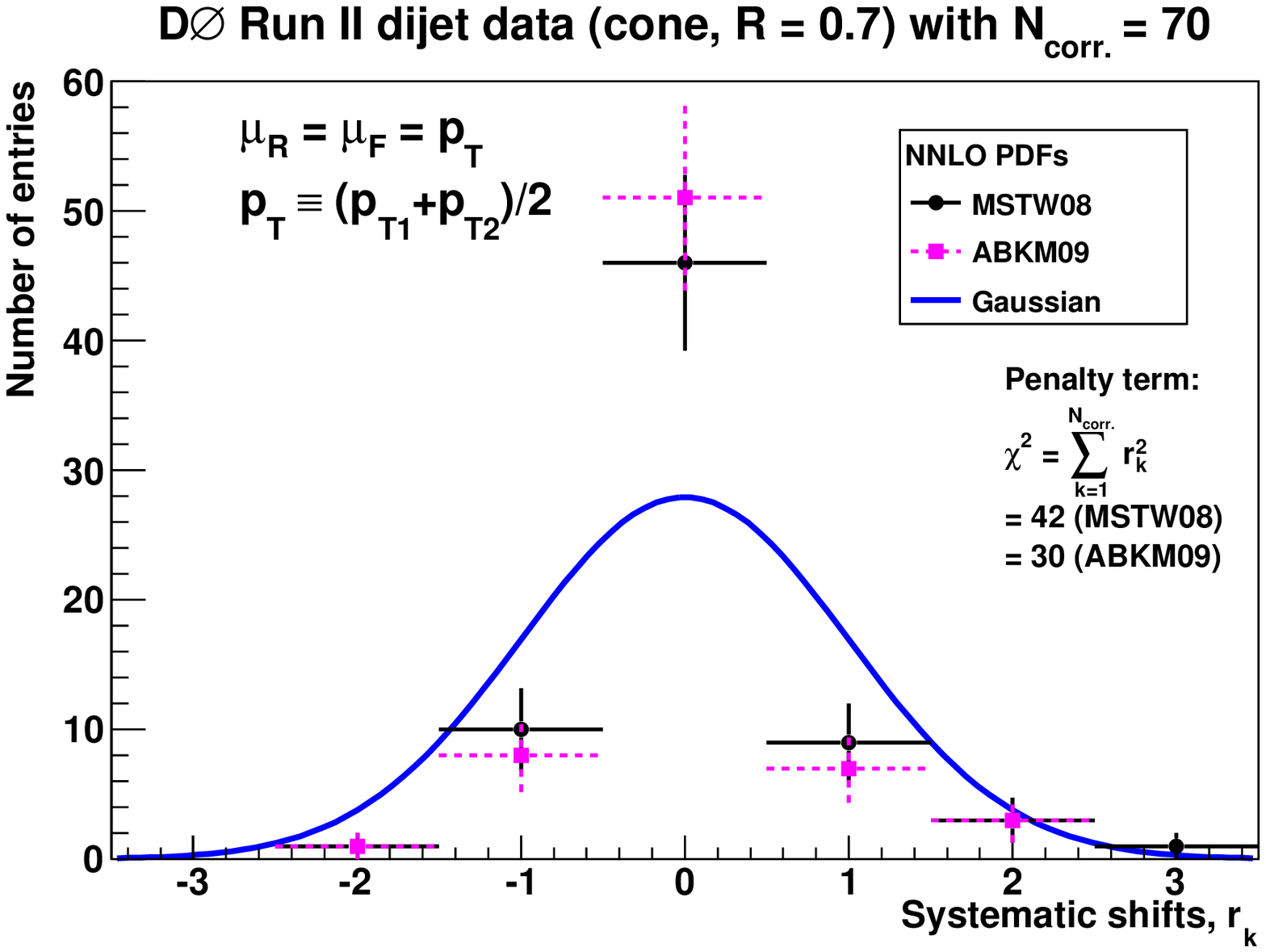}
  \end{minipage}
  \caption{Distributions of the systematic shifts, $r_k$ given by eq.~\eqref{eq:rksolution}, for each of the four Tevatron data sets on jet production, with theory predictions calculated using either MSTW08 or ABKM09 NNLO PDFs, compared to the expectation of a Gaussian distribution with unit width.}
  \label{fig:plotrk}
\end{figure}
In figure~\ref{fig:plotrk} we show the similar distributions of the systematic shifts, $r_k$, again for all four Tevatron Run II data sets on jet production.  We show the expected behaviour of a Gaussian distribution with unit width and the penalty $\chi^2$ term simply given by the sum of the $r_k^2$ values.  For the inclusive jet data, the systematic shifts for MSTW08 show the expected Gaussian behaviour, with small penalty terms $\sum_{k=1}^{N_{\rm corr.}}r_{k}^2 < N_{\rm corr.}$.  On the other hand, the systematic shifts for ABKM09 deviate substantially from Gaussian behaviour, with much larger penalty terms, in particular for the CDF inclusive jet data using the $k_T$ algorithm shown in figure~\ref{fig:plotrk}(a).  The systematic shifts for the dijet data shown in figure~\ref{fig:plotrk}(d) have a much narrower distribution than the expected Gaussian behaviour for both MSTW08 and ABKM09, suggesting that the systematic errors are overestimated, are non-Gaussian, or are not independent (or a combination of these three explanations).  Note that the number of systematic sources ($N_{\rm corr.}=70$) for the dijet data is much greater than for any of the inclusive jet data sets.  Indeed, this allows the value of $\chi^2$ for the description of data by an identically zero theory prediction to be lower than for some of the PDF sets; see table~\ref{tab:d0dijet}.

The presentation of the results in figures~\ref{fig:plotpulls} and \ref{fig:plotrk} enables a separation between contributions to the $\chi^2$ definition, eq.~\eqref{eq:chisqcorr}, from uncorrelated and correlated errors, respectively.  This allows a more informed assessment of the fit quality compared to the more traditional $\chi^2$ definition of eq.~\eqref{eq:chisqcov} in terms of the experimental covariance matrix, used by the ABKM and NNPDF fitting groups; see also appendix B.3 of ref.~\cite{Pumplin:2002vw} and section 4 of ref.~\cite{Lai:2010vv}.

\subsection{Other jet cross sections\texorpdfstring{ from collider experiments}{}} \label{sec:otherjets}

The D{\O} Collaboration has recently made a measurement~\cite{Abazov:2011ub} of the three-jet differential cross section as a function of the invariant mass of the three jets with the largest transverse momentum in an event.  An exercise has been carried out similar to the one presented here, where the $\chi^2$ has been evaluated for different PDF (and $\alpha_S$) choices and scale choices $\mu_R=\mu_F=\mu=\{p_T/2,p_T,2p_T\}$, where the mean jet $p_T\equiv(p_{T1}+p_{T2}+p_{T3})/3$.  The trend is that MSTW08 and NNPDF2.1 are favoured, as for the inclusive jet study presented here, while ABKM09 is worse, and CT10 and HERAPDF1.0 are still poorer.  We have followed a similar approach to that of ref.~\cite{Abazov:2011ub} by evaluating the $\chi^2$ only for the central PDF fit, without accounting for PDF uncertainties.  Since the Tevatron jet data provide by far the most direct constraint on the high-$x$ gluon distribution, the agreement of the central PDF fit is more important and relevant than obtaining agreement only within possibly large PDF uncertainties.  However, the potential choice of scales for the three-jet cross section is even broader than for the dijet cross section.

The LHC data on jet production~\cite{Chatrchyan:2011qt,Chatrchyan:2011me,ATLAS:jetsmeasured,ATLAS:jetscomparison} are becoming more precise and show some sensitivity to the PDF choice.  However, these data are still being understood and are not presented with separated correlated systematic uncertainties which would allow a quantitative $\chi^2$ comparison.  Moreover, the general sensitivity is to lower $x_T\sim2p_T/\sqrt{s}$, and so less relevant for Higgs production at the Tevatron.  Isolated photon production at the LHC may also provide a direct constraint on the gluon distribution~\cite{Ichou:2010wc}.  The HERA jet data are less sensitive to the gluon distribution at high $x$ values, being more of a constraint for $x\sim 0.001$--$0.1$, and there is no NNLO calculation, or any approximation such as the 2-loop threshold corrections available for the Tevatron inclusive jet data.

\subsection{Summary} \label{sec:summary}

Comparison with Tevatron jet data is subtle because of the large correlated systematic uncertainties.  The systematic shifts, eq.~\eqref{eq:datashift}, can compensate for inadequacies in the theory calculation.  The traditional $\chi^2$ definition in terms of the experimental covariance matrix, eq.~\eqref{eq:chisqcov}, can hide such systematic shifts.  In particular, we find that the Tevatron jet data need to be normalised downwards by typically between 3-$\sigma$ and 5-$\sigma$ (see appendix~\ref{sec:freenorm}) to achieve the best agreement with some PDF sets, particularly the ABKM09 predictions.  Even if the luminosity shift is artificially constrained, the other systematic shifts move by large amounts for the inclusive jet data, incompatible with the Gaussian expectation.  No such problems are observed for the MSTW08 predictions.  It can also be seen from the plots in ref.~\cite{Alekhin:2011cf} that the unshifted Tevatron jet data lie significantly above the theory predictions even after including these data in variants of the ABKM09 fit.  Constraining the Tevatron luminosity shifts, for example, so that the predicted $W$ and $Z$ cross sections agreed with Tevatron data, would increase the constraining power of the Tevatron jet data and thereby very likely give a larger $\alpha_S$ and high-$x$ gluon distribution than the current studies of Alekhin, Bl\"umlein and Moch (ABM)~\cite{Alekhin:2011cf}.  Even with the existing treatment, the NNLO Tevatron $gg\to H$ cross section for $M_H=165$~GeV goes up by $\{15,12,17,11\}\%$ when including the \{CDF $k_T$~\cite{Abulencia:2007ez}, CDF Midpoint~\cite{Aaltonen:2008eq}, D{\O} inclusive~\cite{Abazov:2008hua}, D{\O} dijet~\cite{Abazov:2010fr}\} data set in variants of the ABKM09 fit~\cite{Alekhin:2011cf}.  The dijet data has a potentially wider range of allowed scale choices than the inclusive jet data.  We conclude that the data on \emph{inclusive} jet production therefore provide the cleanest probe of different PDF sets.

\section{Value of strong coupling \texorpdfstring{$\alpha_S$ }{}from DIS} \label{sec:alphaS}
There is a common lore (see, for example, ref.~\cite{Alekhin:2011gj}) that DIS-only fits prefer low $\alpha_S(M_Z^2)$ values, but ref.~\cite{Martin:2009bu} showed that not all DIS data sets prefer low $\alpha_S(M_Z^2)$ values.  In particular, this was found to be true only for BCDMS data, and for E665 and SLAC $ep$ data, while NMC, SLAC $ed$ and HERA data preferred high $\alpha_S(M_Z^2)$ values within the context of the global fit~\cite{Martin:2009bu}.  (See also the recent NNPDF study at NLO using an ``unbiased'' PDF parameterisation~\cite{Lionetti:2011pw}.)

It is well known that $\alpha_S$ is highly \emph{anticorrelated} with the low-$x$ gluon distribution through scaling violations of HERA data: $\partial F_2/\partial\ln(Q^2)\sim \alpha_S\,g$.  Then $\alpha_S$ is \emph{correlated} with the high-$x$ gluon distribution through the momentum sum rule; see, for example, figure 14(b) of ref.~\cite{Martin:2009bu}.  Restrictive gluon parameterisations, without the negative small-$x$ term allowed by MSTW~\cite{Martin:2009iq}, can therefore bias the extracted $\alpha_S$ value.  For example, the default MSTW08 NNLO fit obtained $\alpha_S(M_Z^2) = 0.1171\pm 0.0014$, while imposing the restriction of a positive input gluon at $Q_0^2=1$~GeV$^2$ gave a best-fit $\alpha_S(M_Z^2) = 0.1157$, but with a $\chi^2$ worse by 63 units for the global fit to 2615 data points~\cite{Martin:2009bu}.\footnote{The values for the $\chi^2$ increase of 80 at NLO and 63 at NNLO were erroneously interchanged in ref.~\cite{Martin:2009bu}.}

What is $\alpha_S$ from only DIS data in the MSTW08 NNLO fit?\footnote{Studies prompted by question from G.~Altarelli, December 2010.}  Recall that the global fit gave $\alpha_S(M_Z^2) = 0.1171\pm 0.0014$~\cite{Martin:2009bu}.  To expand on the studies made in ref.~\cite{Martin:2009bu}, we performed a new NNLO DIS-only fit, which gave a best-fit $\alpha_S(M_Z^2) = 0.1104$, but with an input gluon distribution which went negative for $x>0.4$ due to lack of any data constraint.  This implies a negative charm structure function, $F_2^{\rm charm}$, and a terrible description ($\chi^2/N_{\rm pts.}\sim 10$ including correlated systematic errors) of Tevatron jet data using the obtained PDFs.  A DIS-only fit fixing the high-$x$ gluon parameters to prevent such bad behaviour gave $\alpha_S(M_Z^2) = 0.1172$, i.e.~very similar to the global fit.  However, a NNLO fit which imposed the condition of the positive low-$x$ gluon, which stopped the gluon from going negative at high $x$ values, and which also omitted the Tevatron jet data, gave $\alpha_S(M_Z^2)=0.1139$, rather closer to the ABKM09 value.  The very low value of $\alpha_S(M_Z^2) = 0.1104$ found in the DIS-only fit is due to the dominance of BCDMS data.  We can show this explicitly by removing the BCDMS data from the DIS-only fit, then the best-fit $\alpha_S(M_Z^2)$ moves from $0.1104$ to $0.1193$.  Repeating the \emph{global} fit with BCDMS data removed gives $\alpha_S(M_Z^2) = 0.1181$, i.e.~a change by less than the quoted experimental uncertainty of $\pm0.0014$.  The conclusion is that the Tevatron jet data are vital to pin down the high-$x$ gluon, giving a smaller low-$x$ gluon and therefore a larger $\alpha_S$ in the global fit compared to a DIS-only fit, at the expense of some deterioration in the fit quality of the BCDMS data.\footnote{The low $y$ data points from BCDMS are strongly affected by the energy scale uncertainty of the scattered muon.  It has been advocated to impose a cut of $y>0.3$ on the BCDMS data, which caused $\alpha_S(M_Z^2)$ to increase by about 0.004 in a fit to only BCDMS data and by about 0.002 in a combined fit to H1 and BCDMS data~\cite{Adloff:2000qk}.}  The benefits of including the Tevatron jet data to obtain sensible results in a simultaneous fit of PDFs and $\alpha_S$ therefore greatly outweighs any disadvantage such as lack of complete NNLO corrections.

The only input DIS value to the current world average $\alpha_S(M_Z^2)$~\cite{Bethke:2009jm} is the BBG06 value~\cite{Blumlein:2006be}, which is from a non-singlet analysis and therefore \emph{in principle} free of assumptions made about the gluon distribution.  A value of
\begin{equation}
  \alpha_S(M_Z^2)=\left\{0.1148^{+0.0019}_{-0.0019}, 0.1134^{+0.0019}_{-0.0021}, 0.1141^{+0.0020}_{-0.0022}\right\}
\end{equation}
was obtained at \{NLO, NNLO, N$^{3}$LO\}, by fitting proton and deuteron structure functions, $F_2^p$ and $F_2^d$, for $x\ge 0.3$ (assuming only valence quarks, neglecting the singlet contribution), and the less precise $F_2^{\rm NS}=2(F_2^p-F_2^d)$ for $x<0.3$.  However, using the MSTW08 NNLO central fit, contributions other than valence quarks are found to make up about $10\%$ ($2\%$) of $F_2^p$ at $x=0.3$ ($x=0.5$).  As an exercise we performed the MSTW08 NNLO DIS-only fit just to $F_2^p$ and $F_2^d$ for $x>0.3$ (comprising 282 data points, 160 of these from BCDMS), which gave $\alpha_S(M_Z^2)=0.1103$ ($0.1130$) without (with) the singlet contribution included.  This is even lower than the BBG06 value presumably due to lack of the $y>0.3$ cut on BCDMS data applied in the BBG06 analysis.  The low value of $\alpha_S(M_Z^2)$ found by BBG06~\cite{Blumlein:2006be} is therefore due to both dominance of BCDMS data and by what we conclude is the unjustified neglect of the singlet contribution to $F_2^p$ and $F_2^d$ for $x\ge 0.3$.  Given that it was argued above that the Tevatron jet data are needed to pin down the high-$x$ gluon, we conclude that an extraction of $\alpha_S(M_Z^2)$ only from inclusive DIS data is not meaningful, and the closest possible to a reliable extraction is the MSTW08 NNLO combined analysis of DIS, Drell--Yan and jet data~\cite{Martin:2009iq,Martin:2009bu}:
\begin{equation}
  \alpha_S(M_Z^2) = 0.1171\quad\pm0.0014\text{ (68\% C.L.)}\quad\pm0.0034\text{ (90\% C.L.)}.
\end{equation}
This value is the only NNLO determination, from a simultaneous fit with PDFs, which is in agreement with the current world average $\alpha_S(M_Z^2)=0.1184\pm0.0007$~\cite{Bethke:2009jm}; see figure~\ref{fig:asmzvalues}(b).

\section{Treatment of NMC data\texorpdfstring{ and stability to low $Q^2$ data}{}} \label{sec:nmcdata}

A recent claim has been made~\cite{Alekhin:2011ey} that the bulk of the MSTW08/ABKM09 difference in both the extracted $\alpha_S(M_Z^2)$ value and the $gg\to H$ predictions is explained by the treatment of NMC data~\cite{Arneodo:1996qe}.  The differential cross section for DIS of charged leptons off nucleons, $\ell N\to \ell X$, neglecting the nucleon and lepton masses, and assuming single-photon exchange, is
\begin{equation} \label{eq:d2sigma}
  \frac{{\rm d}^2\sigma}{{\rm d}x\,{\rm d}Q^2} \simeq \frac{4\pi\alpha^2}{x\,Q^4}\left[1-y+\frac{y^2/2}{1+R(x,Q^2)}\right]F_2(x,Q^2),
\end{equation}
where $R=\sigma_L/\sigma_T\simeq F_L/(F_2-F_L)$ is the ratio of the $\gamma^* N$ cross sections for longitudinally and transversely polarised photons, $Q^2$ is the photon virtuality, $x$ is the Bjorken variable and $y\simeq Q^2/(x\,s)$ is the inelasticity (with $\sqrt{s}$ the $\ell N$ centre-of-mass energy).  The ABKM09~\cite{Alekhin:2009ni} analysis fitted the NMC differential cross sections directly, calculating $F_L$ to $\mathcal{O}(\alpha_S^2)$ and including empirical higher-twist corrections.  The MSTW08~\cite{Martin:2009iq} analysis instead fitted the NMC $F_2$ values corrected for $R$, where~\cite{Arneodo:1996qe}
\begin{equation}
R(x,Q^2) = \begin{cases}R_{\rm NMC}(x)&\mbox{if }x<0.12\\R_{1990}(x,Q^2)&\mbox{if }x>0.12\end{cases}.
\end{equation}
Here, $R_{\rm NMC}(x)$ was a ($Q^2$-independent) value extracted from NMC data, while $R_{1990}(x,Q^2)$ was a $Q^2$-dependent empirical parameterisation of SLAC data dating from 1990~\cite{Whitlow:1990gk}.  By replacing the NMC differential cross-section data by NMC $F_2$ data, ABM~\cite{Alekhin:2011ey} find that their best-fit $\alpha_S(M_Z^2)$ moves from 0.1135 to 0.1170 and their $gg\to H$ cross sections at the Tevatron and LHC move closer to the MSTW08 values.  ABM~\cite{Alekhin:2011ey} therefore conclude that the use of NMC $F_2$ data in the MSTW08 fit rather than the differential cross section is the main reason for the higher $\alpha_S(M_Z^2)$ and Higgs cross sections obtained with MSTW08.

\begin{figure}
  \centering
  \begin{minipage}{0.5\textwidth}
    (a)\\
    \includegraphics[width=\textwidth]{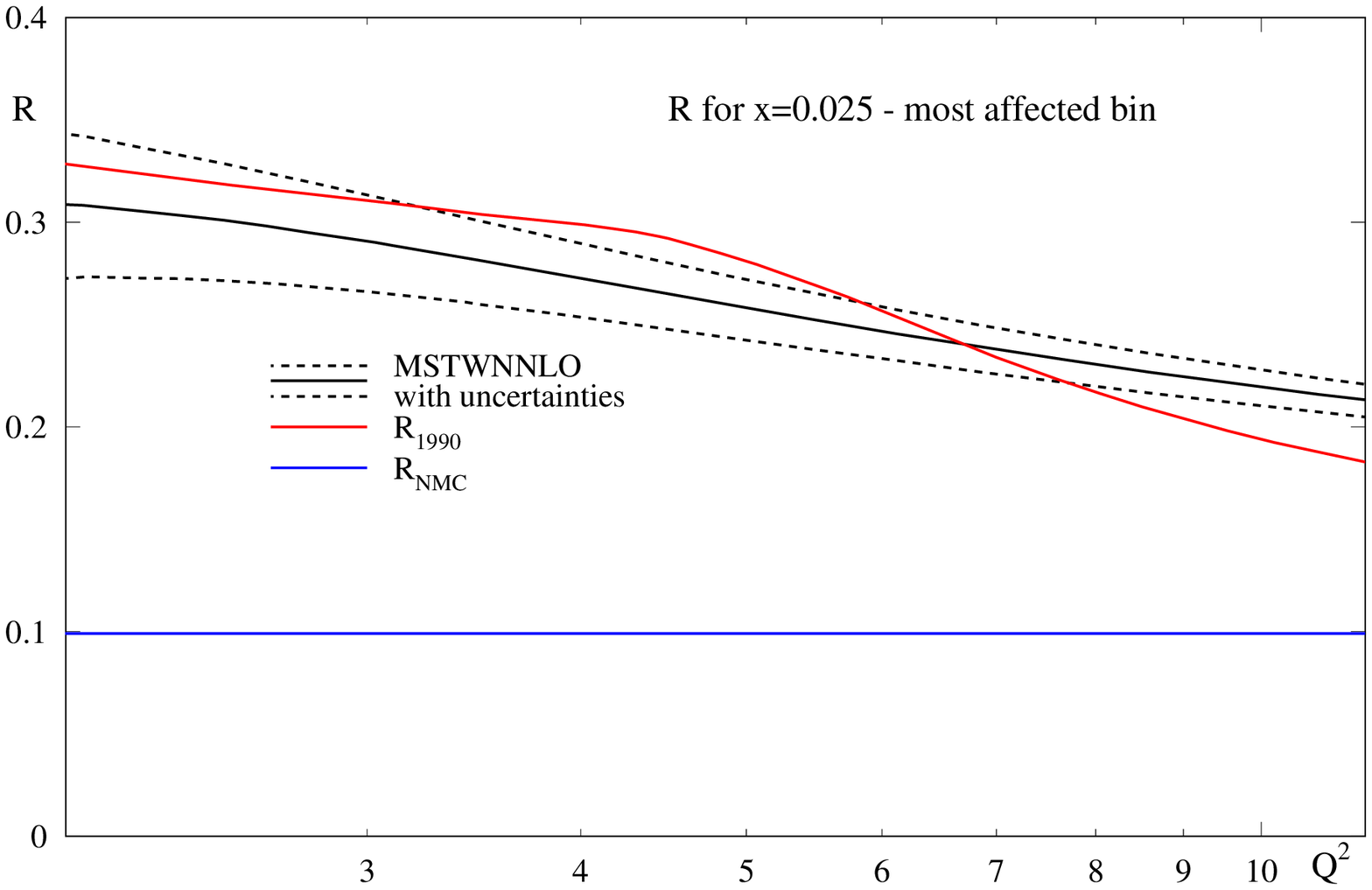}
  \end{minipage}%
  \begin{minipage}{0.5\textwidth}
    (b)\\
    \includegraphics[width=\textwidth]{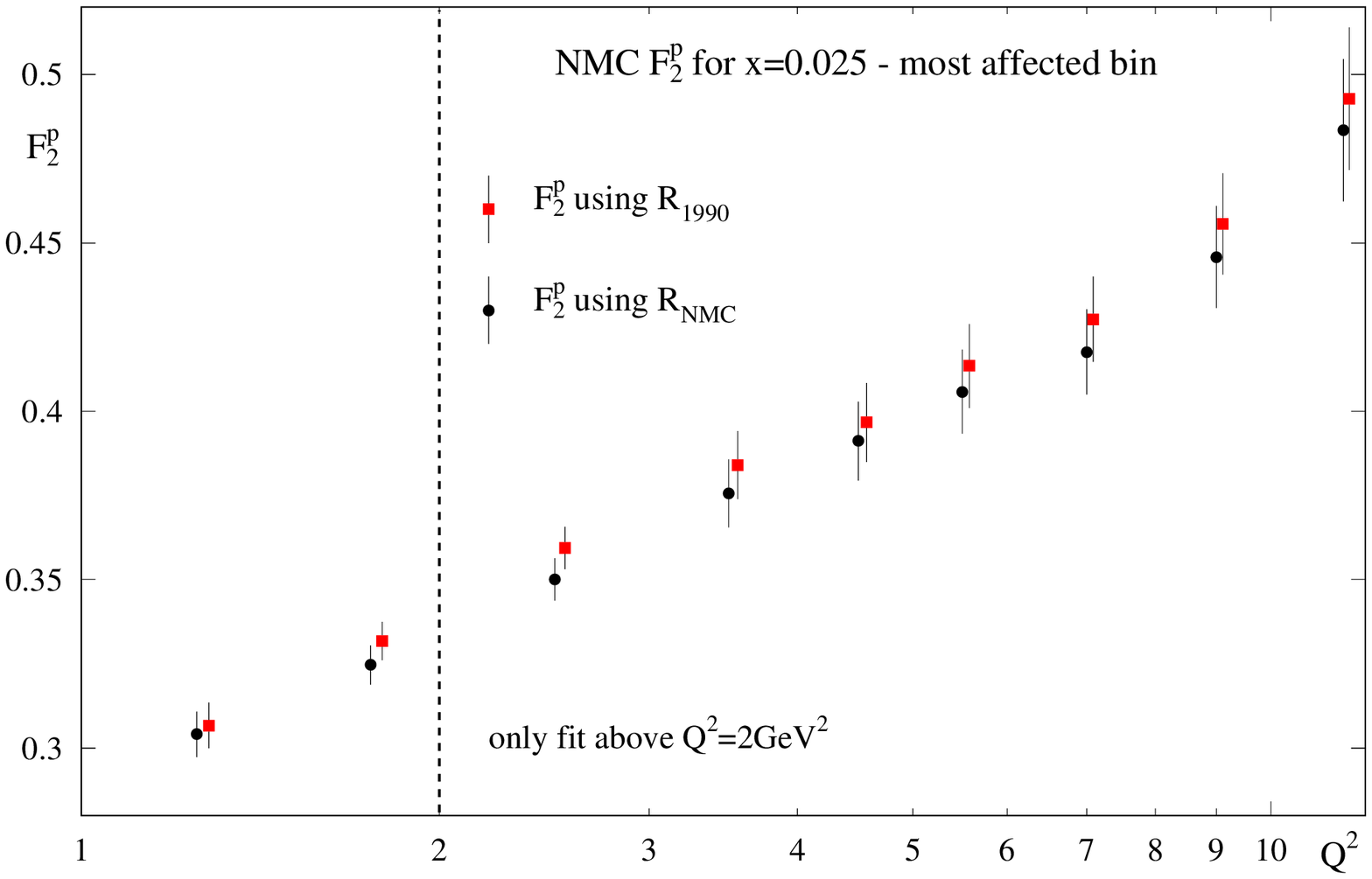}
  \end{minipage}
  \caption{(a)~$R=\sigma_L/\sigma_T\simeq F_L/(F_2-F_L)$ versus $Q^2$ (in units of GeV$^2$) at $x=0.025$ comparing the $Q^2$-independent $R_{\rm NMC}$ extraction~\cite{Arneodo:1996qe}, the $Q^2$-dependent SLAC $R_{1990}$ parameterisation~\cite{Whitlow:1990gk}, and the MSTW08 NNLO calculation including 1-$\sigma$ PDF uncertainties~\cite{Martin:2009iq}.  (b)~$F_2$ versus $Q^2$ (in units of GeV$^2$) at $x=0.025$ comparing the two NMC extractions~\cite{Arneodo:1996qe} using either $R_{\rm NMC}$ or $R_{1990}$.}
  \label{fig:nmc}
\end{figure}
We agree that it is more consistent to fit directly to the NMC differential cross-section data, so here we respond to this rather dramatic assertion made by ABM~\cite{Alekhin:2011ey}, which would obviously be very worrying if correct.  However, rather than repeat the MSTW08 analysis by fitting the NMC differential cross sections, we note that the original NMC paper~\cite{Arneodo:1996qe} made an alternative extraction of $F_2$ values using the SLAC $R_{1990}$ parameterisation~\cite{Whitlow:1990gk}.  In figure~\ref{fig:nmc}(a) we compare $R_{\rm NMC}$ with $R_{1990}$ in the most affected bin of $x=0.025$, i.e.~a low $x$ value where there are a reasonable number (7) of NMC data points surviving the cut on $Q^2\ge2$~GeV$^2$ and where the difference between $R_{\rm NMC}$ and $R_{1990}$ is at its largest.  Recall that a low $x$ value means a high $y$ value and from eq.~\eqref{eq:d2sigma} the correction term from $R$ is only important at large $y$.  In figure~\ref{fig:nmc}(a) we also show the MSTW08 NNLO prediction, including PDF uncertainties at 68\% C.L., with $F_L$ calculated to $\mathcal{O}(\alpha_S^3)$ and without any higher-twist corrections.  We see that it gives a good description of the SLAC $R_{1990}$ parameterisation, with any differences being very much smaller than those between $R_{\rm NMC}$ and $R_{1990}$.  We note that NMC/BCDMS/SLAC $F_L$ data are included in the MSTW08 fit and are well-described at NNLO but less well at NLO (see figure~5 of ref.~\cite{Martin:2009bu}), so the $\mathcal{O}(\alpha_S^3)$ coefficient functions are needed for a good description and the larger MSTW08 $\alpha_S(M_Z^2)$ perhaps explains why there is less room for higher-twist corrections, contrary to the findings of the ABM analysis.  Nevertheless, figure~\ref{fig:nmc}(a) demonstrates that fitting the alternative NMC $F_2$ data extracted using the SLAC $R_{1990}$ parameterisation will give very similar results to fitting the NMC differential cross sections.  In fact, given that $R_{1990}$ in figure~\ref{fig:nmc}(a) generally has a slightly steeper $Q^2$ dependence than the MSTW08 parameterisation, using this will slightly overestimate the true impact of fitting the NMC differential cross sections.  In figure~\ref{fig:nmc}(b) we compare the two different NMC $F_2$ extractions, again for the most affected bin of $x=0.025$, and we see that there is little difference, certainly nothing that seems likely to change $\alpha_S(M_Z^2)$ by $0.0035$ in a fit where it is constrained with an uncertainty of about $0.0014$ by over 2000 other data points.

\begin{table}
\centering
\begin{tabular}{|l|c|c|c|}
\hline
NNLO PDF & $\alpha_S(M_Z^2)$ & $\sigma_H$ at Tevatron & $\sigma_H$ at 7 TeV LHC\\ \hline
{\bf MSTW08} & {\bf 0.1171} & {\bf 0.342~pb} & {\bf 7.91~pb} \\ \hline
Use $R_{1990}$ for NMC $F_2$ & $0.1167$ & $-0.7\%$ & $-0.9\%$ \\
Cut NMC $F_2$ ($x<0.1$) & $0.1162$ & $-1.2\%$ & $-2.1\%$ \\
Cut all NMC $F_2$ data & $0.1158$ & $-0.7\%$ & $-2.1\%$ \\ \hline
Cut $Q^2<5$~GeV$^2$, $W^2<20$~GeV$^2$ & $0.1171$ & $-1.2\%$ & $+0.4\%$ \\
Cut $Q^2<10$~GeV$^2$, $W^2<20$~GeV$^2$ & $0.1164$ & $-3.0\%$ & $-1.7\%$ \\ \hline
Fix $\alpha_S(M_Z^2)$ & $0.1130$ & $-11\%$ & $-7.6\%$ \\ 
Input $xg>0$, no jets & $0.1139$ & $-17\%$ & $-4.9\%$ \\ \hline
ABKM09 & 0.1135 & $-26\%$ & $-11\%$ \\
\hline
\end{tabular}
\caption{Effect of NMC treatment on $\alpha_S(M_Z^2)$ and Higgs cross sections ($M_H=165$~GeV).  We also show the effect of raising the cuts imposed on the DIS data compared to the default of removing data with $Q^2<2$~GeV$^2$ and $W^2<15$~GeV$^2$.  Finally, we show the effect of simply fixing $\alpha_S(M_Z^2)$ to be close to the ABKM09 value, or performing a fit with a positive-definite input gluon distribution and no jet data, and we compare directly to ABKM09.\label{tab:nmc}}
\end{table}
\begin{figure}
  \centering
  \includegraphics[width=\textwidth]{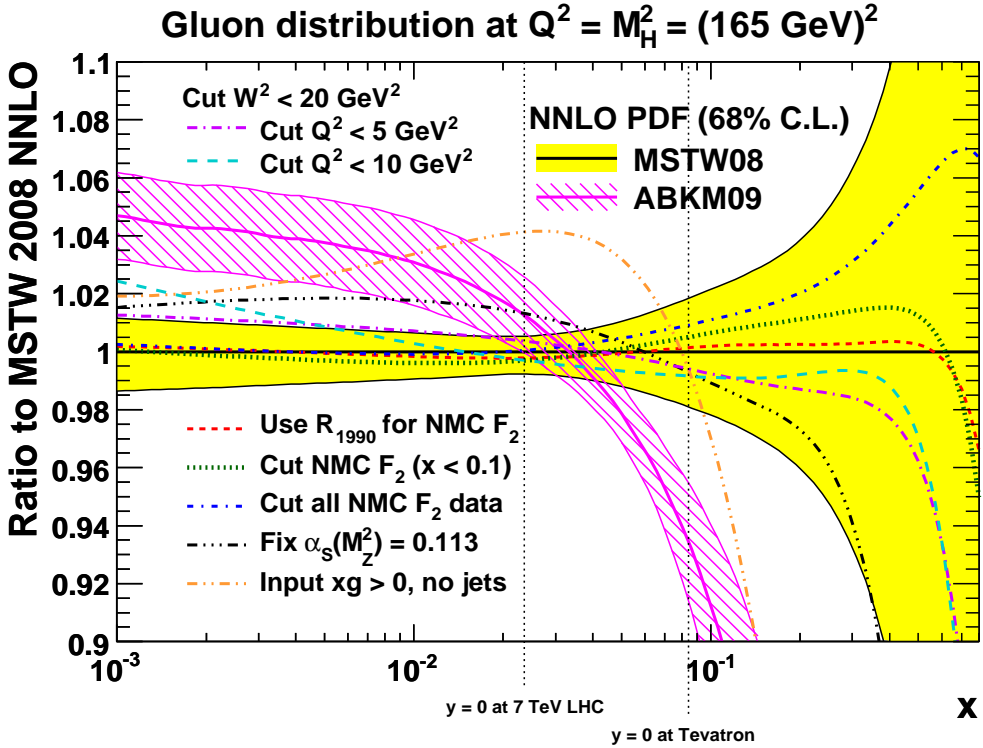}
  \caption{Effect of NMC treatment on the gluon distribution at a scale $Q^2=(165~{\rm GeV})^2$.  The values of $x=M_H/\sqrt{s}$ relevant for central production (assuming $p_T^H=0$) of a Standard Model Higgs boson of mass $M_H=165$~GeV at the Tevatron and LHC are indicated.  We also show the effect of raising the cuts imposed on the DIS data compared to the default of removing data with $Q^2<2$~GeV$^2$ and $W^2<15$~GeV$^2$.  Finally, we show the effect of simply fixing $\alpha_S(M_Z^2)$ to be close to the ABKM09 value, or performing a fit with a positive-definite input gluon distribution and no jet data, and we compare directly to ABKM09.}
  \label{fig:nmcgluon}
\end{figure}
In table~\ref{tab:nmc} we show the effect of repeating the MSTW08 NNLO fit with the NMC $F_2$ data extracted using $R_{1990}$ on $\alpha_S(M_Z^2)$ and the Higgs cross sections (for $M_H=165$~GeV) at the Tevatron and LHC, and in figure~\ref{fig:nmcgluon} we show the change in the gluon distribution at the corresponding scale.  We make other fits either cutting the NMC $F_2$ data for $x<0.1$, above which the $R$ correction in eq.~\eqref{eq:d2sigma} is very small indeed, or completely removing all NMC $F_2$ data.  In all cases there is very little change in $\alpha_S(M_Z^2)$, the gluon distribution, and the Higgs cross section.  We conclude that the treatment of NMC data cannot explain the difference between the MSTW08 and ABKM09 results.  Similar stability has been found by the NNPDF group~\cite{NNPDF:2011we}, but in a less relevant study at NLO with fixed $\alpha_S$.

The cuts on DIS data are not explicitly given in the ABKM09 paper~\cite{Alekhin:2009ni}, but the previous AMP06 paper~\cite{Alekhin:2006zm} mentions that DIS data are removed with $Q^2<2.5$~GeV$^2$ and $W^2<(1.8~{\rm GeV})^2=3.24~{\rm GeV}^2$, compared to the MSTW08 fit which removes DIS data with $Q^2<2$~GeV$^2$ and $W^2<15~{\rm GeV}^2$.  The much weaker cut on the hadronic invariant mass (squared), $W^2\simeq Q^2(1/x-1)$, clearly explains why higher-twist corrections are more important in the ABKM09 analysis.  To investigate the possible effect of neglected higher-twist corrections on the MSTW08 NNLO fit we raised the cuts to remove DIS data with $W^2<20$~GeV$^2$ and either $Q^2<5$~GeV$^2$ or $Q^2<10$~GeV$^2$.  The results are shown in table~\ref{tab:nmc} and figure~\ref{fig:nmcgluon}.  The changes in $\alpha_S$, the gluon distribution and the Higgs cross sections are generally small and within uncertainties, although with the strongest $Q^2$ cut there is no data constraint below $x=10^{-4}$ and little just above, so the PDFs differ but have large uncertainties at low $x$ values.\footnote{We also investigated the effect of increasing the cuts on $W^2$ and $Q^2$ in variants of the MSTW NLO fit.  The changes were slightly bigger, with $\alpha_S(M_Z^2)$ changing from 0.1202 to 0.1192 and 0.1175 with $Q^2$ cuts of 5 and 10~GeV$^2$, respectively.  Similarly, the changes in PDFs and cross-section predictions are generally slightly greater at NLO than at NNLO, i.e.~as expected there is some improved stability at higher orders.}

In table~\ref{tab:nmc} and figure~\ref{fig:nmcgluon} we show the results of the MSTW08 NNLO fit with a fixed $\alpha_S(M_Z^2) = 0.113$~\cite{Martin:2009bu} (slightly below the ABKM09 value), and even in this case the gluon distribution and Higgs cross sections move only part of the way towards the ABKM09 result, as already seen in figure~\ref{fig:gghvsasmznnlo}.  The MSTW08 input gluon parameterisation is \cite{Martin:2009iq}
\begin{equation} \label{eq:inputgluon}
  xg(x,Q_0^2=1~{\rm GeV}^2) \;=\; A_g\,x^{\delta_g}\,(1-x)^{\eta_g}\,(1+\epsilon_g\,\sqrt{x}+\gamma_g\,x)\;+\;A_{g^\prime}\,x^{\delta_{g^\prime}}\,(1-x)^{\eta_g^\prime},
\end{equation}
compared to the much more restrictive functional forms of the other NNLO fits, namely:
\begin{align}
  \text{ABKM09~\cite{Alekhin:2009ni}:}&\quad xg(x,Q_0^2=9~{\rm GeV}^2) = A_g\,x^{\delta_g}\,(1-x)^{\eta_g}\,x^{\gamma_g\,x}, \\
  \text{JR09~\cite{JimenezDelgado:2008hf}:}&\quad xg(x,Q_0^2=0.55~{\rm GeV}^2) = A_g\,x^{\delta_g}\,(1-x)^{\eta_g}, \\
  \text{HERAPDF1.0~\cite{HERA:2009wt}:}&\quad xg(x,Q_0^2=1.9~{\rm GeV}^2) = A_g\,x^{\delta_g}\,(1-x)^{\eta_g}.
\end{align}
The normalisation $A_g$ is determined from the momentum sum rule constraint, leaving 7 free parameters for MSTW08 compared to only 3 for ABKM09 and only 2 for JR09 and HERAPDF1.0 (although the value of $Q_0^2$ is optimised in the case of JR09).  In the lack of any direct data constraint on the high-$x$ gluon distribution, the other fits are therefore constrained by the form of the input parameterisation, avoiding the pathological behaviour of the negative high-$x$ gluon distribution seen for the MSTW08 NNLO DIS-only fit described in section~\ref{sec:alphaS}.  As already mentioned in that section, in an attempt to mimic the ABKM09 fit we performed a variant of the MSTW08 NNLO fit without jet data and with the second term of eq.~\eqref{eq:inputgluon} set to zero.  The $\epsilon_g$ and $\gamma_g$ parameters were fixed in the fit iteration before the high-$x$ gluon distribution went negative.  The results of this fit are shown in table~\ref{tab:nmc} and figure~\ref{fig:nmcgluon} and it goes some way towards reproducing the high-$x$ gluon of the ABKM09 fit and the corresponding Tevatron $gg\to H$ prediction, certainly closer than we come with other modifications.  Finally, we then investigated the effect of using NMC data corrected using $R_{1990}$ rather than $R_{\rm NMC}$ in this fit.  Similar to our default fit all changes were at the percent level, or less, so we do not explicitly show them, although the gluon does move marginally closer again to that of ABKM09.

Other differences between the two analyses are that ABKM09 used the NMC data for separate muon beam energies, whereas MSTW08 used the NMC data averaged over beam energies, which reduces the maximum effect of the change in $R$ for a particular data point, i.e.~at a given $x$ and $Q^2$, a data point at high $y$, and so very sensitive to $R$ at a low beam energy, is at lower $y$ for a higher beam energy.  In the case of the averaged NMC data, correlated systematic uncertainties are unavailable, so the MSTW08 fit simply added errors (other than normalisation) in quadrature similar to the simple $\chi^2$ form of eq.~\eqref{eq:chisqtot}.  As with the Tevatron jet data, deficiencies in the theory calculation may be hidden, without much trace, by large systematic shifts implicit in the $\chi^2$ definition, eq.~\eqref{eq:chisqcov}, similar to that used in the ABKM09 analysis.  We conclude that the greater sensitivity to the treatment of NMC data found by ABM~\cite{Alekhin:2011ey} is due to a variety of reasons, but perhaps most significantly, the inclusion of higher-twist corrections due to the weaker cuts on DIS data, and, as we have repeatedly emphasised, the lack of additional constraints provided by the Tevatron jet data to pin down the high-$x$ gluon distribution.

\section{Conclusions} \label{sec:conclusions}

The anomalously large higher-order QCD corrections to Higgs production at the Tevatron and LHC, via the dominant production channel of gluon--gluon fusion through a top-quark loop, mandate the use of (at least) NNLO calculations, together with corresponding NNLO PDFs and $\alpha_S$ values.  The Tevatron Higgs cross section, in particular, requires knowledge of the gluon distribution at large $x\gtrsim 0.1$ where constraints from DIS or Drell--Yan data are weak and the only direct constraint comes from Tevatron inclusive jet production.  The MSTW08 fit~\cite{Martin:2009iq} is currently the only public NNLO PDF set including the Tevatron jet data, and is used in the analyses of the Tevatron~\cite{CDF:2010ar,CDF:2011gs} and LHC~\cite{LHCHiggsCrossSectionWorkingGroup:2011ti} experiments, while other NNLO PDF fitting groups (ABKM09~\cite{Alekhin:2009ni}, JR09~\cite{JimenezDelgado:2008hf,JimenezDelgado:2009tv}, HERAPDF1.0~\cite{HERA:2009wt}) choose to omit it, finding quite different results for the predicted Higgs cross sections.  This common choice to use only the MSTW08 set, and not the other publicly available NNLO PDF sets, has faced a barrage of recent criticism~\cite{Baglio:2010um,Baglio:2010ae,Baglio:2011wn,Alekhin:2010dd,Alekhin:2011ey,Alekhin:2011cf}, which we have responded to in detail in this paper.  We summarise our main findings below:
\begin{itemize}
  \item We do \emph{not} recommend that the (experimental) PDF+$\alpha_S$ uncertainty be supplemented with an additional \emph{theoretical} uncertainty on $\alpha_S$ when calculating uncertainties on predicted cross sections, contrary to the approach taken in refs.~\cite{Baglio:2010um,Baglio:2010ae}.
  \item The claim~\cite{Baglio:2011wn} that the HERAPDF1.0 NNLO set with $\alpha_S(M_Z^2)=0.1145$ lowers the Higgs cross section compared to MSTW08 by $\approx40\%$ for $M_H\approx 160$~GeV at the Tevatron is due to a mistake in the calculation, and therefore the conclusions in the published version of ref.~\cite{Baglio:2011wn} are flawed.  On the other hand, the observed 25\% reduction with the central value of ABKM09 is still a serious problem and we give evidence in this paper that the ABKM09 set is not consistent enough with existing Tevatron data to be used for the calculation of Higgs cross sections.
  \item Comparison with Tevatron jet data is subtle because of the large correlated systematic uncertainties and the need to make choices in luminosity which are consistent with the predictions for $W$ and $Z$ cross sections.  The traditional $\chi^2$ definition in terms of the experimental covariance matrix, eq.~\eqref{eq:chisqcov}, can hide large systematic shifts, which can compensate for inadequacies in the theory calculation.  In particular, we find that the Tevatron jet data need to be normalised downwards by typically between 3-$\sigma$ and 5-$\sigma$ to achieve the best agreement with the ABKM09 (and some HERAPDF) predictions; see appendix~\ref{sec:freenorm}.  Even if the luminosity shift is artificially constrained, the other systematic shifts move by large amounts for the inclusive jet data, incompatible with the Gaussian expectation.  No such problems are observed for the MSTW08 predictions and good agreement is found with all Run II inclusive jet data, and also with the dijet data if taking a larger scale choice than for the inclusive jet data.
  \item We have demonstrated that the MSTW08 fit is stable to the treatment of NMC $F_2$ data, unlike the ABKM09 fit~\cite{Alekhin:2011ey}, most likely because of the averaging over muon beam energies, because the Tevatron jet data pin down the high-$x$ gluon distribution, and also due to the stronger cuts reducing the need for large higher-twist corrections.  Moreover, the MSTW08 NNLO determination of the strong coupling $\alpha_S$ is compatible with the world average value, unlike other NNLO determinations shown in figure~\ref{fig:asmzvalues}(b).
\end{itemize}
We conclude that the current Tevatron Higgs exclusion bounds~\cite{CDF:2010ar,CDF:2011gs} are robust, at least with respect to the treatment of PDFs and $\alpha_S$ in the calculation of the Higgs cross section.  Similar remarks hold for the Higgs cross sections at the LHC recently calculated in ref.~\cite{LHCHiggsCrossSectionWorkingGroup:2011ti}.

\appendix
\section{\texorpdfstring{Appendix: $\chi^2$ tables with unrestricted}{Unrestricted} luminosity shifts} \label{sec:freenorm}

For completeness, in tables~\ref{tab:cdfkt_freenorm}, \ref{tab:cdfmid_freenorm}, \ref{tab:d0incl_freenorm} and \ref{tab:d0dijet_freenorm} we show $\chi^2/N_{\rm pts.}$ values \emph{without} the restriction in the luminosity shifts of $|r_{\rm lumi.}|\le 1$ imposed in the main tables given in section~\ref{sec:tevjets}.  Recall from eq.~\eqref{eq:datashift} that a positive value of $r_{\rm lumi.}$ means a downwards shift in the luminosity, so we choose to give in brackets the values of ``$-r_{\rm lumi.}$'', i.e.~negative numbers correspond to downwards shifts in the luminosity.  In the table captions we give the $\chi^2$ values with an identically zero theory prediction ($T_i\equiv 0$) just to illustrate an extreme case of how large downwards luminosity shifts can partially accommodate an inadequate theory prediction.

\begin{table}
  \centering
  \begin{tabular}{|l|l|l|l|}
    \hline
    NLO PDF (with NLO $\hat{\sigma}$) & $\mu=p_T/2$ & $\mu=p_T$ & $\mu=2p_T$ \\
    \hline
    MRST04 & 1.05 ({\it$+$1.25}) & 0.94 ($+$0.02) & 0.77 ({\it$-$1.83}) \\
    MSTW08 & 0.75 ($+$0.32) & 0.68 ($-$0.88) & 0.63 ({\it$-$2.69}) \\
    CTEQ6.6 & 1.03 ({\it$-$2.47}) & 1.04 ({\bf$-$3.49}) & 0.99 ({\bf$-$4.75}) \\
    CT10 & 0.99 ({\it$-$1.64}) & 0.92 ({\it$-$2.69}) & 0.86 ({\bf$-$4.10}) \\
    NNPDF2.1 & 0.74 ($-$0.33) & 0.79 ({\it$-$1.60}) & 0.80 ({\bf$-$3.12}) \\
    HERAPDF1.0 & 1.52 ({\bf$-$4.07}) & 1.57 ({\bf$-$5.21}) & 1.43 ({\bf$-$6.22}) \\
    HERAPDF1.5 & 1.48 ({\bf$-$3.85}) & 1.52 ({\bf$-$5.00}) & 1.39 ({\bf$-$6.03}) \\
    ABKM09 & 1.03 ({\bf$-$3.49}) & 1.01 ({\bf$-$4.53}) & 1.05 ({\bf$-$5.80}) \\
    GJR08 & 1.14 ({\it$+$2.47}) & 0.93 ({\it$+$1.25}) & 0.79 ($-$0.50) \\
    \hline \multicolumn{4}{c}{} \\ \hline
    NNLO PDF (with NLO+2-loop $\hat{\sigma}$) & $\mu=p_T/2$ & $\mu=p_T$ & $\mu=2p_T$ \\
    \hline
    MRST06 & 2.80 ({\it$+$2.23}) & 1.20 ({\it$+$1.34}) & 1.03 ($+$0.53) \\
    MSTW08 & 1.39 ($+$0.35) & 0.69 ($-$0.45) & 0.97 ({\it$-$1.30}) \\
    HERAPDF1.0, $\alpha_S(M_Z^2)=0.1145$ & 2.37 ({\it$-$2.65}) & 1.48 ({\bf$-$3.64}) & 1.29 ({\bf$-$4.12}) \\
    HERAPDF1.0, $\alpha_S(M_Z^2)=0.1176$ & 2.24 ($-$0.48) & 1.13 ({\it$-$1.60}) & 1.09 ({\it$-$2.23}) \\
    ABKM09 & 1.53 ({\bf$-$4.27}) & 1.23 ({\bf$-$5.05}) & 1.44 ({\bf$-$5.65}) \\
    JR09 & 0.75 ($+$0.13) & 1.26 ($-$0.61) & 2.20 ({\it$-$1.22}) \\
    \hline
  \end{tabular}
  \caption{Values of $\chi^2/N_{\rm pts.}$ for the CDF Run II inclusive jet data using the $k_T$ jet algorithm~\cite{Abulencia:2007ez} with $N_{\rm pts.}=76$ and $N_{\rm corr.}=17$, for different PDF sets and different scale choices $\mu_R=\mu_F=\mu=\{p_T/2,p_T,2p_T\}$.  The $\chi^2$ values are calculated accounting for all 17 sources of correlated systematic uncertainty, using eq.~\eqref{eq:chisqcorr}, including the 5.8\% normalisation uncertainty due to the luminosity determination.  No restriction is imposed on the shift in normalisation and the optimal value of ``$-r_{\rm lumi.}$'' is shown in brackets, where the data points are shifted as $D_i\to D_i(1-0.058\,r_{\rm lumi.})$; see eq.~\eqref{eq:datashift}.  Values of $|r_{\rm lumi.}|\in[1,3]$ are shown in \emph{italics} and values $|r_{\rm lumi.}|>3$ are shown in \textbf{bold}.  If the theory prediction was identically zero, then $\chi^2/N_{\rm pts.}=3.43$ with $r_{\rm lumi.}=15.1$.\label{tab:cdfkt_freenorm}}
\end{table}
\begin{table}
  \centering
  \begin{tabular}{|l|l|l|l|}
    \hline
    NLO PDF (with NLO $\hat{\sigma}$) & $\mu=p_T/2$ & $\mu=p_T$ & $\mu=2p_T$ \\
    \hline
    MRST04 & 2.14 ({\it$+$1.40}) & 2.01 ($+$0.02) & 1.57 ({\it$-$1.38}) \\
    MSTW08 & 1.52 ({\it$+$1.05}) & 1.40 ($-$0.31) & 1.15 ({\it$-$1.74}) \\
    CTEQ6.6 & 1.93 ({\it$-$1.46}) & 1.90 ({\it$-$2.50}) & 1.58 ({\bf$-$3.41}) \\
    CT10 & 1.75 ($-$0.63) & 1.67 ({\it$-$1.76}) & 1.39 ({\it$-$2.82}) \\
    NNPDF2.1 & 1.69 ($+$0.30) & 1.56 ({\it$-$1.01}) & 1.40 ({\it$-$2.20}) \\
    HERAPDF1.0 & 2.49 ({\it$-$2.84}) & 2.45 ({\bf$-$3.86}) & 2.11 ({\bf$-$4.54}) \\
    HERAPDF1.5 & 2.39 ({\it$-$2.68}) & 2.36 ({\bf$-$3.72}) & 2.05 ({\bf$-$4.42}) \\
    ABKM09 & 1.52 ({\it$-$2.05}) & 1.53 ({\bf$-$3.10}) & 1.38 ({\bf$-$4.04}) \\
    GJR08 & 2.02 ({\it$+$2.60}) & 1.75 ({\it$+$1.18}) & 1.52 ($-$0.26) \\
    \hline \multicolumn{4}{c}{} \\ \hline
    NNLO PDF (with NLO+2-loop $\hat{\sigma}$) & $\mu=p_T/2$ & $\mu=p_T$ & $\mu=2p_T$ \\
    \hline
    MRST06 & 2.72 ({\it$+$2.83}) & 2.07 ({\it$+$1.14}) & 2.11 ($+$0.12) \\
    MSTW08 & 1.66 ({\it$+$1.54}) & 1.39 ($+$0.06) & 1.62 ($-$1.00) \\
    HERAPDF1.0, $\alpha_S(M_Z^2)=0.1145$ & 2.20 ({\it$-$1.15}) & 1.99 ({\it$-$2.45}) & 2.04 ({\bf$-$3.06}) \\
    HERAPDF1.0, $\alpha_S(M_Z^2)=0.1176$ & 2.08 ($+$0.63) & 1.76 ($-$0.97) & 1.96 ({\it$-$1.78}) \\
    ABKM09 & 1.63 ({\it$-$2.42}) & 1.73 ({\bf$-$3.50}) & 1.93 ({\bf$-$4.15}) \\
    JR09 & 1.57 ($+$0.87) & 2.05 ($-$0.55) & 2.81 ({\it$-$1.44}) \\
    \hline
  \end{tabular}
  \caption{Values of $\chi^2/N_{\rm pts.}$ for the CDF Run II inclusive jet data using the cone-based Midpoint jet algorithm~\cite{Aaltonen:2008eq} with $N_{\rm pts.}=72$ and $N_{\rm corr.}=25$, for different PDF sets and different scale choices $\mu_R=\mu_F=\mu=\{p_T/2,p_T,2p_T\}$.  The $\chi^2$ values are calculated accounting for all 25 sources of correlated systematic uncertainty, using eq.~\eqref{eq:chisqcorr}, including the 5.8\% normalisation uncertainty due to the luminosity determination.  No restriction is imposed on the shift in normalisation and the optimal value of ``$-r_{\rm lumi.}$'' is shown in brackets, where the data points are shifted as $D_i\to D_i(1-0.058\,r_{\rm lumi.})$; see eq.~\eqref{eq:datashift}.  Values of $|r_{\rm lumi.}|\in[1,3]$ are shown in \emph{italics} and values $|r_{\rm lumi.}|>3$ are shown in \textbf{bold}.  If the theory prediction was identically zero, then $\chi^2/N_{\rm pts.}=2.44$ with $r_{\rm lumi.}=10.2$.\label{tab:cdfmid_freenorm}}
\end{table}
\begin{table}
  \centering
  \begin{tabular}{|l|l|l|l|}
    \hline
    NLO PDF (with NLO $\hat{\sigma}$) & $\mu=p_T/2$ & $\mu=p_T$ & $\mu=2p_T$ \\
    \hline
    MRST04 & 1.76 ({\it$+$1.58}) & 1.34 ($-$0.09) & 0.98 ({\it$-$1.77}) \\
    MSTW08 & 1.40 ({\it$+$1.05}) & 1.08 ($-$0.55) & 0.85 ({\it$-$2.25}) \\
    CTEQ6.6 & 1.52 ({\it$-$1.61}) & 1.25 ({\it$-$2.88}) & 1.01 ({\bf$-$4.02}) \\
    CT10 & 1.39 ($-$0.66) & 1.11 ({\it$-$2.02}) & 0.90 ({\bf$-$3.35}) \\
    NNPDF2.1 & 1.41 ($+$0.37) & 1.23 ({\it$-$1.22}) & 0.95 ({\it$-$2.67}) \\
    HERAPDF1.0 & 1.55 ({\it$-$2.16}) & 1.38 ({\bf$-$3.51}) & 1.07 ({\bf$-$4.52}) \\
    HERAPDF1.5 & 1.63 ({\it$-$1.98}) & 1.45 ({\bf$-$3.35}) & 1.12 ({\bf$-$4.40}) \\
    ABKM09 & 1.25 ({\it$-$1.90}) & 1.04 ({\bf$-$3.20}) & 0.89 ({\bf$-$4.44}) \\
    GJR08 & 1.72 ({\it$+$2.14}) & 1.34 ($+$0.53) & 0.98 ({\it$-$1.05}) \\
    \hline \multicolumn{4}{c}{} \\ \hline
    NNLO PDF (with NLO+2-loop $\hat{\sigma}$) & $\mu=p_T/2$ & $\mu=p_T$ & $\mu=2p_T$ \\
    \hline
    MRST06 & 2.92 ({\it$+$2.66}) & 1.70 ({\it$+$1.31}) & 1.25 ($+$0.44) \\
    MSTW08 & 1.87 ({\it$+$1.34}) & 1.23 ($+$0.09) & 1.08 ($-$0.87) \\
    HERAPDF1.0, $\alpha_S(M_Z^2)=0.1145$ & 2.11 ($-$0.82) & 1.52 ({\it$-$2.03}) & 1.14 ({\it$-$2.61}) \\
    HERAPDF1.0, $\alpha_S(M_Z^2)=0.1176$ & 2.28 ($+$0.94) & 1.50 ($-$0.49) & 1.11 ({\it$-$1.23}) \\
    ABKM09 & 1.48 ({\it$-$2.33}) & 1.13 ({\bf$-$3.35}) & 1.02 ({\bf$-$4.03}) \\
    JR09 & 1.84 ($+$0.63) & 1.61 ($-$0.60) & 1.50 ({\it$-$1.35}) \\
    \hline
  \end{tabular}
  \caption{Values of $\chi^2/N_{\rm pts.}$ for the D{\O} Run II inclusive jet data using a cone jet algorithm~\cite{Abazov:2008hua} with $N_{\rm pts.}=110$ and $N_{\rm corr.}=23$, for different PDF sets and different scale choices $\mu_R=\mu_F=\mu=\{p_T/2,p_T,2p_T\}$.  The 2-loop threshold corrections are included only for the NNLO PDFs.  The $\chi^2$ values are calculated accounting for all 23 sources of correlated systematic uncertainty, using eq.~\eqref{eq:chisqcorr}, including the 6.1\% normalisation uncertainty due to the luminosity determination.  No restriction is imposed on the shift in normalisation and the optimal value of ``$-r_{\rm lumi.}$'' is shown in brackets, where the data points are shifted as $D_i\to D_i(1-0.061\,r_{\rm lumi.})$; see eq.~\eqref{eq:datashift}.  Values of $|r_{\rm lumi.}|\in[1,3]$ are shown in \emph{italics} and values $|r_{\rm lumi.}|>3$ are shown in \textbf{bold}.  If the theory prediction was identically zero, then $\chi^2/N_{\rm pts.}=1.84$ with $r_{\rm lumi.}=12.3$.\label{tab:d0incl_freenorm}}
\end{table}

\begin{table}
  \centering
  \begin{tabular}{|l|l|l|l|}
    \hline
    NLO PDF (with NLO $\hat{\sigma}$) & $\mu=p_T/2$ & $\mu=p_T$ & $\mu=2p_T$ \\
    \hline
    MRST04 & 5.79 ({\bf$+$3.58}) & 4.52 ({\it$+$1.73}) & 2.75 ($+$0.01) \\
    MSTW08 & 3.03 ({\it$+$2.82}) & 2.25 ({\it$+$1.11}) & 1.56 ($-$0.61) \\
    CTEQ6.6 & 5.41 ($+$0.62) & 4.85 ($-$0.92) & 3.31 ({\it$-$2.14}) \\
    CT10 & 4.74 ({\it$+$1.33}) & 4.06 ($-$0.24) & 2.68 ({\it$-$1.62}) \\
    NNPDF2.1 & 2.59 ({\it$+$2.41}) & 1.93 ($+$0.79) & 1.47 ($-$0.75) \\
    HERAPDF1.0 & 2.05 ($-$0.66) & 2.17 ({\it$-$2.06}) & 1.94 ({\bf$-$3.11}) \\
    HERAPDF1.5 & 1.90 ($-$0.46) & 1.97 ({\it$-$1.85}) & 1.73 ({\it$-$2.93}) \\
    ABKM09 & 1.49 ($-$0.17) & 1.39 ({\it$-$1.58}) & 1.21 ({\it$-$2.83}) \\
    GJR08 & 10.2 ({\bf$+$4.48}) & 7.79 ({\it$+$2.78}) & 5.30 ({\it$+$1.07}) \\
    \hline \multicolumn{4}{c}{} \\ \hline
    NNLO PDF (with NLO $\hat{\sigma}$) & $\mu=p_T/2$ & $\mu=p_T$ & $\mu=2p_T$ \\
    \hline
    MRST06 & 7.87 ({\bf$+$3.19}) & 6.55 ({\it$+$1.42}) & 4.07 ($-$0.13) \\
    MSTW08 & 2.34 ({\it$+$1.96}) & 1.80 ($+$0.42) & 1.31 ({\it$-$1.12}) \\
    HERAPDF1.0, $\alpha_S(M_Z^2)=0.1145$ & 2.61 ($-$0.09) & 2.55 ({\it$-$1.44}) & 2.31 ({\it$-$2.51}) \\
    HERAPDF1.0, $\alpha_S(M_Z^2)=0.1176$ & 2.71 ({\it$+$1.53}) & 2.31 ($-$0.04) & 1.95 ({\it$-$1.37}) \\
    ABKM09 & 1.35 ({\it$-$1.36}) & 1.40 ({\it$-$2.56}) & 1.31 ({\bf$-$3.60}) \\
    JR09 & 3.27 ({\it$+$1.75}) & 2.55 ($+$0.33) & 1.88 ({\it$-$1.08}) \\
    \hline
  \end{tabular}
  \caption{Values of $\chi^2/N_{\rm pts.}$ for the D{\O} dijet data using a cone jet algorithm~\cite{Abazov:2010fr} with $N_{\rm pts.}=71$ and $N_{\rm corr.}=70$, for different NLO PDF sets and different scale choices $\mu_R=\mu_F=\mu=\{p_T/2,p_T,2p_T\}$, where $p_T\equiv(p_{T1}+p_{T2})/2$.  Only NLO partonic cross sections are used with the NNLO PDFs, since the 2-loop threshold corrections are only available for the inclusive jet cross section.  The $\chi^2$ values are calculated accounting for all 70 sources of correlated systematic uncertainty, using eq.~\eqref{eq:chisqcorr}, including the 6.1\% normalisation uncertainty due to the luminosity determination.  No restriction is imposed on the shift in normalisation and the optimal value of ``$-r_{\rm lumi.}$'' is shown in brackets, where the data points are shifted as $D_i\to D_i(1-0.061\,r_{\rm lumi.})$; see eq.~\eqref{eq:datashift}.  Values of $|r_{\rm lumi.}|\in[1,3]$ are shown in \emph{italics} and values $|r_{\rm lumi.}|>3$ are shown in \textbf{bold}.  If the theory prediction was identically zero, then $\chi^2/N_{\rm pts.}=2.41$ with $r_{\rm lumi.}=10.4$.\label{tab:d0dijet_freenorm}}
\end{table}

\acknowledgments

We thank A.~D.~Martin and W.~J.~Stirling for numerous discussions, and S.~Alekhin, G.~Hesketh, D.~Waters and M.~Wobisch for useful information.  The work of R.S.T.~is supported partly by the London Centre for Terauniverse Studies (LCTS), using funding from the European Research Council via the Advanced Investigator Grant 267352.

\bibliographystyle{JHEP}
\bibliography{bench7TeV}

\end{document}